%% file: thesis.tex
\documentclass[12pt,oneside]{book}
\include{macros/style}

\include{macros/use_packages}

\usepackage{natbib}
\usepackage{aastex_hack}
\usepackage{deluxetable}
\usepackage[linktoc=all]{hyperref}

\usepackage{booktabs}

\usepackage{amssymb}
\makeatletter
\renewcommand*\l@figure{\@dottedtocline{1}{1.5em}{2.5em}}
\makeatother

\newcommand{\degree}{\ensuremath{^\circ}}
\newcommand{\icarus}{Icarus}

\begin{document}

% Front Matter
\newcommand\thesistitle{From Exoplanets to Quasars: Adventures in Angular Differential Imaging}

\newcommand\nameanddegrees{%
Mara Johnson-Groh \\
B.A., Gustavus Adolphus College, 2014 }
\newcommand\panel{%
\HRule\\\panelist{Dr. Christian Marois}{Co-Supervisor}{Department of Physics \& Astronomy}
\HRule\\\panelist{Dr. Sara Ellison}{Co-Supervisor}{Department of Physics \& Astronomy}
%\HRule\\\panelist{Dr. Andrew Skemer}{Outside Member}{Department of Astronomy \& Astrophysics, University of California Santa Cruz}}
%\HRule\\\panelist{Dr. Outside Member}{Outside Member}{Department of Not Same As Candidate}}
%\HRule\\\panelist{Dr. \Ans{Current Unknown}}{Additional Member}{Department of \Ans{Current Unknown Department}}}
% Titlepage break
}
\newcommand\tpbreak{\\[\baselineskip]}

\newpage
% Suppress numbers on the first page
\thispagestyle{empty}

% Header Footer Setup (for Front matter)
\pagestyle{myheadings}
\pagenumbering{roman}
\fancypagestyle{plain}{%
\fancyhf{}
\fancyhead[R]{\thepage}
\renewcommand{\headrulewidth}{0pt}
\renewcommand{\footrulewidth}{0pt}
}

% Front Matter
\pagebreak
{
\centering
\thesistitle
\tpbreak
by
\tpbreak
\nameanddegrees
\tpbreak
% One needs to make adjustments for a Master's Thesis
A Thesis Submitted in Partial Fulfilment of the \\
Requirements for the Degree of
\tpbreak
MASTER OF SCIENCE
\tpbreak
in the Department of Physics \& Astronomy\\
\vfill
\begin{tabular}{cc}
& \copyright\ Mara Johnson-Groh, 2016\\
& \phantom{\copyright} University of Victoria
\end{tabular}
%\begin{center}
%\copyright\ Hannah Broekhoven-Fiene, 2011\\
%\phantom{\copyright} University of Victoria
%\end{center}
\tpbreak
All rights reserved. This thesis may not be reproduced in whole or in part, by \\
\hfill photocopying or other means, without the permission of the author. 
\hfill
}
\pagebreak

\newpage
\TOCadd{Supervisory Committee}

{
\centering
\thesistitle
\tpbreak
by
\tpbreak
\nameanddegrees
\tpbreak
}

\newcommand\panelist[3]{\noindent #1, #2\\\noindent(#3)\tpbreak}
\vfill
\noindent Supervisory Committee
\tpbreak
\panel
\vfill
\pagebreak

\newpage
\TOCadd{Abstract}

%\noindent \textbf{Supervisory Committee}
%\tpbreak
%\panel

\begin{center}
\textbf{\huge Abstract}
\end{center}
%\phantom{wtf}

Angular differential imaging provides a novel way of probing the high contrast of our universe.  Until now, its applications have been primarily localized to searching for exoplanets around nearby stars.  This work presents a suite of applications of angular differential imaging from the theoretical underpinning of data reduction, to its use characterizing substellar objects to a new application looking for the host galaxies of damped Lyman $\alpha$ systems which are usually lost in the glare of ultra-bright quasars along the line of sight.  

The search for exoplanets utilizes angular differential imaging and relies on complex algorithms to remove residual speckles and artifacts in the images.  One such algorithm, the Template Locally Optimized Combination of Images (TLOCI), uses a least-squares method to maximize the signal-to-noise ratio and can be used with variable parameters, such as an input spectral template, matrix inversion method, aggressivity and unsharp mask size.  Given the large volume of image sequences that need to be processed in any exoplanet survey, it is important to find a small set of parameters that can maximize detections for any conditions.  Rigorous testing of these parameters were done with on-sky images and model inserted planets to find the optimal combination of parameters.  Overall, a standard matrix inversion, along with two to three input templates, a modest aggressivity of 0.7 and the smallest unsharp mask was found to be the best choice to balance optimal detection.

Beyond optimizations, TLOCI has been used in conjunction with angular differential imaging to characterize substellar objects in our local solar neighbourhood.  In particular, the star HD 984 was imaged as a part of the Gemini Planet Imager Exoplanet Survey.  Although previously known to have a substellar companion, new imaging presented here in the H and J bands help further characterize this object.  Comparisons with a library of brown dwarf spectral types found a best match to HD 984 B of a type M7$\pm$2.  Orbital fitting suggests an 18~AU (70 year) orbit, with a 68\% confidence interval between 12 and 27 AU.  Object magnitude was used to find the luminosity, mass and temperature using DUSTY models.  

Although angular differential imaging has proven its value in high contrast imaging, it has largely remained in the field of substellar object detection, despite other high contrast regimes in which it could be applied.  One potential application is outside the local solar neighbourhood with studies of damped Lyman $\alpha$ systems, which have struggled to identify host galaxies thought to be caused by systems seen in the spectra of bright quasars.  Work herein presents the first application of angular differential imaging to finding the host galaxies to damped Lyman $\alpha$ systems.  Using ADI we identified three potential systems within 30kpc of the sightline of the quasar and demonstrate the potential for future imaging of galaxies at close separations.  

In summary, this thesis presents a comprehensive look at multiple aspects of high contrast angular differential imaging.  It explores optimizations with a data reduction algorithm, implementations characterizing substellar objects, and new applications imaging galaxies.

%\end{center}

\TOCadd{Contents}\tableofcontents
\TOCadd{List of Tables}\listoftables
\setcounter{lofdepth}{2}
\TOCadd{List of Figures}\listoffigures

\newpage
\TOCadd{Acknowledgements}

\phantom{wtf}

\begin{center}
\textbf{\huge Acknowledgements}
\end{center}

Much thanks goes to Christian Marois and Sara Ellison for their supervision and support.  Research within is based on observations obtained at the Gemini Observatory, which is operated by the Association of Universities for Research in Astronomy, Inc., under a cooperative agreement with the NSF on behalf of the Gemini partnership: the National Science Foundation (United States), the National Research Council (Canada), CONICYT (Chile), Ministerio de Ciencia, Tecnolog\'{i}a e Innovaci\'{o}n Productiva (Argentina), and Minist\'{e}rio da Ci\^{e}ncia, Tecnologia e Inova\c{c}\~{a}o (Brazil). We acknowledge and respect the native peoples on whose traditional territories the Gemini North Telescope stands and whose historical relationships with the land and Mauna a W\={a}kea (Mauna Kea) continue to this day.  We further acknowledge and respect the \underline{W}S'ANEC', Songhees, and Esquimalt peoples of the Coast Salish Nation on whose traditional territories the University of Victoria stands.  Additional thanks to the comments from Luc Simard, Lise Christensen, Nissim Kanekar, Jon Willis, and Karun Thanjavur, and help from Eric Nielsen. 
\\

%\begin{verse}
%In starry skies, long years ago, \\
%I found my Science; heart aglow \\
%I watched each night unfold a maze\\ 
%Of mystic suns and worlds ablaze, \\
%That spoke: "Know us and wiser grow."\\
%\end{verse}
%\attrib{Sterling Bunch, \textit{In Starry Skies}}

%----------------------------------------------------------------------------------------------------------------------

%\phantom{wtf}

%\noindent I would like to personally thank:
%\begin{description}
%\item[Christian Marois and Sara Ellison]
%\item[Brenda Matthews,]
	%for their supervision and support.

%\item[The UVic astrograds and my friends,]
	%for providing much needed relief and comraderie.

%\item[Andy Pon, James DiFrancesco, Kaushi Bandara, Lisa Glass, Stephen Gwyn and Ben Hendricks]
	%for the many drives up and down the observatory hill.

%\item[Canadian Space Agency,]
	%for funding DEBRIS research.

%\end{description}

%\begin{flushright}
%\textit{I believe I know the only cure, which is to make
%one's centre of life inside of one's self, not
%selfishly or excludingly, but with a kind of
%unassailable serenity-to decorate one's inner house
%so richly that one is content there, glad to welcome
%any one who wants to come and stay, but happy all
%the same in the hours when one is inevitably alone.}
%\\
%Edith Wharton \\
%\end{flushright}

\newpage
\TOCadd{Dedication}

\phantom{wtf}

\begin{center}
\textbf{\huge Dedication}
\end{center}

\begin{center}
To the Giants' shoulders
\end{center}

\begin{center}
%Just hoping this is useful!

\end{center}

%\input frontmatter/co

% Header Footer Setup
\newpage
\pagestyle{myheadings}
\pagenumbering{arabic}
\fancypagestyle{plain}{%
\fancyhf{}
\fancyhead[R]{\ifnum\thepage=1\relax\else\thepage\fi}
\renewcommand{\headrulewidth}{0pt}
\renewcommand{\footrulewidth}{0pt}
}

\newpage

\chapter{Introduction}
For millennia, humankind has observed the night sky, but only recently has technology progressed sufficiently to truly probe the depths of our universe.  Even still, there are many parts of it that remain elusive to see.  In fact, some of the brightest objects in our universe and our night sky are the greatest hindrances to seeing fainter companion objects.  These high contrast regions require special observational techniques and data reduction methods to resolve faint, hidden objects.  By untangling the light of these bright objects, we can see more of our universe and have a better understanding of the worlds around us.

\section{Techniques of Direct Imaging}
High contrast direct imaging is primarily associated with the field of exoplanets where scientists work to image the small, dim companions to stars.  Given the challenges of direct imaging, an entire subdivision of exoplanetary studies has emerged to overcome the difficulties.  Specialized optics systems have been designed to combat atmospheric effects and complex computer algorithms have been written to suppress image noise.  In order to appreciate the complexities of direct imaging it is necessary to understand these techniques.  

\subsection{Adaptive Optics}
Since the dawn of the telescope, astronomers have been limited by atmospheric seeing.  As light from the cosmos propagates through the Earth's turbulent atmosphere, it passes through areas of inhomogeneous index of refractions which distorts the incoming wavefront.  These distortions typically limit seeing resolutions to one arcsecond at optical wavelengths \citep{hickson14}.  In order to push to higher resolutions, a technology termed adaptive optics (AO) was developed to restore for the distorted wavefront.  

\begin{figure}[h!]
\epsscale{.55}
\plotone{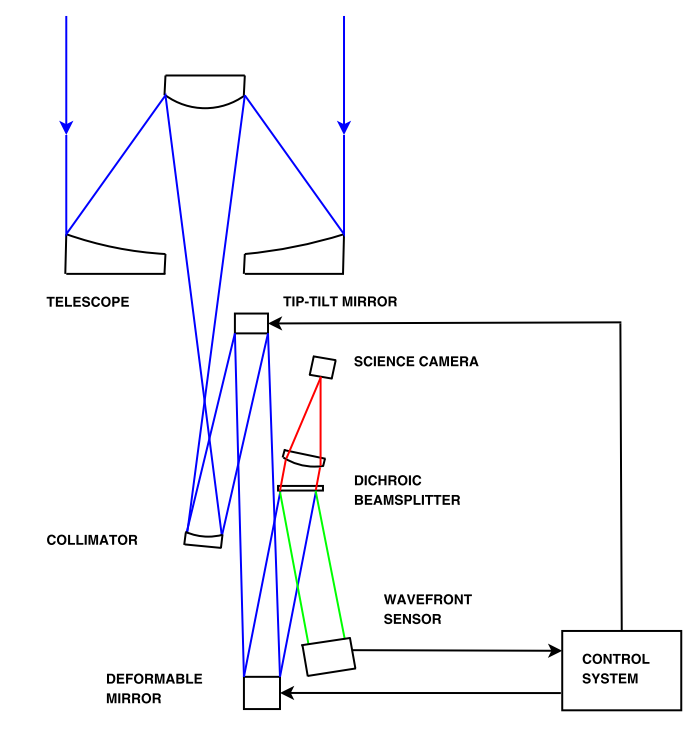}
\caption[Adaptive Optics Diagram]{Schematic of a simple AO system.  Incoming light paths shown in blue.  Light from the science object is shown in red and light from the reference star in green. Diagram from \citet{hickson14}.}
\label{ao}
\end{figure}

Compensating for atmospheric seeing was first proposed by \citet{babcock53} but it was not until the 1980s that advances in technology made AO feasible.  At a fundamental level, AO works by using a single reference star to sense the atmospheric turbulence at kHz speeds, and a deformable mirror (DM) to adjust for the wavefront's deformation caused by that turbulence.  Phase input read from the reference star is corrected by the DM which has a grid of tiny actuators to create an uneven surface.  When the wavefront from the science target hits the DM, its wavefront phase is corrected by the irregularities of the DM and a nearly flat wavefront is reflected.  Corrections from the reference star are continually applied by the DM on timescales less than the atmospheric variations allowing for continual phase corrections.  In order to have the highest success with corrections it is best to have the science object close to the reference star so that atmospheric variations along the different lines of sight will be correlated.  A schematic of a simple AO system is shown in Figure~\ref{ao}.
%---------------------------------------------------------
%---------------------------------------------------------

\begin{figure}[h!]
\epsscale{.55}
\plotone{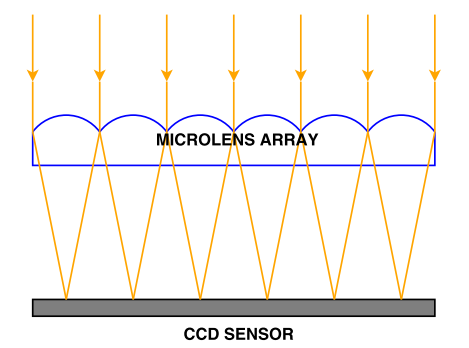}
\caption[Shack-Hartmann Wavefront Sensor]{Diagram of a Shack-Hartmann wavefront sensor. Figure from \citet{hickson14}.}
\label{sensor}
\end{figure}

To characterize the wavefront distortions, sensors have been developed to account for the minute corrections that need to be applied to the DM.  One of the most common types is the Shack-Hartmann sensor, which is composed of an array of small lenslets.  Each lenslet focuses the incoming light from the reference star onto a CCD (see Figure~\ref{sensor}).  If the wavefront is distorted, the focused image will be off-centre. The dispersion of the position of the focal spot from the centre of the lenslet is used to calculate the local wavefront tilt. The position of each focused image is corrected by the DM in a feedback loop until each imaged is centred.  
 %---------------------------------------------------------
%---------------------------------------------------------

 In some fields-of-view, there simply is not a suitable guide star for AO.  In these cases, laser guide stars (LGS) can be used to create artificial stars by reflecting light off of the mesosphere. Lasers tuned to the D$_2$ line of atomic sodium (589nm) are used to excite the sodium layer at an altitude of 90km.  The sodium layer's height is ideal since most of the atmospheric turbulence happens at lower layers. (Indeed, atmospheric models at Cerro Pach\'{o}n show 64\% of the total turbulence is caused by the ground layer \citep{hickson14}.)   Unfortunately, LGS cannot sense the tilt components of atmospheric turbulence since any deflection of the beam as it propagates upwards is negated by the opposite deflection it gets going down.  Consequentially, at least one natural guide star is always needed. 
 
\subsection{Angular Differential Imaging}\label{adi}
The contrast improved through adaptive optics can be further enhanced through various post-processing techniques.  In adaptive optics corrected images, speckles, which are caused by noise from the stellar point spread function (PSF), are the largest inhibitor to planet detection in imaging \citep{marois03}.  Speckles are caused by rapid (1--10ms) fluctuations in the atmosphere and instrumental imperfections which evolve on longer time scales.   Several methods have been developed to remove the PSF and suppress speckles, including simultaneous spectral differential imaging \citep[SSDI,][]{racine,marois00} which relies on the spectral differences between a planet and a star in polychromatic images, and angular differential imaging \citep[ADI,][]{marois06} which uses the field-of-view rotation of an altitude/azimuth telescope to distinguish speckles from planets.   

The ADI technique works by disabling the instrument rotator on an altitude/azimuth telescope so that the field-of-view (FOV) slowly rotates during a sequence of images (see Figure~\ref{adi}).  The star used by the  wavefront sensor is locked in by the AO system to provide an unmoving guide.  A reference PSF is created for each image of the target sequence of images by combining all the other images of the sequences acquired at different angles. Reference PSF images are then subtracted from each of the individual frames which are subsequently all rotated to north up and median combined (see Figure~\ref{psfsub}).  Any off-axis objects will have been averaged out of the reference PSF and will mainly remain unaffected by the process.  The reference PSF has the added benefit of subtracting off any sky flux and ghost images from internal optics.  In order for the off-axis objects to remain unchanged through the PSF subtraction, it is necessary for those objects to have a high amount of rotation (at least twice the full width at half maximum) so as to not contribute to the median reference PSF.  Because of this, it is preferable to take image sequences as the object transits the local meridian.   Given enough field-of-view rotation, these images can be combined to create a reference PSF and suppress quasi-static speckles by up to two orders of magnitude.

%----------------------------------------------------------------------------------------------------------------------
% Figure: ADI
\begin{figure}[h!]
  \centering
  \begin{tabular}[b]{@{}p{0.44\textwidth}@{}}
    \includegraphics[width=1.\linewidth]{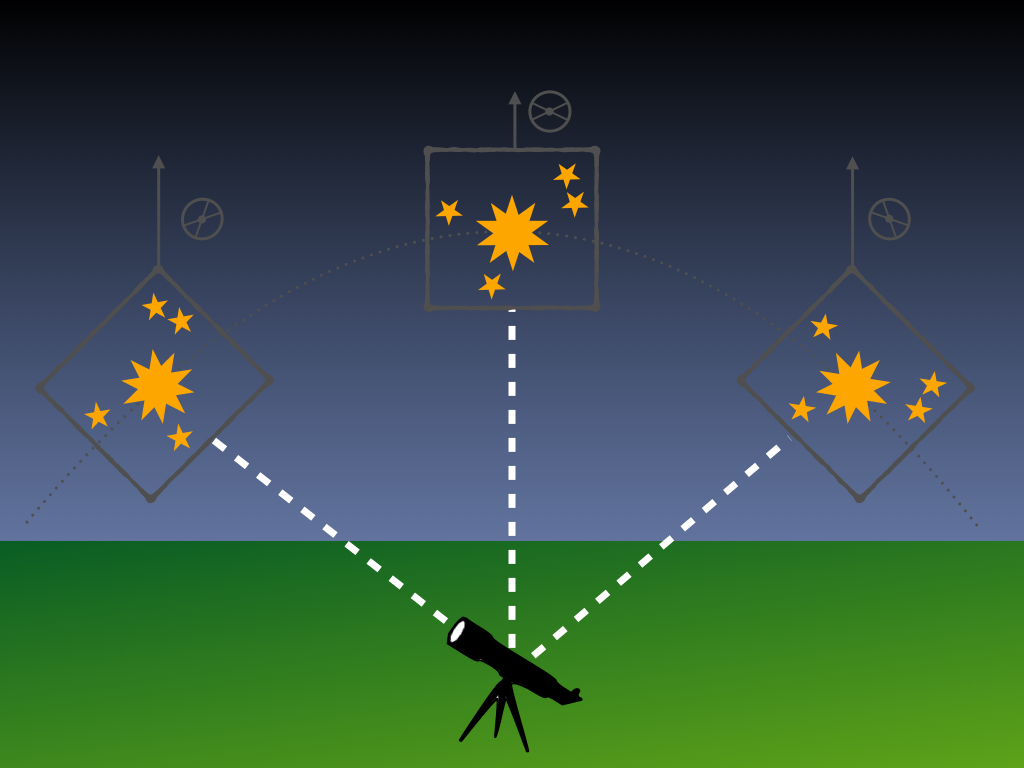} \\
    \centering\small (a)  Normal Telescope Rotation
  \end{tabular}%
  \quad
  \begin{tabular}[b]{@{}p{0.45\textwidth}@{}}
    \includegraphics[width=.978\linewidth]{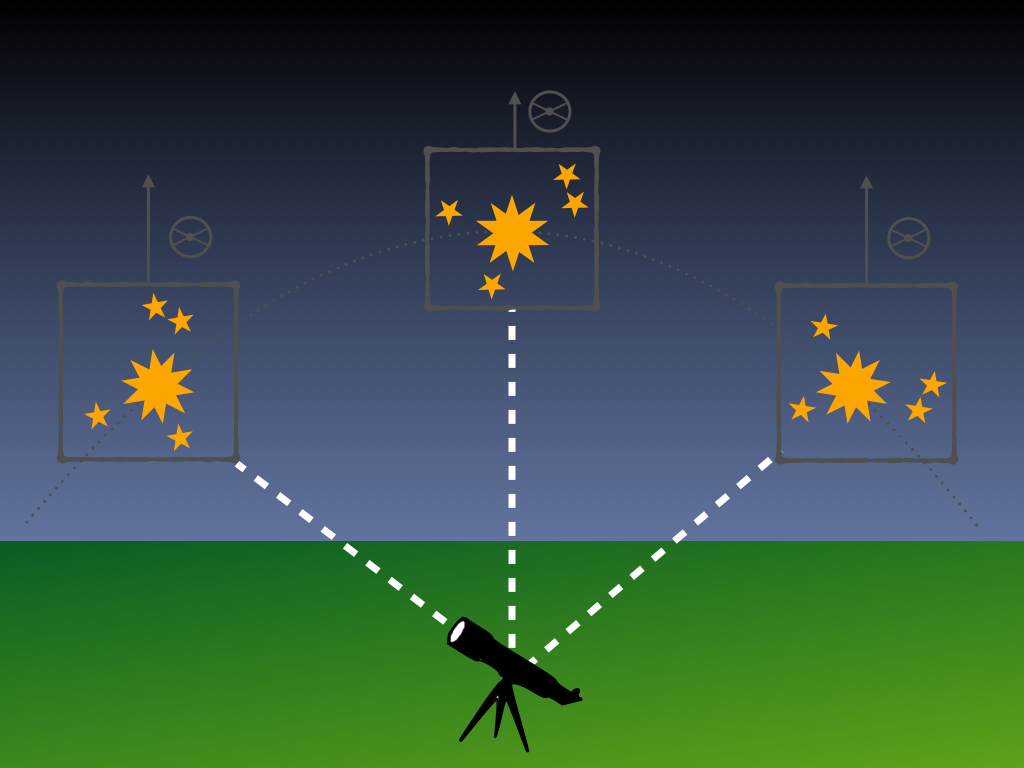} \\
    \centering\small (b) ADI without Telescope Rotation
  \end{tabular}
  \caption[Angular Differential Imaging]{In normal observation mode, the instrument rotates during the exposures so all images align (a).  In ADI, the instrument rotation is disabled so that all images are at different angles (b).The $\bigotimes$ symbol indicates instrument axis. }
  \label{adi}
\end{figure}
%----------------------------------------------------------------------------------------------------------------------
 %---------------------------------------------------------
\begin{figure}[h!]
\epsscale{.85}
\plotone{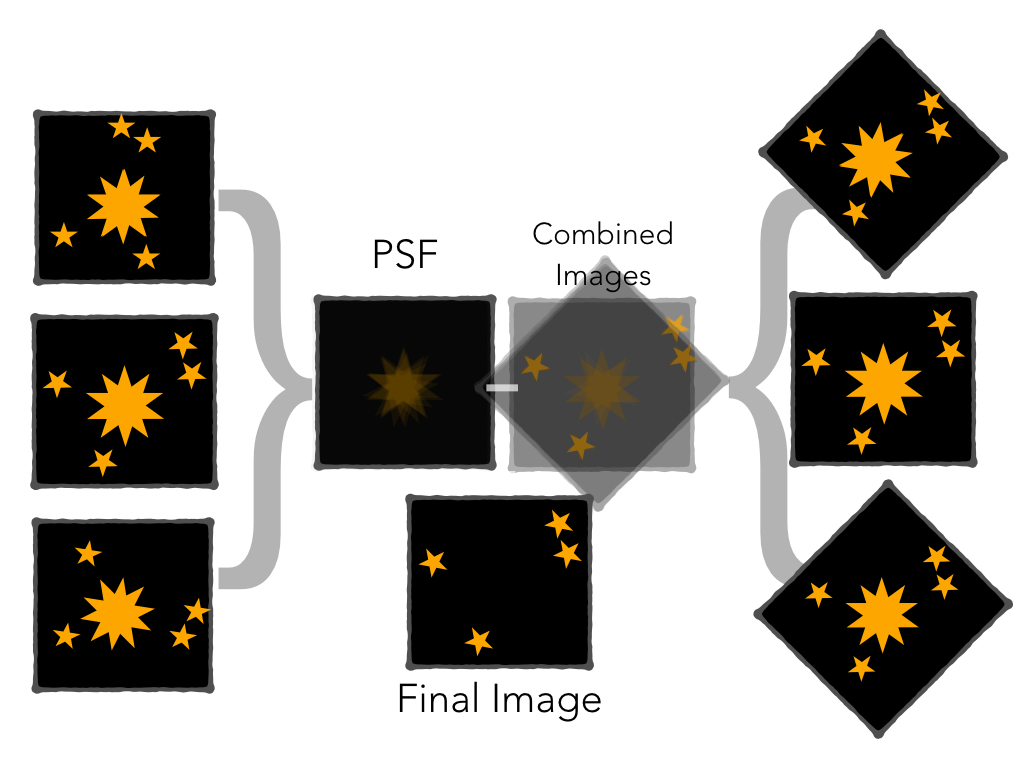}
\caption[Point Spread Function Subtraction]{A PSF can be easily created from ADI images by combining the unrotated exposures (left).  This master PSF can then be subtracted from the images that have been combined after rotating (right) to create a final image with near perfect PSF subtraction. Note that the rotation of the field objects in the images represents the rotation of the objects as you would see for resolved, extended objects (e.g. galaxies).  The PSF of these objects would not rotate. }
\label{psfsub}
\end{figure}
%---------------------------------------------------------

\subsection{Point Spread Function Subtraction Techniques}\label{psftech}
Although methods like ADI and the use of AO can greatly remove the central PSF at small angular separations and other problematic artefacts, additional data reduction techniques are required at low impact parameters.  Many advanced PSF subtraction techniques exist, such as LOCI \citep{lafreniereb}, SOSIE \citep{marois10}, KLIP  \citep{soummer}, TLOCI \citep{marois14}, ANDROMEDA \citep{cantalloube}, and LLSG \citep{gomez}. Some, such as the Locally Optimized Combination of Images \citep[LOCI,][]{lafreniereb}, have been developed with techniques like ADI in mind.  At the heart of these techniques is a least-squares minimization.  LOCI, for example, creates a series of annuli, each with a library of reference PSF images which are combined such that their subtraction minimizes speckle noise in the final image.  This minimization is achieved through the inversion of a covariance matrix created from the reference images.  %Similarly, KLIP organizes the reference images into eigenvectors for optimization.  

One variation on these least-squares methods is TLOCI, the Template LOCI \citep{marois14}.  TLOCI was developed to maximize the signal-to-noise ratio in angular and spectral differential imaging data to find exoplanets by using least-squares to combine a set of reference PSF images to subtract speckle noise, similar to LOCI.  However, TLOCI also uses an input template spectrum to maximize detection of planets having a similar spectrum.  TLOCI solves one issue affecting spectral differential imaging \citep[SDI,][]{marois00}, wherein planet flux can vary substantially over wavelength (most notably, methane absorption at 1.60 $\mu$m).  Using LOCI with SDI can result in significant planet self-subtraction at wavelengths with higher flux.  By weighting with an input template, this self-subtraction can be minimized and an optimal balance between speckle subtraction and companion self-subtraction can be found that maximizes the companion signal to noise ratio (SNR).

%----------------------------------------------------------------------------------------------------------------------

\section{Overview}
Although high contrast imaging is a challenging endeavour, the hard work of dedicated individuals has opened a door into a new realm of our universe.  As we will discover, it is not only exoplanets that high contrast imaging benefits.  We begin with Chapter~\ref{tloci} wherein the PSF subtraction algorithm TLOCI is optimized to find new exoplanets.  From these theoretical beginnings, we will then apply TLOCI to characterizing the substellar object HD 984 B in Chapter~\ref{bd}.  In Chapter~\ref{dla} we step back from the local universe to apply ADI to damped Lyman alpha systems in the pursuit of imaging galaxies hidden under the light of nearby bright quasars.  A summary of these adventures in high contrast angular differential imaging are presented in Chapter~\ref{sum}.

\newpage

\chapter{Hide and Seek: Optimizing the TLOCI Algorithm for Exoplanet Detection}\label{tloci}

Directly detecting exoplanets requires complex computer algorithms to sufficiently suppress image noise and distinguish the planet signal.  One such algorithm, TLOCI, has many parameters that can be used to enhance the SNR of planets.  In order to maximize planet detection, parameter settings need to be optimized to yield the maximum SNR.  Given the large volume of image sequences that need to be processed in any exoplanet survey, it is important to find a small set of parameters that can maximize detections for any conditions.  This chapter presents the systematic testing of various parameters like input spectrum type, matrix inversion method, aggressivity, and unsharp masking, to determine the best parameters to apply generally and efficiently when looking for new planets.

%----------------------------------------------------------------------------------------------------------------------
%----------------------------------------------------------------------------------------------------------------------

%----------------------------------------------------------------------------------------------------------------------
\section{Introduction}

The idea that we are not alone in the cosmos is not a new one.  In the late 1500s, Giordano Bruno, an Italian philosopher, wrote in his treatise \textit{On the Infinite Universes and Worlds}, ``Since it is well that this world doth exist, no less good is the existence of each one of the infinity of other worlds" \citep{bruno}.  Two centuries later, Isaac Newton purposed in his \textit{Principia} that fixed stars were the centres of systems similar to our sun's \citep{newton}.   Star-planet systems are at such high contrast and small separation that only recently has technology advanced enough for these speculations to be vindicated.  This section explores the various methods used to detect the systems these great thinkers proposed.  

\subsection{Exoplanets}\label{exoplanets}

The International Astronomical Union (IAU) defines a planet as ``a celestial body that (a) is in orbit around the Sun, (b) has sufficient mass for its self-gravity to overcome rigid body forces so that it assumes a hydrostatic equilibrium (nearly round) shape, and (c) has cleared the neighbourhood around its orbit"\footnote{\url{http://www.iau.org/static/resolutions/Resolution_GA26-5-6.pdf}}. These qualifications help distinguish planets from smaller objects (e.g. dwarf planets and satellites) and larger bodies (e.g. brown dwarfs). Exoplanets are simply planets outside of our own solar system.  

There are two main camps of planet formation: core accretion \citep{pollack96,marley07} and disk instability \citep{boss97,boss06}.    In core accretion, sub-micron-sized dust and small solid grains in a protoplanetary disk clump into centimeter-sized particles which eventually gravitationally aggregate into kilometre-sized bodies \citep{dangelo}.  These objects, or planetesimals, collide and grow larger into protoplanets.  Once the protoplanet is large enough, gases begin to accumulate in an envelope around the core. Core accreation can form planets from one to a few Myr, although it must take place before 10Myr, the  typical lifetime of a disk before the gas disperses from photoevaporation and stellar winds \citep{haisch,wyatt}.  Alternatively, disk instabilities can form planets in just one to a few disk orbital periods \citep{gammie,rice}.  In this top-down scenario, gravitational instabilities in the gas of the protoplanetary disk fragment and clumps form.  Most of the gas is accumulated immediately, with heavier elements settling to form a sedimentary core afterwards \citep{dangelo}.  Disk instability is most effective in explaining the formation of giant planets at high separations ($>100$ AU) where core accretion timescales are inefficient. 

Observations of exoplanets, particularly multi-planet architecture, are key in deciphering planet formation mechanisms.  For the giant planets found with radial velocity, population synthesis models as well as statistical analysis of planet frequency, mass and radius, all point to core accretion as the likely formation mechanism \citep{mordasini09,howard10,boruki11}.  Early results from high contrast imaging surveys indicate that core accretion is also the dominant formation scenario \citep{janson11}.  Furthermore, this group also find that few planets are found at large separations, where disk instability is likely to form giant planets and brown dwarfs.  However, there are planets in these high separation regimes \citep[e.g.][]{kalas08,marois08b} where formation by instability is a possibility, though by no means is this the only formation mechanism.  High orbital separation planets could also form via outward migration \citep{crida} or planet-planet interactions \citep{veras}.  \citet{udry} find a substantial number of multi-planet systems are in mean-motion resonances, which is likely to occur only from migrations. 

\begin{figure}[h!]
\epsscale{.51}
\plotone{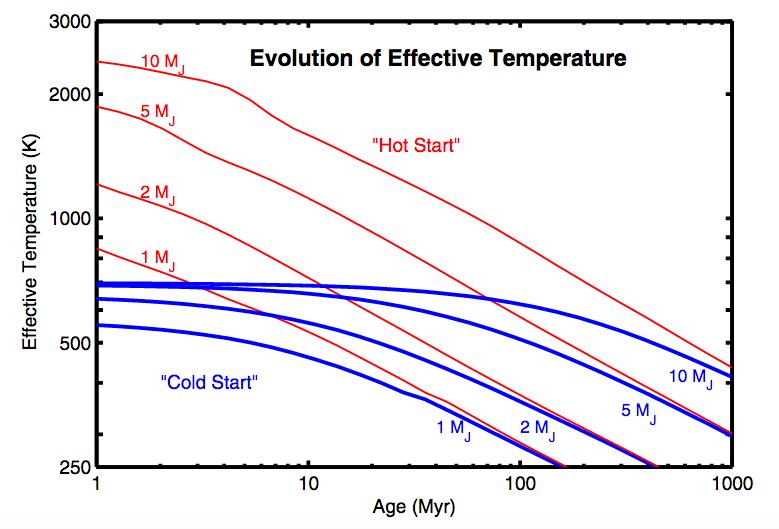}
\caption[Hot and Cold Start Temperature Evolution Tracks]{Difference in effective temperature for hot (red) and cold (blue) start models.  Young planets have substantially different initial temperature which can be used to distinguish their formation mechanism.  Figure from \citet{spiegel12}.}
\label{hotcold}
\end{figure}

Models of post-formation cooling may help further constrain initial conditions.  Core accretion, also known as a `cold-start' formation, tends to retain much less entropy from the initial disk conditions than disk instability which is also called a `hot-start', likely due to the accretion shock which forms around the protoplanet's boundary through which material must infall \citep{marley07}.  If the age of the object is well constrained, it may be possible to distinguish these hot and cold start models through observables like effective temperature, luminosity and spectrum that are each effected by the entropy \citep{spiegel12}.   Figure~\ref{hotcold} shows the time evolution differences between hot and cold start models which can be observed with effective temperature.  Direct imaging campaigns, which target young systems, can use the planet's effective temperature and luminosity to help determine the object's formation mechanism.  

% %-------------------------------------------
% \begin{figure}[b!]
%\epsscale{.51}
%\plotone{Figures/coolingtrack.png}
%\caption[Exoplanet Cooling Tracks]{Cooling tracks for hot-start planets from $1-20$ M$_{\mathrm{Jup}}$ computed in \citet{marleau}.}
%\label{hot}
%\end{figure}
%%-------------------------------------------
 %-------------------------------------------
%-------------------------------------------

\begin{figure}[b!]
\epsscale{.51}
\plotone{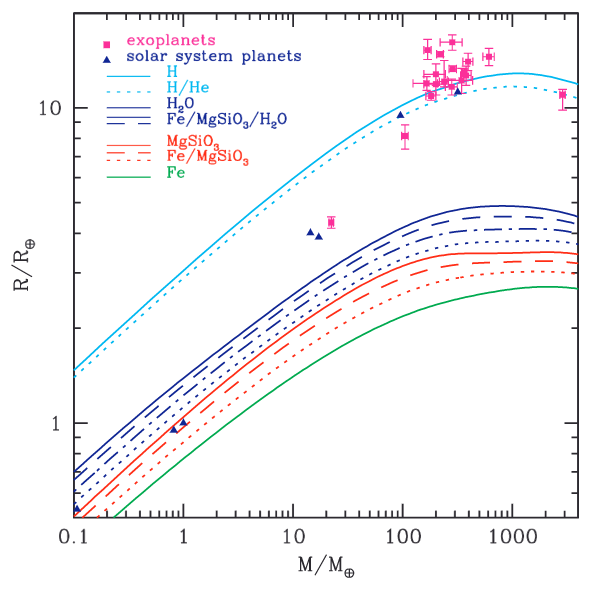}
\caption[Exoplanet Composition Models]{Compositional models for exoplanets shown via mass-radius relationship.  Soild lines indicate models of homogenous composition whereas dashed lines represent differentiated models. Known planets shown for comparison. Figure from \citet{seager}.}
\label{composition}
\end{figure}

Planets are often described as being small and rocky or gaseous giants.  The composition of rocky exoplanets is largely extrapolated from models of the solar system planets \citep{seager,zeng}.  Oxygen, iron, magnesium and silicon constitute 95\% of the Earth's mass and combined with sulphur, calcium, aluminum and nickel, they comprise 99.9\% \citep{javoy}.  Given their extreme diversity, exoplanets are believed to form from the same base elements but in a wide range of ratios \citep{seager}.  Since exoplanet composition can only be inferred through mass and radius, there is some degeneracy in models.  Models describing carbon-dominant planets can also describe water and silicon based compositions \citep{seager}.  Figure~\ref{composition} shows various compositional models for exoplanets of different radii and masses.  
 
 %-------------------------------------------
%-------------------------------------------.  

Gas giants are composed primarily of hydrogen and helium with a small amount of heavier elements \citep{spiegel}. Some giants, like Saturn, are known to have heavy-element cores, but the fraction of gas giants possessing these types of interiors is uncertain \citep{pollack77}.  However, there does seem to be a trend of more metals in planets orbiting super-solar metallicity stars \citep{guillot}.   The outer most layer, above the convective interior, is the atmosphere.  Spectra of exoplanets' atmospheres are key in understanding their gas composition, temperature profile and gravity \citep{marley15}.

 Exoplanets occupy a low temperature extension of the OBAFGKM system used to classify stars, the MK system, which adds types M, L, T and Y.  These classifications are identified using temperature and chemical signatures.  M types range $2000-3000$ K and have TiO, MgH and H$_2$O absorption lines \citep{kirkpatrick}.  At the low end from $1400-2100$ K, L types are characterized by strong alkali metal lines (\ion{K}{1}, \ion{Na}{1}, \ion{Cs}{1}) and strong metal hydride bands (FeH, CrH, MgH, CaH). T types ($1200-1400$K) show prominent methane (CH$_4$) absorption which condensates at temperatures less than 1400 K.  The Y type has been proposed to classify objects with strong ammonia (NH$_4$) features which appear at temperatures below $500-600$K.  Brown dwarfs, which have similar atmospheres and spectral features, are also classified with the MK system.  
 
The atmospheric composition of planets has been found to  be much more complex than the interior composition.  Fundamentally, the atmospheric composition is dictated by carbon and oxygen \citep{marley15}.  Many other key absorbers (H$_2$O, CO and CH$_4$) are dependent on the amount of C and O available in the atmosphere.  As the ratio of C/O approaches one, condensation of oxides and silicates and C-based compounds become more common \citep{lodders03}.  Condensates, formed by the interaction of these compounds, are responsible for forming clouds.  L dwarfs are known for their iron and silicate clouds but these disappear at the L/T transition, which some attribute to the breaking up of clouds into patches \citep{morley12}.  Patchy cloud structures has been seen in spatial resolved maps of a L/T transition brown dwarf  \citep{crossfield14}.  The formation of clouds is the greatest obstacle shrouding our understanding of planetary atmospheres.  Fortunately, cooler planets tend to have fewer clouds.

\begin{figure}[h!]
\epsscale{1}
\plotone{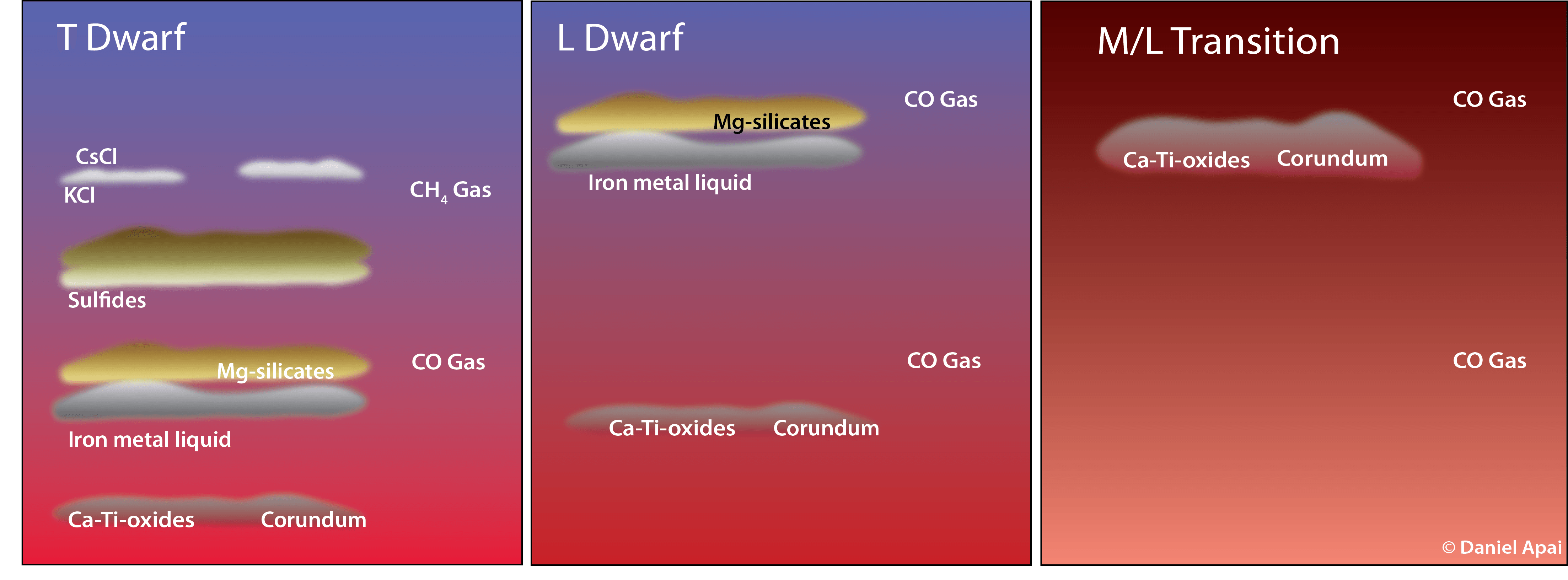}
\caption[Exoplanet Atmosphere Cartoon]{Different exoplanet categories show different spectral signatures.  Cooler T-types have fewer high clouds allowing for light to escape from deeper levels, giving these planets a bluer tone.  Figure by Daniel Apai. }
\label{atmpsoheres}
\end{figure}

The cooler T dwarfs form condensates of CH$_4$, Cr, MnS, Na$_2$S, ZnS, and KCl \citep{morley12}.  The formation of these compounds can lead to `rainout' where interactions between compounds causes them to condensate and fall out of the upper atmosphere.  Condensates seen in T dwarf spectra show evidence of this type of atmospheric chemistry \citep{morley12}.  The dissipation of the high altitude clouds in T dwarfs allows flux to escape from deeper layers which can be seen as a trend towards blue in a J--K colour diagram (see Figure~\ref{atmpsoheres} for a visual explanation).
 %-------------------------------------------
%-------------------------------------------.
  Beyond the compounds that form through condensation, ultraviolet radiation from the host star can photochemically split compounds which can then form more complex molecules.  Modelling of these atmospheres can be extremely complex and some models use upwards of 3,000 species \citep{lodders02}.

\subsection{Exoplanet Detection Methods}
As the success of the Kepler and High Accuracy Radial Velocity Planet Searcher (HARPS) missions have shown, detecting a wide range exoplanets is now relatively easy.  Kepler alone has identified over 2,000 confirmed exoplanets and nearly 5,000 candidates as of May, 2016 \citep[NASA Exoplanet Archive,][]{akeson}.  

The first exoplanets were discovered with radial velocity.  As early as 1988, a potential 1.7 M$_{\mathrm{Jup}}$ planet was discovered with this technique, though it wasn't confirmed until much later \citep{campbell}.  The radial velocity technique assess the reflex motion of a star about the system's centre of mass in order to detect periodic perturbations that would suggest the presence of a secondary mass \citep{carroll}.  This technique is best for finding planets with a high planet-to-star mass ratio on a small orbit.  A cooler, one solar mass star is ideal; stars above F6 have blurred spectral lines due to fast rotation and young, hot stars are more prone to stellar activity, like sunspots, which makes radial velocity measurements difficult \citep{saar,wright,barnes}.  Typically, for high-accuracy detections, high resolution spectra (R $=\lambda / \Delta \lambda \sim 50,000- 100,000$) are obtained with \'{e}chelle spectrographs \citep{fischer}.  While radial velocity has been successful in detecting many new exoplanets, it has the disadvantage of only providing lower mass estimates.  This is because the reflex motion can only be measured for movement towards or away from Earth.  As a result, any orbital inclination, $i$, is only visible along $sin(i)$, making the mass measurement a function of M$*sin(i)$. When $i$ is large, and the system is nearly edge-on, the mass measurement is close to accurate; however, if the inclination is large then the true mass is much higher than the estimate.  

Undoubtedly, the most prolific producer of planets is the transit method, wherein a planet travels in front of the star, attenuating a fraction of the star's light \citep{borucki,moutou,morton16}.  Again, this method naturally favours large planets around small stars at close orbits.  Transits are a favourable approach to detecting planets as they allow for a high amount of system characterization.  In addition to orbital period, transits can also offer information on the planet's radius, inclination, and atmosphere through transit spectroscopy \citep{charbonneau}.  Unfortunately, the chance alignment necessary for transits is low, with only a 10\% possibility for low orbits and a significant decline out to higher orbits \citep{beatty}.  This probability is also a function of stellar type, which is analogous to stellar radius.  When  transits are combined with the radial velocity method, it is possible to determine the planet's density, which is useful in learning about the physical structure of planets and their formation \citep{weiss}.  

In the past two decades since the first exoplanet was discovered \citep{campbell,latham,mayor}, huge advances have progressed the field to a point where it is now possible to directly images exoplanets \citep[e.g.][]{chauvin,marois08b,kalas08,lafreniere08,lagrange09,thalmann,marois10b,todorov,carson,kuzuhara,rameau,bailey,macintosh15}.  The emergence of direct imaging opened a new exoplanet parameter space for exploration.  Direct imaging generally relies on infrared thermal emission from the planet itself or visible light reflected from the host star.  Unlike previous detection methods, the intrinsic brightness of host stars favours detections at wide separations.  In our local solar neighbourhood this separation extends from tens to hundreds of AU.  It is thought that at these high orbital separations, planets are likely shaped by disk instabilities, planet-planet scattering and other migrational channels \citep{bowler}.  Imaging is also independent of inclination, allowing a more complete search of systems.   However, the large contrasts between the relative brightness of a planet to its host star can be a hindrance and makes direct imaging a challenging task and only a handful have been discovered with this method despite exhaustive searches like the VLT and MMT Simultaneous Differnetial Imager Survey (45 stars), Gemini Deep Planet Survey (85 stars), NaCo Survey of Young Nearby Austral Stars (88 stars), NaCo Survey of Young Nearby Dusty Stars (59 stars),  Strategic Exploration of Exoplanets and Disks with Subaru ($\sim$ 500 stars), Gemini NICI Planet-Finding Campaign (230 stars), International Deep Planet Search ($\sim$ 300 stars), Planets Around Low-Mass Stars Survey (78 stars), NaCo-LP Survey (86 stars), and ongoing surveys like Project 1640, LBTI Exozodi Exoplanet Common Hunt, Gemini Planet Imager Exoplanet Survey, and Spectro-Polarimetric High-contrast Exoplanet REsearch Survey \citep{bowler}.  Unlike radial velocity searches which require only a minute of observation per star, direct imaging takes approximately an hour, with additional follow-up time required for candidate confirmation or rejection.

\subsection{The Gemini Planet Imager Exoplanet Survey}\label{gpies}
The Gemini Planet Imager Exoplanet Survey (GPIES) campaign is an ongoing survey of over 600 nearby stars with nearly 900 hours dedicated time on the 8-meter Gemini South Telescope at Cerro Pach\'{o}n.  Using bright guide stars (I $<$ 9.5 mag) for AO corrections, the Gemini Planet Imager (GPI) can produce diffraction limited images from $0.9-2.4$ microns \citep{macintosh08}.  GPI is sensitive to planets from $0.2-1$ arcseconds from their parent star, a region that is complementary to that which is accessible to the Doppler shift method.  

The GPI system is composed of an AO system, a coronagraph mask, a calibration interferometer, an integral field spectrograph (IFS), and an opto-mechanical superstructure.  The coronagraph is an apodized-pupil Lyot coronagraph \citep[APLC,][]{soummer09} which combines a Lyot coronagraph with an apodization function \citep{macintosh08}. The classic Lyot coronagraph has a hard-edged mask to block most of the light of the star as well as a Lyot stop which blocks diffracted light.  The apodizer further reduces any diffraction to improve the contrast.  The calibration interferometer, or CAL system, works to sense wavefront quasi-static errors to provide corrections for the AO system.  Corrections are made from central on-axis light that has been diffracted outside the pupil.
%A beamsplitter funnels 20\% of off-axis light to the CAL system for the high-order calibration interferometer. 
 A separate low-order wavefront sensor composed of a Shack-Hartmann wavefront sensor with seven sub-apertures is used for tip/tilt and low-frequency aberrations.  These corrections are fed to the AO system once per second. The GPI science instrument is an IFS which allows for coarse spectral resolution across the FOV.  The GPI IFS uses a lenslet array which disperses the image into a grid with sufficient space for the spectra to be dispersed between each point.  The disperser is a prism which allows for low resolution spectra (R $\sim$ 45 for the H band, $\sim$ 35 for the Y band and $\sim$70 for the K band).  Atmospheric molecules like water vapour, carbon monoxide, and methane are also visible at theses wavelengths, allowing for the detailed characterization of any discovered exoplanet atmosphere, including clouds, as well as the effective temperature and bolometric luminosity.  GPI can also be used in a polarimetric mode to sense polarized light and in this case a Wollaston prism is used which can separate the orthogonal polarizations of light.  The opto-mechanical superstructure is the physical housing which keeps GPI together.  

GPIES aims to look at giant exoplanets since smaller bodies would have insufficient detection contrasts. Even a Jupiter-mass planet seen in reflected light with a contrast of $10^{-9}$ is not detectable by the current class of ground-based instruments, and finding an Earth-Sun analogue would require an additional order of magnitude of contrast.  The luminosity ratio between a star and a planet in reflected light is dependant on the stellar spectral type, the planet-star separation, the planet's mass, radius, and age.  Luckily, young planets are much brighter.  A 10-million year old Jupiter analogue around a solar-type star would have a contrast ratio of a few times $10^{-7}$, which is detectable with GPI \citep{burrows97}.  Since planet-star contrast is also a function of planet temperature, the limiting temperature for detection with GPI is  $\sim$300 K \citep{macintosh08}.  
%The GPI survey has selected predominantly young ($<150$ Myr), cool stars to exploit this contrast advantage. 
 Additionally, GPI uses near infrared bands (Y, J, H and K) which are sensitive to the excess heat a planet radiates. This heat is left over from formation and gravitational contraction.    
 
Many exoplanet surveys, including GPIES, target stars in young moving groups, comoving associations of stars.   These young systems are advantageous for study because their ages can be highly constrained with isochrones and lithium depletion measurements \citep{mentuch}.  Since mass estimation is so highly dependant on stellar age, this allows for more precise characterization of planets in these moving groups.  Mass estimates are also highly dependant on initial conditions and formation methods, as well as, though to a lesser extent,  star variability among other factors \citep{bowler}.

GPI team members observe at Gemini South approximately five nights every month.  Throughout each night, when observing conditions are good, stars are selected for observation as they transit the local meridian in order to maximize FOV rotation.  On summit, the acquired images are processed in real-time with the GPI data reduction pipeline \citep{perrin} to ensure data quality.  Bad images, such as those taken with a misaligned coronagraph or vibration and smearing effects due to wind, can immediately be flagged for removal from the final dataset.  Additionally, the wavelength solution can be corrected manually for images in which the flexure solution is misaligned.  All data is automatically uploaded to a Dropbox account where it is accessible by all members of the GPI collaboration for further processing.  

GPI, as of April 2016, had observed 272 targets out of 621.  At nearly halfway through the campaign, there has only been one confirmed exoplanet discovery, though a handful of other objects of interest are under further scrutiny.  Though the lack of discoveries has been surprising (predicted exoplanet yields from RV statistic extrapolations exceeded fifty), this level of non-detection has also been telling \citep{graham}.  The paucity of large mass planets at large separations from their host stars is indicative of how commonly gas giant planets form at these distances, which can help differentiate planet formation models.  GPI is not alone in its struggle to find new worlds; the Spectro-Polarimetric High-contrast Exoplanet REsearch instrument \citep[SPHERE, ][]{beuzit}, which runs a twin campaign, has found no exoplanets to date.  Occurrence rates for planets $5-13$ M$_{Jup}$ found to date between $30-300$ AU are astonishingly low at only $0.6^{+0.7}_{-0.5}\%$ \citep{bowler}.  Indeed, it seems rather remarkable that only a few giant planets have been discovered through imaging in what has been revealed as akin to searching for a needle in the cosmic haystack.  Even though GPI has not detected many new exoplanets, it has been successful in finding multiple new disks and binaries \citep{hung,kalas}.   

\subsection{Template Locally Optimized Combination of Images}
As we have seen in Section~\ref{psftech}, there are many approaches to PSF subtraction.  In this study we focus our optimization on TLOCI \citep{marois14}.  TLOCI is a complete IDL data processing package start to finish.  To begin, it registers and spatially magnifies the images, since images at different wavelengths will be at slightly different diffraction sizes, normalizes the flux to flatten the stellar spectrum and applies an unsharp mask.  For images from the Gemini Planet Imager Exoplanet Survey (see Section~\ref{gpies}), the registration and magnification are derived from the positions of the four calibration satellite spots \citep{marois06,siv,wang}.  The satellite spots from each slice of each data cube are combined to make a reference PSF.  

To subtract the PSF, reference images ($I^{REF}$) are created for each individual image ($I_i$) by making a linear combination of the images:
\begin{equation}
I^{REF}_i = \sum_k (c_{(k,i)} \times I_k) ,
\label{eq:loci}
\end{equation}
where $c_{(k,i)}$ are various weights assigned to each reference images as derived from:
\begin{equation}
Ac_i=b_i ,
\label{eq:matrix}
\end{equation}
where $A$ is a covariance matrix comprised of the reference images and $b_i$ is a correlation vector of the images to subtract with all the reference images.  To find $c_i$, matrix $A$ is inverted and multiplied to the correlation vector $b_i$ in a least-squares fit to minimize residual noise in the final reference subtracted image.  Reference images that are closer in rotation than $\frac{1}{2} \lambda /D$ are rejected from the set so as to minimize potential self-subtraction.  Several approaches to the matrix inversion are possible.  A standard inversion can be used, but if the the reference image archive is too large and composed of only partially correlated images, noise can contaminate the minimization.  Alternatively, a single value decomposition (SVD) can be used to truncate and invert the matrix.  The SVD cutoff method calculates a diagonalized matrix from the correlation matrix to find the eigenvalues.  Any eigenvalues below the cutoff threshold are set to zero to avoid fitting noise.  Additionally a non-negative least-squares (NNLS) inversion can be used which forces the coefficients to be positive to avoid large oscillating positive/negative coefficients.  A reference image is created in annuli to maximize the SNR as a function of separation, as self-subtraction (or throughput) and noise varries with separation.

After the optimized coefficients have been found though the least-squares fit, the reference images is generated and subtracted.  The program then simulates the throughput of an artificial planet, created from the template PSF (obtained from the four calibration spots) and wavelength flux normalized to a planet spectrum template, at different annuli in the image.  This throughput correction is applied to the final image. These steps are repeated for all annuli and all wavelengths for each image in the sequence.  Each image is then rotated to north up and median combined.  This image is then collapsed by using a template spectrum (generally a brown dwarf spectrum) to create a weighted mean final 2D image over all wavelengths.

\subsection{Overview}\label{over}  
  
Although the TLOCI framework has already proven to be an effective algorithm for exoplanet and debris disk detection \citep[e.g.][]{macintosh15, hung,lagrange}, it has many parameters, which when optimized, have the potential of increasing SNR further, allowing the detection of dimmer, less massive planets.  For example, TLOCI can be run with three different matrix inversion methods.  Understanding the matrix's effect on planet SNR can help us choose the method most likely to detect planets.  Another parameter, the subtraction aggressivity, is used to compromise between noise subtraction and planet flux preservation. Optimizing aggressivity can reduce the chance of self-subtraction and thus increase detectability.  

In order to determine the parameters most likely to yield high SNR, various comparison tests can be run.  The aim of this study was to maximize the parameters for use with data from the Gemini Planetary Imager  and this chapter presents the optimized results.  Section~\ref{methods} presents the methods of the study.  Parameter tests are described in Section~\ref{tests}, including details on the matrix inversion tests (\ref{matrix}), aggressivity tests (\ref{agg}), and unsharp mask tests (\ref{id}).  Conclusions are in Section~\ref{tconc}.
%----------------------------------------------------------------------------------------------------------------------
%----------------------------------------------------------------------------------------------------------------------

% Data:

\section{Methods}\label{methods}

\begin{figure}[h]
  \centering
  \begin{tabular}[b]{@{}p{0.65\textwidth}@{}}
    \includegraphics[width=1.0\linewidth]{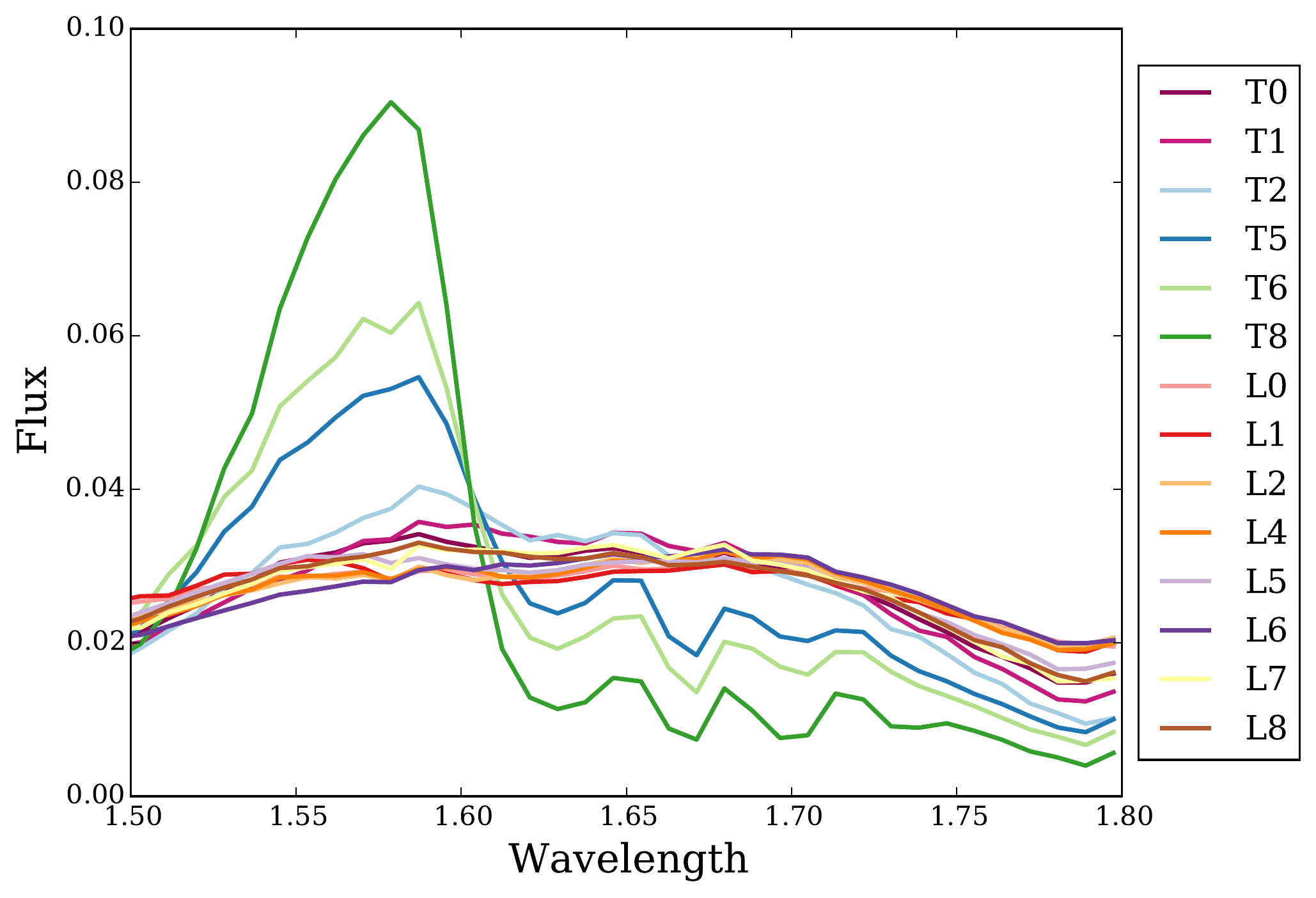} \\
  \end{tabular}%

  \caption[Template for H Band]{Template for H band generated from brown dwarf spectra.}
  \label{templates}
\end{figure}

As the name implies, TLOCI requires an input template spectrum.  A library of such spectra were created from the NIRSPEC \citep{mclean98} Brown Dwarf Spectroscopic Survey \citep[BDSS,][]{mclean} given brown dwarfs' spectral similarities with giant exoplanets \citep{faherty}. TLOCI templates from the brown dwarf spectra were produced for H ($1.50-1.79$ microns), J ($1.12-1.35$ microns), K1 ($1.90-2.19$ microns), K2 ($2.12-2.38$ microns) and Y ($0.95-1.14$ microns) bands, though only the H band templates were tested in these simulations, as this is the band used in the detection phase with the GPI exoplanet survey. Templates were created for T8, T6, T5, T2, T1, T0, L8, L7, L6, L5, L4, L2, L1, L0 spectra but only a subset of seven (T8, T6, T2, T0, L8, L4, L0) were used in tests.  To create the templates, the spectra were binned in 37 wavelength steps to match the GPI data resolution and each bin was normalized by the number of contributing spectral points and by the total flux of the spectrum.  Figure~\ref{templates} shows templates for select bands.

%--------------------------------------------------------
% Figure: Spectra Templates

%--------------------------------------------------------

\begin{figure}[h]
  \centering
  \begin{tabular}[b]{@{}p{0.45\textwidth}@{}}
    \includegraphics[width=1.\linewidth]{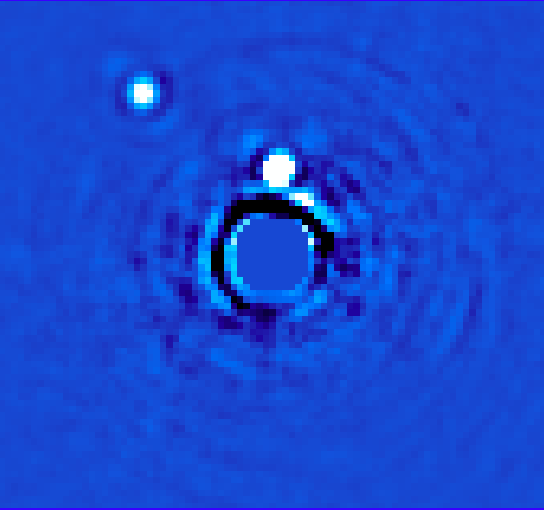} \\
    \centering\small (a) 
  \end{tabular}%
  \quad
  \begin{tabular}[b]{@{}p{0.45\textwidth}@{}}
    \includegraphics[width=1.\linewidth]{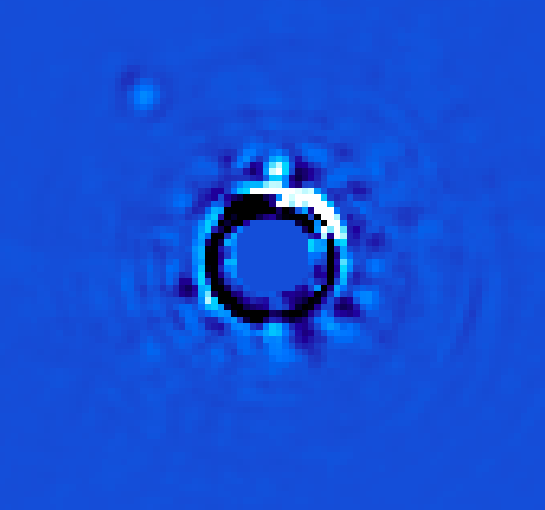} \\
    \centering\small (b) 
  \end{tabular}
  \caption[Differences in Input Templates]{Differences in Input Templates -- Both images have two simulated T8 planets at a separation of 0.2" and 0.45" from the star at relative brightnesses of $4\times10^{-4}$ and $3\times10^{-5}$ respectively.  Image (a), which uses a T8 input spectrum, shows a much higher SNR for the recovery of both planets than image (b) which uses a L0 input spectrum.}
  \label{tlociexample}
\end{figure}

To test various parameters, simulated planets were inserted into a sequence of reduced GPI data cubes.  The main sequence of images used was AF Lep, a sequence selected for being archetypal of GPI data and one previously determined not to host any planets.  It has a field of rotation of 25\degree.  A fake planet was inserted at a separation of 0.2" and one at 0.45" from the star at relative brightnesses of 4x$10^{-4}$ and 3x$10^{-5}$, respectively, in each of the individual data frames of each data cube.  Two planets were used to look for any differences caused by the location of the planet, since a closer planet would have less rotational compensation in the final stacked images.  This is also why the inner planet required a higher flux in order for its signal to be distinguished from the noise at that annulus. After the insertion of the fake planet, the images were processed with TLOCI set with various parameters to test the recovery of the simulated planet.  An example of an image with simulated planets recovered with two different input templates is shown in Figure~\ref{tlociexample}.

The SNR of the planet was used as the indicator of the effectiveness of the trial parameters.  A convolved image is created by taking the image and convolving it with a six pixel, normalized circular kernel.  The convolution multiplies each pixel by its neighbours, weighting by the kernel.  This helps emboss the edges of the planet so as to be easier to detect.  Using a convolved image, the signal of the planet was selected as the maximum pixel value within a circle (seven pixel diameter) drawn around the simulated planet.  The noise was calculated by creating an annulus at the radius of the planet with a width of two pixels, but masking out a circular area slightly larger than the planet (10 pixels) to ensure the planet's flux did not contaminate the noise sample.  The standard deviation on this region was taken as the noise.  The SNR was then calculated as the maximum planet flux divided by this noise.  

%annulus width = 8 pix, mask diameter = 10pix

%----------------------------------------------------------------------------------------------------------------------
% Figure 1: GPI Images w/ Planets 
%----------------------------------------------------------------------------------------------------------------------

\section{Parameter Tests}\label{tests}

%Subsection - Identification Tests
\subsection{Identification Tests}\label{id}

\begin{figure}[h]
  \centering
  \begin{tabular}[b]{@{}p{0.45\textwidth}@{}}
    \includegraphics[width=1.0\linewidth]{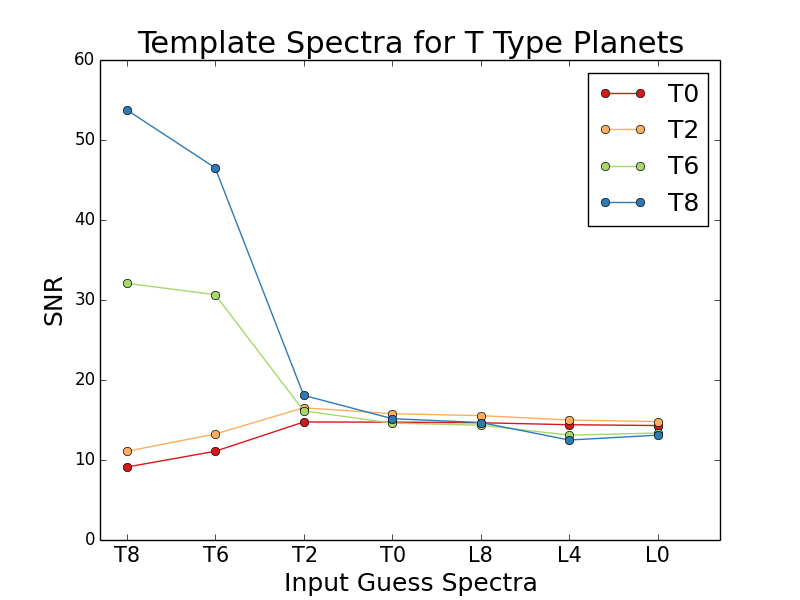} \\
    \centering\small (a)  T Types
  \end{tabular}%
  \quad
  \begin{tabular}[b]{@{}p{0.45\textwidth}@{}}
    \includegraphics[width=1.0\linewidth]{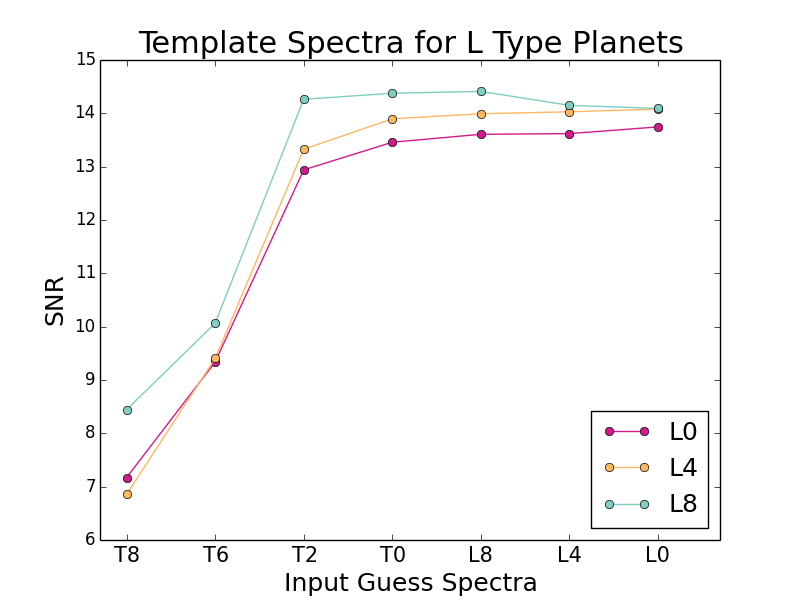} \\
    \centering\small (b) L Types
  \end{tabular}
  \caption[Template recovery for L and T type planets]{Template recovery results showing high similarities among all L types as well as early T types (T2, T0).  This indicates only two templates, such as a T8 and T2, are needed for exoplanet detections.}
  \label{idtypes}
\end{figure}

Of primary interest to planet detection is being able to detect any type of planet.  Since TLOCI requires and input spectrum, it is important that that spectrum be able to find a wide range of planets since an undiscovered planet will not have a known spectral type.  Using the library of templates created, a series of identification tests were conducted to maximize the number of discoverable planet types with the least number of templates.  A simulated planet was inserted into the sequence at 0.45" and TLOCI was run with seven different input spectra in order to see the recovery of the inserted planet for different spectra.  From these tests it apparent that at least a methane and a dusty spectra are required to sufficiently detect any type of planet (see Figure~\ref{idtypes}).  The T8 and T2 spectra are selected for this purpose.  For additional assurance in planet detection, a flatter dusty spectra, like the L8, can be used to compliment the T8 and T2.  
%----------------------------------------------------------------------------------------------------------------------
% Figure2: ID Tests

%-------------------------------------------------------------------------------------------------------

%Subsection - Matrix Tests
\subsection{Matrix Inversion Tests}\label{matrix}

\begin{figure} [h]
  \centering
  \begin{tabular}[b]{@{}p{0.45\textwidth}@{}}
    \includegraphics[width=1.05\linewidth]{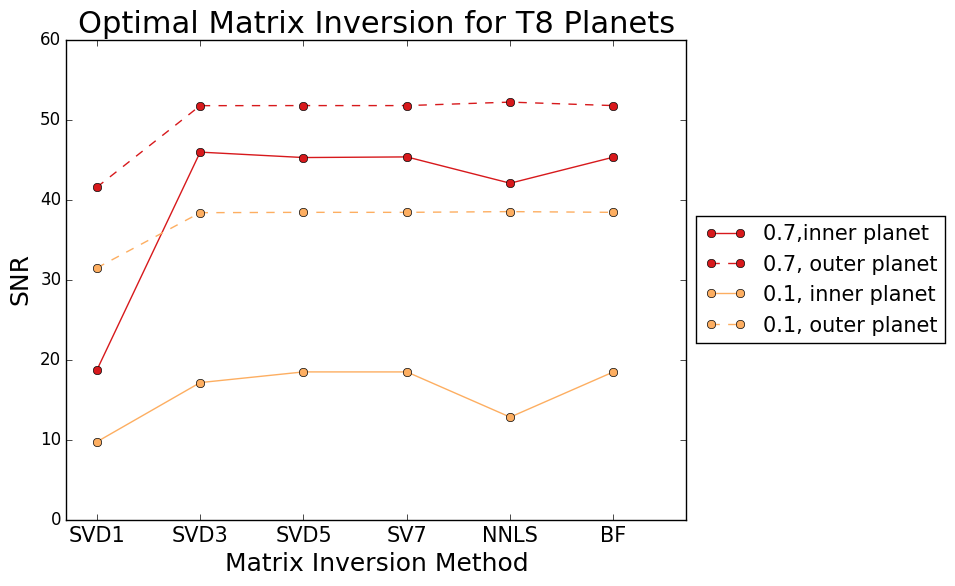} \\
    \centering\small (a)  T8 Planets
  \end{tabular}%
  \quad
  \begin{tabular}[b]{@{}p{0.45\textwidth}@{}}
    \includegraphics[width=1.05\linewidth]{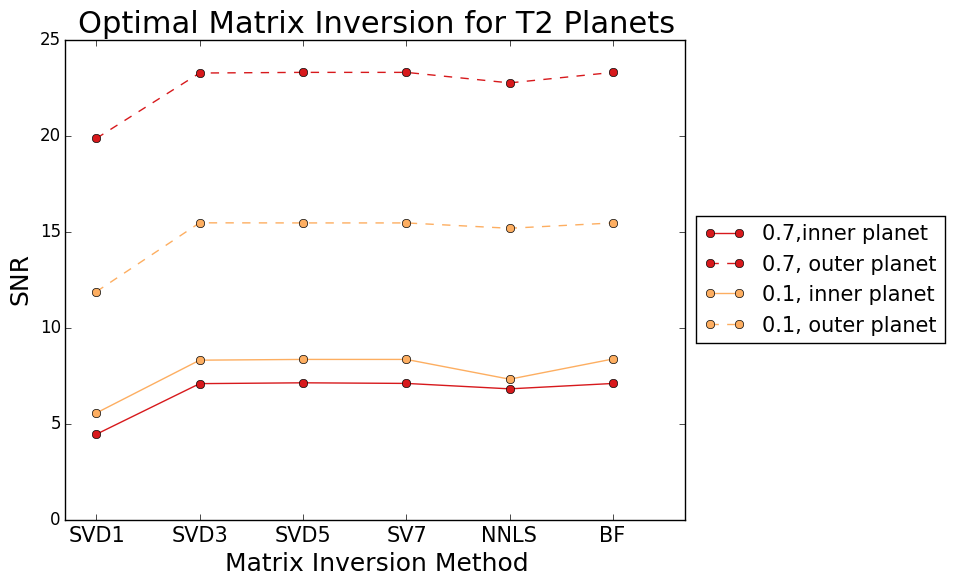} \\
    \centering\small (b) T2 Planets
  \end{tabular}
  \caption[Matrix Inversion Method Comparison]{Tests comparing different matrix inversion methods find very little SNR difference between methods.  An SVD with cut-off $10^{-1}$ and NNLS are clearly inferior methods but there is no obvious preferred method.  Dashed lines indicate the outer planet and solid the inner.  Red shows tests with an aggressivity of 0.7 and orange shows tests with an aggressivity of 0.1.}
  \label{matrixtests}
\end{figure}

The most basic of the parameters varied in this study was the matrix inversion type.  TLOCI is able use one of three types of matrix inversion methods: standard inversion, NNLS, and SVD inversion.  The cut-off value for eigenvalues using the SVD invert method can be varied.  For testing different matrix methods, four cut off ratios to the larger eigenvalue were used: $10^{-1}$, $10^{-3}$, $10^{-5}$, $10^{-7}$. For  two planet types (T8, T2), TLOCI was run seven times to test each matrix type, but using the same, matching spectral input template each time.  This testing sequence was repeated once for an aggressivity of 0.1 and once at 0.7.  Results are shown in Figure~\ref{matrixtests}.  In general, using a SVD matrix inversion with a cutoff at $10^{-1}$  should be avoided.  Using a NNLS inversion similarly seems to lower the SNR.  Given the relatively small difference between the remaining methods, the standard method was selected for the rest of the trials for its efficiency in computing time.  

%----------------------------------------------------------------------------------------------------------------------
% Figure 1: Plot

%----------------------------------------------------------------------------------------------------------------------

%Subsection - Agg Tests
\subsection{Aggressivity Tests}\label{agg}

If references are not selected carefully, self-subtraction can remove significant planet flux, hindering detection.  In order to limit self-subtraction, a parameter known as aggressivity can be changed.  A less aggressive reduction minimizes self-subtraction but cannot fully attenuate noise, making detections at low impact parameter more difficult.  Higher aggressivity allows more self-subtraction but also minimizes noise.  Physically the aggressivity is determined by the number of reference images used.  In the most aggressive reduction (1) all the reference images are allowed, whereas the least aggressive reduction (0) allows fewer images.

\begin{figure}[h]
\epsscale{.6}
\plotone{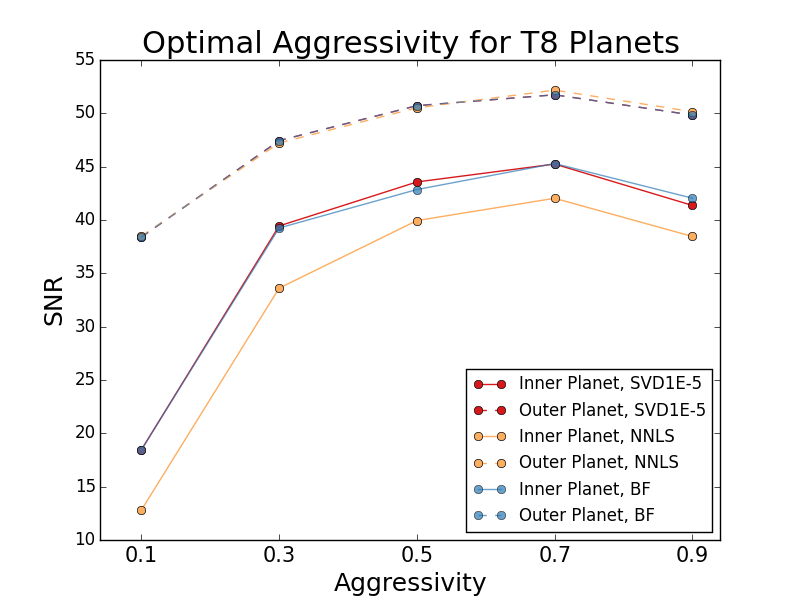}
\caption[Aggressivity Results for T8 Planets]{Aggressivity results using AF Lep and T8 planets. The optimal aggressivity in all cases is 0.7.}
\label{agg-aa}
\end{figure}

\begin{figure}[h!]
\epsscale{.6}
\plotone{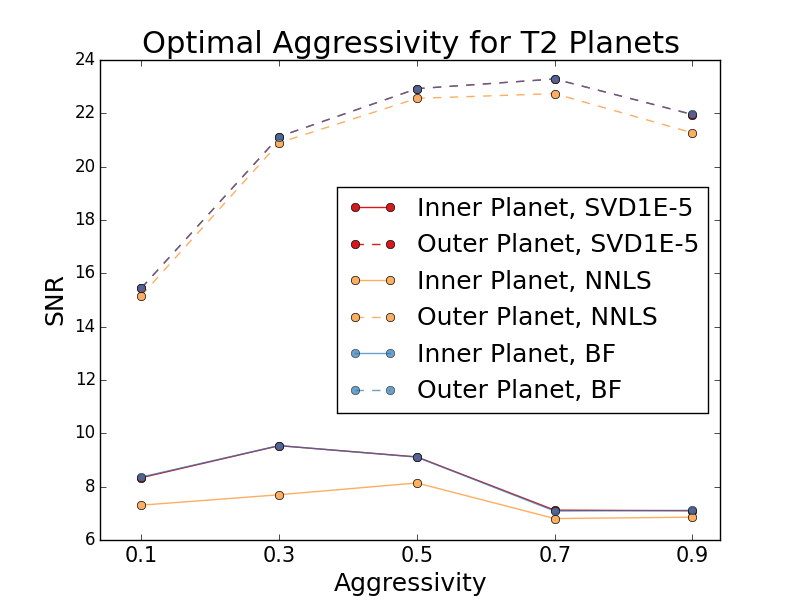}
\caption[Aggressivity Results for T2 Planets]{Aggressivity results using AF Lep and T2 planets.  Unlike the results from the T8 planet, the inner planets benefit by a lower aggressivity.}
\label{agg-a}
\end{figure}

In order to find the optimal aggressivity, five aggressivities were tested with four planet types (L8, L0, T8, T2) and three matrix inversions (SVD $10^{-5}$, NNLS, standard).  Planet and input templates were matched for each trial.  See Figures~\ref{agg-aa} and \ref{agg-a} for results from T8 and T2 trials.  In most cases, an aggressivity of 0.7 yielded the highest SNR, although L8 and T2 inner planets did better with a lower aggressivity, peaking at 0.3.  
%----------------------------------------------------------------------------------------------------------------------
% Figure 1: GPI Images w/ Planets - not working

%----------------------------------------------------------------------------------------------------------------------

\begin{figure}[h!]
\epsscale{.7}
\plotone{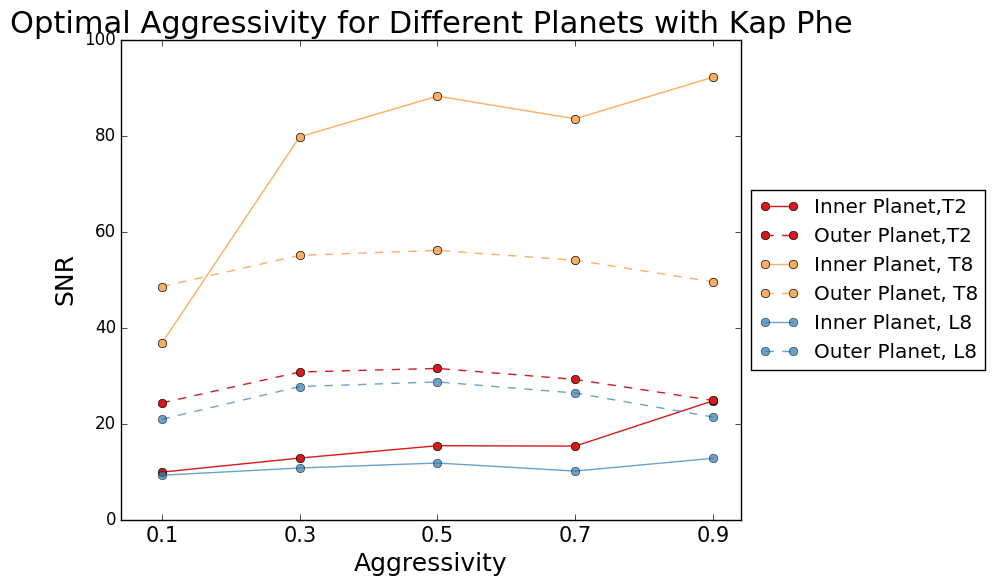}
\caption[Aggressivity Results for Kap Phe]{Aggressivity results using Kap Phe.  Unlike AF Lep, inner planets yield a higher SNR with high aggressivity, and outer planets peak at a lower aggressivity.}
\label{agg-b}
\end{figure}

In order to test the dependance on image conditions, another sequence (Kap Phe) was tested.  This sequence had a much larger field of view rotation, 47\degree.  Using the standard matrix inversion method, the same set of tests as before were run varying aggressivity using T2, T8 and L8 planets.  This time, all the inner planets had maximum SNR at an aggressivity of 0.9, an outer planets a maximum at 0.5 (see Figure~\ref{agg-b}).  From these results, it seemed aggressivity optimization was linked not only with planet type, but also with field of view rotation of the images.  

%----------------------------------------------------------------------------------------------------------------------
% Figure 1: Kap Phe Agg
%----------------------------------------------------------------------------------------------------------------------

\begin{figure}[h!]
\epsscale{.7}
\plotone{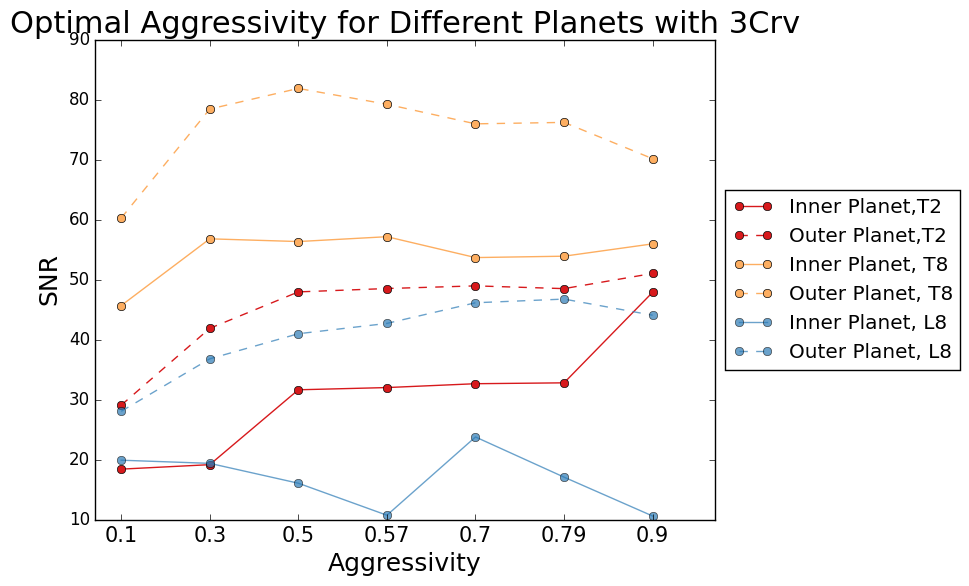}
\caption[Aggressivity Results for 3Crv]{Using an image with intermediate field of view rotation does not yield predictable SNR results.  In this case, there is no trend in aggressivity preference, although 0.57 consistantly does well with the exception of the L8 inner planet.  As seen here, and in Figures \ref{agg-aa} and \ref{agg-a} and \ref{agg-b}, aggressivity optimization is dependant on other variables such as field of view rotation and image conditions.}
\label{agg-c}
\end{figure}

To test this correlation between aggressivity and rotation, a linear relation was fit between optimal aggressivity and field of view rotation for both the inner and outer planets using the results from AF Lep and Kap Phe.  This relation was tested with a third sequence, 3Crv, with an intermediate rotation.  The relation predicted that the new sequence, with a rotation of 35\degree, would see maximum aggressivity of 0.57 for T2 and L8 planets, and 0.79 for a T8 planet.  None of the SNR for any of the planets matched this prediction, as shown by Figure~\ref{agg-c}.  It seems instead that any correlation is much more complicated than a simple linear relation.  The aggressivity is dependant not only on the planet type, but also the amount of rotation and the speckle time stability during the sequence.  
%----------------------------------------------------------------------------------------------------------------------
% Figure 1: 2Crv Agg
%----------------------------------------------------------------------------------------------------------------------

%Subsection - unsharp Mask Tests
\subsection{Unsharp Mask Tests}\label{id}

\begin{figure}[h!]
\epsscale{.7}
\plotone{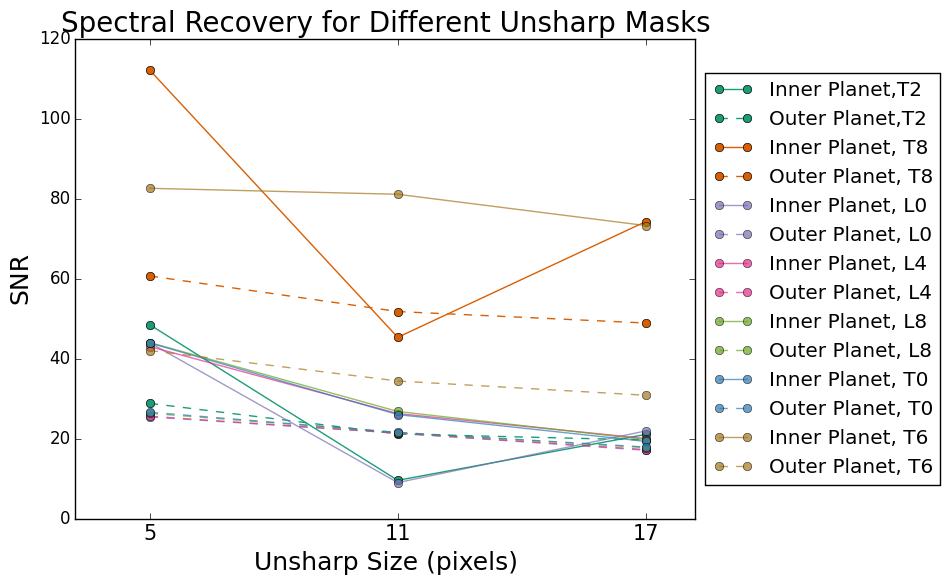}
\caption[Unsharp Mask Results]{Three different unsharp mask sizes were tested. Smaller sizes correlate to higher SNR in all cases. With a few spectra, a mask of 11 pixels was found to be significantly worse for inner planets (T8, T2, L0).}
\label{unsharpmask}
\end{figure}

Before PSF subtraction, an unsharp mask is applied at each pixel location in the image to remove low-frequency noise.  The area of pixels used by the unsharp mask can be changed.   Normally the mask size is set to 11 pixels.  Trials were run using smaller and larger masks of 5 and 17 pixels. A size of 5 pixels yielded the highest SNR in all cases (see Figure~\ref{unsharpmask}). 
%----------------------------------------------------------------------------------------------------------------------
% Figure 1: unsharp
%---------------------------------------------------------------------------------------------------------------------- 

%----------------------------------------------------------------------------------------------------------------------
\section{Conclusion}\label{tconc}
TLOCI is a complex algorithm and optimizing its performance is similarly multiplicious.  Although maximum signal to noise can be obtained by changing parameters on an image-to-image basis, there appears to be no single set that are widely applicable.  Additionally, image conditions and parameters like the field-of-view of rotation have an affect on the SNR.  Future additions to the TLOCI code should include an automated method to optimize aggressivity in each annulus.  Until such improvements are implemented, the best set of parameters to use, not knowing the image condition or type of planet, are as follows:
\begin{itemize}
 \item A standard matrix inversion.  Although a standard inversion doesn't have a significant advantage in SNR over any other method, it is the fastest to compute so is advantageous in saving time.
\item An aggressivity of 0.7.  Because we will not know the conditions of the image or planet type before discovery, this parameter has the greatest room for improvement.  Considering all the aggressivity tests presented, using an aggressivity of 0.7 has the highest chance of producing the optimal SNR without knowing any image conditions. 
\item An unsharp mask size of 5 pixels is best for all planet spectra and planet locations.  
\item Planet spectra T2, T8 and L8.  When discovering new planets, their spectra will not be known a priori so it is important to use a range of templates that will maximize the detection of any planet type.  However, in the interest of computing time, it is also advantageous not to use all possible types.  From the identification tests, a strong methane planet (T8 type) and a dusty type (T2) are sufficient for detecting reasonably bright planets of any type.  An additional dusty type (L8) can be used for extra security in detection but any more than this is unnecessary.  
 \end{itemize}
While there is no end-all solution, using these parameters can help achieve the highest return on planet detection with reasonable efficiency.  

\newpage

\chapter{Applications in Angular Differential Imaging: Characterizing a Substellar Object }\label{bd}

Now that we have covered the optimization of a PSF subtraction technique, namely TLOCI, we can proceed to applying it to a science objective.  In this chapter, we present the characterization of the substellar object HD 984 B, which was imaged with GPI as a part of the GPIES campaign.

\section{Introduction}
The search for exoplanets through direct imaging has led to many serendipitous detections of brown dwarfs and low-mass stellar companions \citep[e.g.][]{mawet,biller,nielsen}.  These surveys tend to target young, bright stars whose potential companions would still be warm and bright in the infrared.  Brown dwarfs, having higher temperatures for the same age, are naturally easier to see and thus easier to find.
Even though discoveries of substellar companions in exoplanet surveys are only aftereffects of the study design, they are useful in their own right for comparing competing formation models \citep[e.g.][]{perets,bodenheimer,boley}.  Many surveys exclusively targeting brown dwarfs have also been conducted \citep[e.g.][]{epchtein,sartoretti,chauvin03,mccarthy,joergens05,nakajima05,skrutskie,delorme10,kirkpatrick11}.

\subsection{Brown Dwarfs}
As discoveries of all sizes of sub-stellar objects burgeoned, it became necessary to distinguish what was a planet and what was not.  Since there seemed to be no natural break in the continuum of discovered masses, it was more practical to set the boundary at a physical condition limit.  Whereas the criterion for stars is sufficient mass to burn hydrogen ($\sim$78M$_{\mathrm{Jup}}$ at solar metallicity \citep{kumar}), planets were limited to masses less than 13.6 M$_{\mathrm{Jup}}$, the limit for deuterium burning.  The objects sandwiched between stars and planets with too little mass to burn hydrogen, but enough to burn deuterium, were termed brown dwarfs, though are also known by the jocose name `failed stars'.  Although the deuterium burning limit has generally been accepted as the differentiator for brown dwarfs, it is important to acknowledge that it is still an artificial limit, particularly when comparing numbers of substellar objects.  
 
Early surveys of brown dwarfs found them twice as numerous as main sequence stars \citep{reid}, and for a short time they were even considered as a possible solution to the missing mass problem in the solar neighbourhood \citep{hawkins}.  Brown dwarfs have been found at high separations, up to even nearly 7000 AU from another star \citep{deacon}, and some are completely free from a host star.  It is thought that these `free-floating planets' were not ejected from planetary systems but simply formed from the low-mass tail of the star formation spectrum.  Although some have suggested that brown dwarfs form from disk instabilities in the same way that planets are known to \citep[see Section~\ref{exoplanets},][]{stamatellos09}, a more widely accepted theory is that they form in the same manner as stars \citep[e.g.][]{bayo11,scholz,alves,chabrier14}.  This method proposes brown dwarfs form from fragmentation of turbulent clouds which gravitationally contract.  Proto brown dwarfs have even been found with thermal radio jets that young stellar objects are known to form as a result of accretion \citep{morata15}. 

\begin{figure}[h!]
\epsscale{.65}
\plotone{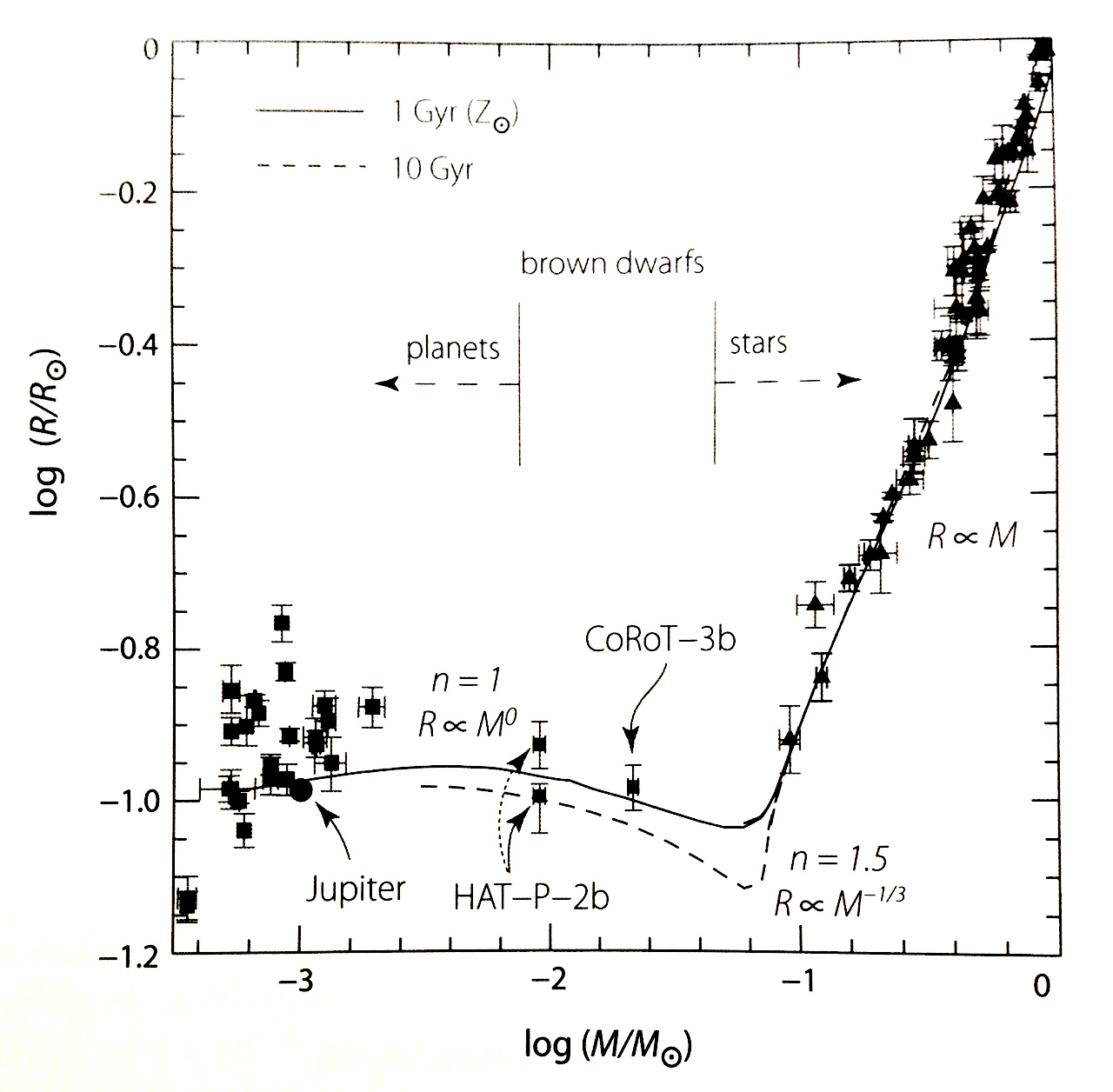}
\caption[Mass-Radius Scaling Diagram]{Mass-radius relation for stellar and sub-stellar objects.  Star radius scales with mass as governed by nuclear-burning physics. In substellar objects, which are governed by Coulomb and electron degeneracy pressure, scaling relations are much less correlated. Image from \citet{perryman}.}
\label{rm}
\end{figure}

During the formation of brown dwarfs, gravitational collapse causes an increase in temperature and density until the core becomes partially degenerate \citep{basri}.  Stability is reached as the free electron degeneracy pressure balances gravitational potential.  Low mass brown dwarfs, like planets, are primarily governed by Coulomb pressure (the electromagnetic repulsion of electrons), whereas high mass brown dwarfs are set by the Pauli exclusion principle, which prohibits fermions from occupying the same quantum state.  The weak dependance of radius on mass, which scales as $R\propto M^{-1/3}$ for electron degeneracy pressure and $R\propto M^{1/3}$ for Coulomb pressure, means that all brown dwarfs have radii of $\sim$1 R$_{\mathrm{Jup}}$ (see Figure~\ref{rm}).  At about two Jupiter masses, as electrons are forced into higher energy levels, the increase in mass and thus pressure leads to higher density making the radius decrease \citep{basri}.  

%---------------------------------------------------------
%---------------------------------------------------------

Brown dwarfs quickly exhaust their limited supply of deuterium at an abundance of 10$^{-5}$ that of hydrogen and cool soon thereafter \citep{basri}.  Brown dwarfs may also fuse lithium but only at masses above $\sim$$63-65$M$_\mathrm{Jup}$ \citep{perryman}.  This means deriving a mass from the luminosity or effective temperature is highly dependant on an age estimate.  Brown dwarfs can be classified as M, L, T and Y types, just as exoplanets are (See Section~\ref{exoplanets}). 

The first confirmed brown dwarf was discovered in 1995 \citep{nakajima95}.  Two decades later, brown dwarf discoveries are in excess of 1,000.  Many of these detections are a result of large surveys like the Deep near-infrared Southern Sky Survey \citep[DENIS, ][]{epchtein} and the Two Micron All Sky Survey \citep[2MASS, ][]{skrutskie}.  Surveys for brown dwarfs continue today and surveys for exoplanets frequently find new brown dwarfs as well.   

\subsection{Overview}
The remainder of this chapter has been adapted from Johnson-Groh et al. (in prep) in which we characterize the substellar object HD 984 B, which was imaged with GPI as apart of the GPIES campaign. Section~\ref{case} presents background information on the star and its companion.  
%We present new observations of HD 984 B at J and H bands using GPI, and updated astrometry, luminosity   measurements and physical characteristics.
  In Section~\ref{obsv} we describe the GPI observations. Basic reductions are explained in Section~\ref{redu} and Section~\ref{psfsub} details the point spread function subtraction technique.  Astrometry is discussed in Section~\ref{astrometry}, orbital fitting in Section~\ref{orbitp}, and spectral and photometric analyses are presented in Section~\ref{analysis}. We conclude in Section~\ref{conc}.

\section{The Case of HD 984}\label{case}
HD 984 is a bright (V $=7.32$), nearby ($d=47.1\pm1.4$pc) $\sim$1.2M$_\sun$ F7V star \citep{hog,van,meshkat} with a temperature of 6315$\pm$89 K \citep{white,casagrande}. Using COND evolutionary tracks, \citet{meshkat} determine the luminosity to be log(L/L$_{\sun}$) = 0.346$\pm$0.027 dex.  \citet{van} report a proper motion for the star of $\mu_{\alpha\ast}, \mu_\delta = 102.79\pm0.78,-66.36\pm0.36$ mas yr$^{-1} $ and a parallax of 21.21$\pm$0.64 mas. With an age estimate of 30 -- 200 Myr (115$\pm$85 Myr at a 95\% confidence level) derived from isochronal age, X-ray emission and rotation \citep{meshkat}, and previous age estimates \citep{wright04,torres}, the star is ideal for direct imaging campaigns to search for young substellar objets.

\citet{meshkat} report the discovery of a bound low-mass companion at a separation of 0.19$\pm$0.02 arcsec (9.0$\pm$1.0AU) based on $L'$, and H+K band data observed in July 2012 using NaCo \citep{lenzen,rousset} and SINFONI \citep{eisenhauer,bonnet} on the Very Large Telescope (VLT).  Comparing the SINFONI spectrum to field brown dwarfs in the NASA Infrared Telescope Facility (IRTF) library, \citet{meshkat} conclude the companion to be a M6.0$\pm$0.5 object \citep{cushing,rayner}.  Their findings are summarized in Table~\ref{meshk}.  \citet{meshkat} note that future observations in the J band could identify low gravity signatures in the companion and further epochs of astrometry could allow for dynamical mass determination.  
\begin{deluxetable}{ccccl}
\tabletypesize{\scriptsize}
\tablewidth{0pt}
\setlength{\tabcolsep}{3pt}
\tablecaption{Discovery Results}
\tablehead{{Property}&{HD 984 B}}
\startdata
\hline
Separation & 0.19$\pm$0.02" (9.0$\pm$1.0AU) \\
T$_{eff}$ & 2777$^{+127}_{-130}$ K \\
log(L$_{\mathrm{Bol}}$/L$_{\sun}$) & $-2.815 \pm 0.024$ \\
$H_{2MASS}$ & 12.58$\pm0.05$ mag \\
$Ks_{2MASS}$ & 12.19 $\pm 0.04$ mag \\
Age & $30-200$ Myr \\
Mass$_{30 Myr}$,DUSTY model & 34$\pm$6M$_{Jup}$\\
Mass$_{200 Myr}$,DUSTY model & 0.10 $\pm$0.01M$_\sun$\\
\enddata
\label{meshk}
\tablecomments{All values from \citet{meshkat}.}
\end{deluxetable}
%----------------------------------------------------------

HD 984 (HIP1134) was observed as one of the targets of the GPIES campaign \citep{macintosh14}.  Although originally imaged without the knowledge of its previously discovered companion, the data taken by GPI was able to contribute to the characterization of the substellar object.

%----------------------------------------------------------------------------------------------------------------------
%----------------------------------------------------------------------------------------------------------------------

% Data:
\section{Observations}\label{obsv}
HD 984 was observed with GPI integral field spectrograph \citep[][]{chilcote,larkin} on August 30, 2015  UT during the GPIES campaign (program GPIES-2015B-01, Gemini observation ID: GS-2015B-Q-500-982) at Gemini South.  The GPI IFS has a FOV of 2.8 x 2.8 arcsec$^2$ with a plate scale of 14.14$\pm$0.01 miliarcseconds/pixel \citep{macintosh, konopacky,larkin}.  Coronographic images were taken in spectral mode in the J and H bands.  Observations were performed when the star was close to the meridian at an average airmass of $\sim$1.1 so as to maximize FOV rotation for angular differential imaging \citep[][]{marois06} and minimize the airmass during observations.  23 exposures of 60 seconds of one coadd each were taken in the H band (1.50 -- 1.80$\mu$m) and 23 exposures, also of 60s and one coadd, were followed up in the J band (1.12 -- 1.35$\mu$m); the J band data was acquired with the H band apodizer due an apodizer wheel mechanical issue. Two H band exposures and five J band exposures were rejected due to unusable data quality.  Total FOV rotation for H band was 15.2$\degree$ and a total rotation of 11.9$\degree$ was acquired with J band. Average DIMM seeing for H and J band sequences was 1.14" and 0.82" respectively.  The windspeed averages for H and J bands were 2.5m/s and 1.9m/s.  Images for wavelength calibration were taken during the daytime and short arc images were acquired just before the sequences to correct for instrument flexure.

%In order to extend the baseline for astrometry, archival data was retrieved from the Keck Observatory Archive (KOA).  Viable images of HD984 taken with the Near Infrared Camera 2 (NIRC2) were found for epochs from 2002 (PI: Zuckerman) and 2009 (PI: Macintosh).  Data from 2002-08-21 was taken in position angle rotator tracking mode where as data from 2009-07-31 was taken utilizing ADI. Both epochs were imaged in the K$_{\mathrm{p}}$ band.   The 2009 epoch had 60 images at 10 second exposures.  Only SOME NUMBER of images with short exposures (0.2 seconds) from 2002 were usable as longer exposure images were oversaturated.  

\section{Reductions}\label{redu}

%\subsection{GPI Reductions}\label{gpiredu}
The images were reduced using the GPI Data Reduction Pipeline \citep{perrin} v1.3.0\footnote{\href{http://docs.planetimager.org/pipeline}{\url{http://docs.planetimager.org/pipeline}}}.  Using primitives in the pipeline, raw images were dark subtracted, argon arc image comparisons were used to compensate for instrument flexure \citep{wolff}, the spectral data cube was extracted from the 2D images \citep{maire}, bad pixels were interpolated in the cube and distortion corrections were applied \citep{konopacky}.   A wavelength solution was calibrated using arc lamp images taken prior to data acquisition.
 %(I THINK THE WAVE SOLN IS DERIVED BY A HIGH SNR CALIBRATION TAKEN DURING THE DAY TIME, THE SHORT ARC IMAGE TAKING JUST BEFORE THE SEQUENCE IS FOR FLEXURE CORRECTION ONLY, CHECK WITH SCHUYLER).  
 Four satellite spots, PSF replicas of the star generated by the pupil plane diffraction grating \citep{siv,marois06},  were used to calculate the location of the star behind the coronagraphic mask for image registration at a common centre, as well as to calibrate the object flux to star flux \citep{wang}.

%\subsection{NIRC2 Reductions}\label{keckredu}
%NIRC2 data was processed separately from the GPI data. For the 2009-07-31 ADI data, there were 60 short exposure images (10 seconds).  A basic reduction was conducted which included dark image subtraction, bad pixel correction and flat fielding.  The images were registered to the star's centre and a PSF was created by taking a median of all the images.  This PSF was subtracted off each exposure before they were rotated to north up and median combined.  No companion was visible in the final combined image.  
%----------------------------------------------------------
% Figure 
%\begin{figure}[H]
%\epsscale{.45}
%\plotone{2009.png}
%\caption{NIRC2 reduced image from 2009 epoch data.  Green circle denotes the separation at which HD984 B would be expected. No companion is visible.}
%\label{keck}
%\end{figure}
%----------------------------------------------------------

%Data from 2002-08-21 was not observed with ADI so a pupil-aligned PSF was used for subtraction.  Only images take at short exposure (0.2 seconds) were used to avoid saturation from the PSF at the location where we expect to see the brown dwarf.   Again, a basic reduction was conducted by subtracting a dark image, correcting bad pixels, and applying flat fielding.  A PSF was created by rotating all images to align the pupil, and taking their combined median.  This was subtracted from the registered images prior to median combining.  No objects were seen after the pupil-aligned PSF had been subtracted. 

\section{PSF Subtraction}\label{psfsub}
\begin{figure}[h!]
  \centering
  \begin{tabular}[b]{@{}p{0.35\textwidth}@{}}
   \includegraphics[width=1.0\linewidth]{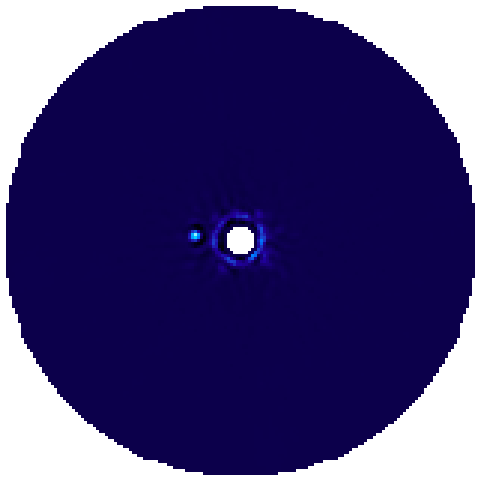} \\
    \centering\small (a)  
  \end{tabular}%
  \quad
  \begin{tabular}[b]{@{}p{0.35\textwidth}@{}}
    \includegraphics[width=1.0\linewidth]{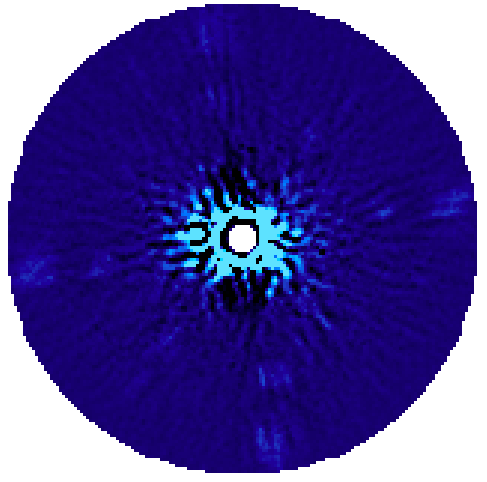} \\
    \centering\small (b) 
  \end{tabular} \\
    \begin{tabular}[b]{@{}p{0.35\textwidth}@{}}
   \includegraphics[width=1.0\linewidth]{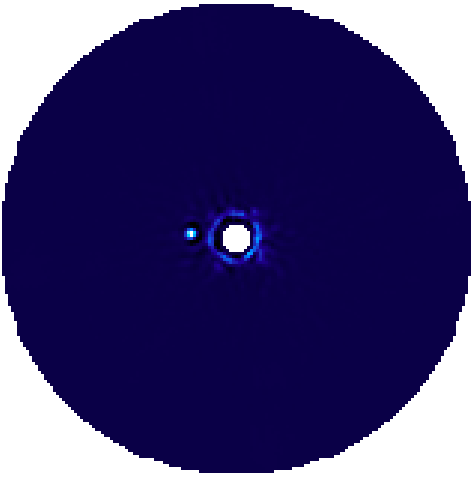} \\
    \centering\small (c)  
  \end{tabular}%
  \quad
  \begin{tabular}[b]{@{}p{0.35\textwidth}@{}}
    \includegraphics[width=1.0\linewidth]{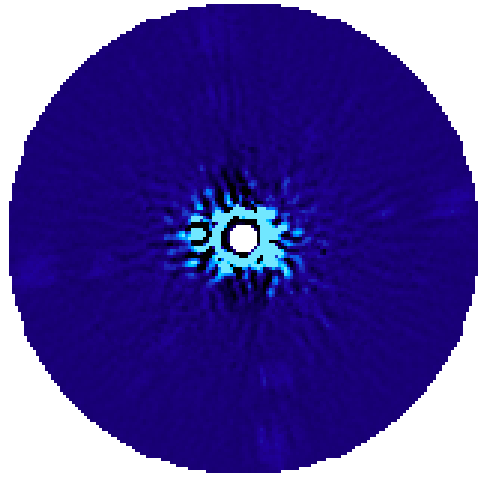} \\
    \centering\small (d) 
  \end{tabular}
  \caption[HD 984 B Final Images]{H band (a \& b) J band (c \& d) final 2D images, post TLOCI using only an ADI subtraction. Two contrast levels are shown to highlight the companion (left) and the image noise (right). North is up and East is left.}
  \label{finalimages}
\end{figure}

After the initial data reduction, each slice of each data cube, which were flux normalized using the average maximum of a Gaussian-fit on the four calibration spots, were spatially magnified to align diffraction-induced speckles (using the pipeline-derived spot positions).  They were then unsharp masked using a 11$\times$11 pixel kernel to remove the seeing halo and background flux, and were PSF subtracted using the TLOCI algorithm \citep{marois14}.  Once all the data cubes had been PSF subtracted, a final 2D image was obtained by performing a weighted-mean of the 37 slices, using the input template spectrum and image noise to maximize SNR. Final images for J and H bands shown in Figure~\ref{finalimages}. While the initial discovery was obtained by performing both an SSDI and ADI subtractions, to avoid the spectral cross-talk bias, only a less aggressive ADI-only subtraction (reference images are selected if they have less than 30\% of the substellar object flux in a 1.5 $\lambda/D$ diameter aperture centred at the object position) was used for spectral and astrometry extractions. 

%----------------------------------------------------------
% Figure: Final Images
%----------------------------------------------------------

%\subsection{SPECTROPHOTOMETRY}\label{spec}

%----------------------------------------------------------------------------------------------------------------------
%---------------------------------------------------------------------------------------------------------------------
\section{Astrometry and Spectral Extraction}\label{astrometry}
Using the TLOCI ADI-only subtracted final combined data cube, the flux and position of the companion were measured relative to the star. As TLOCI uses a training zone that differs from the subtraction zone, the companion is not fitted by the least-squares, thus removing one bias. To minimize the well-known self-subtraction bias, the ADI subtraction was performed using a less aggressive algorithm. To further take into account the self-subtraction bias for both spectral and astrometry extraction, the companion's signal was fitted using a forward model derived from the median-average PSF of the four calibration spots for the entire sequence. For each slice of each data cube, a noiseless image was created with the calibration spot PSF at the approximate location of the companion. These simulated images were then processed using the same steps as the science images to produce the companion forward model. The forward model was then iterated in flux and position to minimize the local residual post subtraction in a $1.5 \lambda/D$ aperture. 
Error bars are derived by adding and extracting the forward model flux at the same separation as the companion, but at nine different position angles.  The one sigma measurement of standard deviation in flux and position of the nine simulated companions are the spectral and astrometric errors. A correction is applied to account for forward model errors. Instead of adding the simulated companions having the same flux as the recovered flux, they are added after normalizing the companion signal by the ratio of the local residual noise inside a 1.5 $\lambda$/D aperture after the forward model best subtraction relative to the noise at the same angular separation calculated away from the companion.  

In the H band, the separation was found to be 216.3$\pm$1.0 mas, and in the J band, 217.9$\pm$0.7 mas.  The position angles for H and J bands were 83.3$\pm$0.3$\degree$ and 83.6$\pm$0.2$\degree$ respectively.  The astrometry is summarized in Table~\ref{astrom}.
%----------------------------------------------------------
\begin{deluxetable}{ccccl}
\tabletypesize{\scriptsize}
\tablewidth{0pt}
\setlength{\tabcolsep}{3pt}
\tablecaption{Astrometry}
\tablehead{{Band} &{Date} &{Separation}&{PA($\degree$)}}
\startdata
\hline
%Kp	&	August 21, 2002&	0.163$\pm$.001"	& 140			& (Archival NIRC2 data)\\
L'	&	July 18,20, 2012&	0.19$\pm$0.02"	&108.8$\pm$3.0 	&(APP data) \\
	&				&	0.208 $\pm$0.023"	&108.9$\pm$3.1 	&(direct imaging) \\
H+K &Sep 9, 2014		&201.6$\pm$0.4 mas	&92.2 $\pm$0.5\\
H	&	Aug 29, 2015	&	216.3$\pm$1.0mas	&83.3$\pm$0.3  \\
J	&	Aug 29, 2015	&	217.9$\pm$0.7mas	&83.6$\pm$0.2 \\
\enddata
\label{astrom}
\tablecomments{Data from 2012 and 2014 epochs from \citet{meshkat}.}
\end{deluxetable}

%----------------------------------------------------------

Since TLOCI outputs the object's spectra relative to the star, the spectra needed to be calibrated to the star prior to normalization. This was done using a custom IDL program and the Pickles stellar spectral flux library \citep{pickles}.  Since the Pickles library did not contain a model spectra for an F7V star, the average between a F5V and F8V star was created.  The spectra for both F5V and F8V models in J and H bands was degraded to GPI resolution and interpolated at the same wavelengths as the GPI wavelength channels.  The Pickles spectra are only scaled in the V band so J and H band stellar spectra were colour corrected and magnitude corrected using Vega as a zero point reference. For each band, the normalized F5V and F8V spectra were averaged to compute a F7V.  These F7V calibrated models were multiplied by the planet-to-star spectra for each band. The final planet spectrum was normalized over the entire band for spectral type comparisons.  

%[IMAGE OF GPI AND MESHKAT SPECTRA]

\section{Orbital Fitting}\label{orbitp}
\begin{figure}[h!]
\epsscale{.85}
\plotone{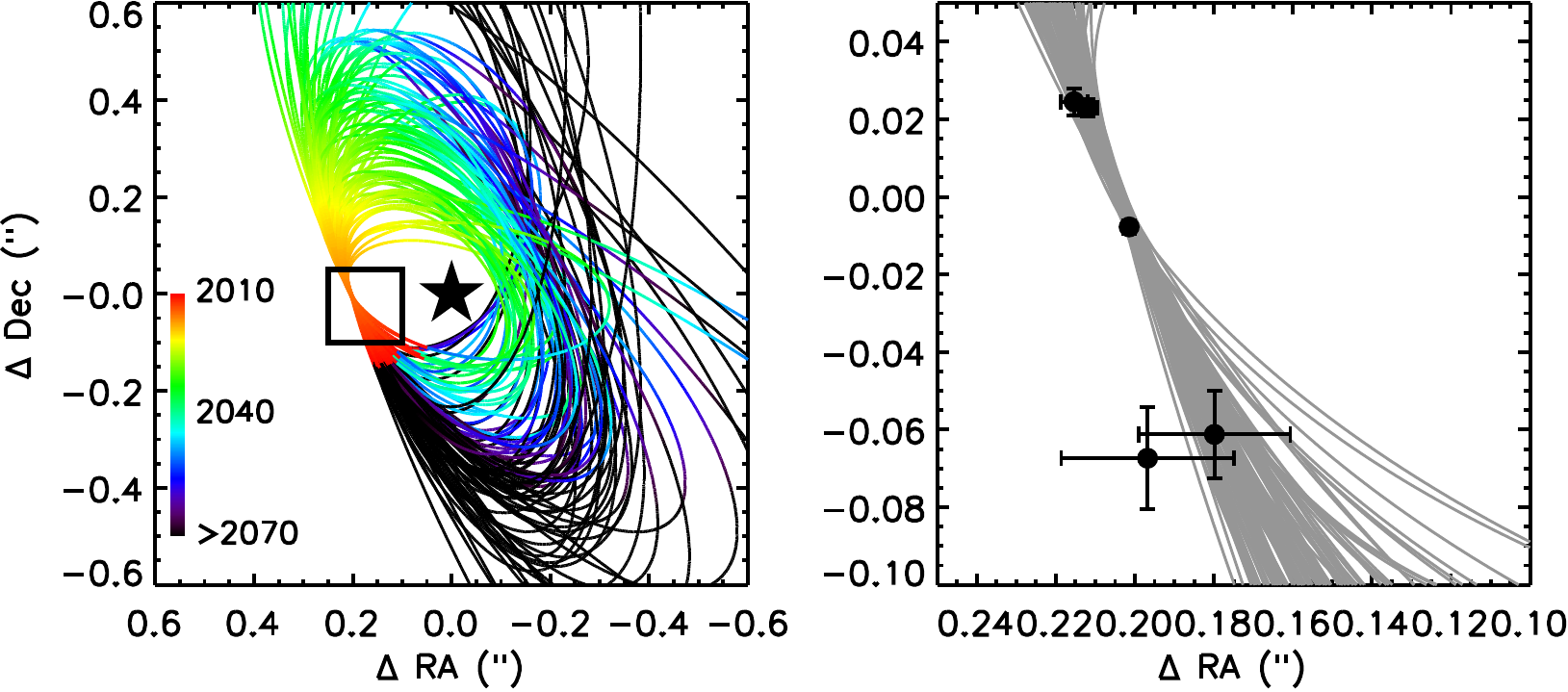}
\caption[Orbital Models for HD 984 B]{Orbital fitting using the NaCo, SINFONI and GPI epochs. Results indicative of an 18~AU (70 year) orbit, with a 68\% confidence interval between 12 and 27 AU and an eccentricity of 0.24 with a 68\% confidence interval between 0.083 and 0.495 and inclination of 118\degree with a 68\% confidence interval between 112\degree and 127\degree. Figure created by collaborator Eric Nielsen.  }
\label{orbit}
\end{figure}

\begin{figure}[h!]
\epsscale{.75}
\plotone{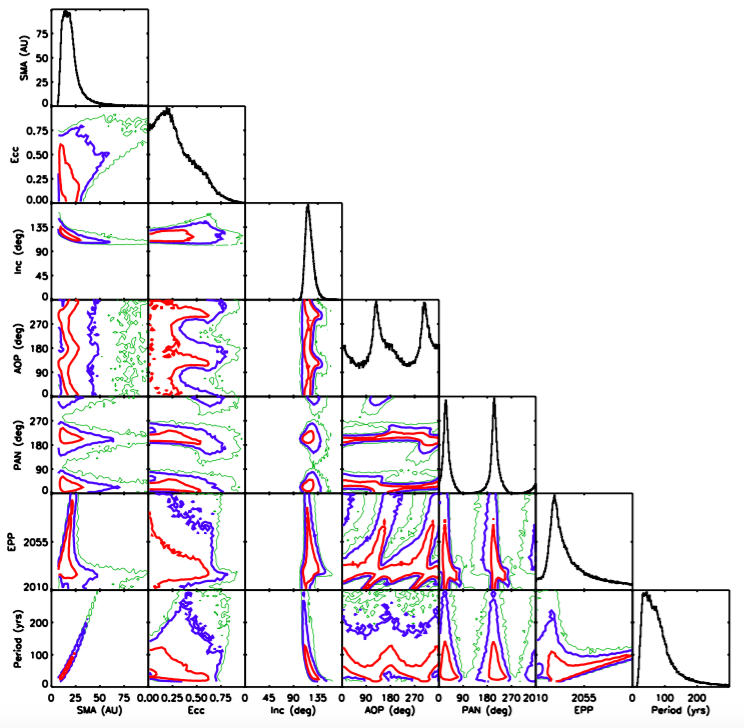}
\caption[Orbital Fitting Confidence Intervals]{Confidence intervals from orbital fitting. 1 (red), 2 (blue), and 3 (green) $\sigma$ contours enclose 68.27\%, 95.45\%, and 99.70\% of all orbital elements. Figure created by collaborator Eric Nielsen.}
\label{intervals}
\end{figure}

We determined the orbital parameter consistent with the full astrometric record of HD~984 using the rejection sampling method previously presented in \citet{derosa16} and \citet{rameau2016}.  Orbital parameters were drawn from the prior distributions except for semimajor axis and position angle of nodes, which are assigned initial values of 1 AU and 0$^\circ$.  These values are then adjusted so that the orbit matches one of the epochs of data, and this reference epoch is rotated through all the available epochs.  Observational errors are taken into account by scaling to Gaussian distributions in separation and PA, centered on the measurement with the 1$\sigma$ of the Gaussians equal to the errors.  The probability associated with the orbit is then computed by calculating the $\chi^2$ statistic for the remaining epochs, and the orbit is accepted if a uniform random variable is less than or equal to $Prob. = e^{-\frac{\chi^2 }{2}}$. Fitted orbits are shown in Figure~\ref{orbit}. The epoch positions are close enough within error bars to allow both clockwise and counterclockwise orbits, though clockwise is preferred. The fitting suggests an 18~AU (70 year) orbit, with a 68\% confidence interval between 12 and 27 AU and an eccentricity of 0.24 with a 68\% confidence interval between 0.083 and 0.495 and inclination of $118\degree$ with a 68\% confidence interval between $112\degree$ and $127\degree$.  Confidence intervals are shown in Figure~\ref{intervals}.  \citet{meshkat} do not perform an orbital fit in their analysis, but from their two epochs believe the system to have a non-zero inclination. 

%----------------------------------------------------------
% Figure 
%----------------------------------------------------------
%----------------------------------------------------------
% Figure 
%----------------------------------------------------------

%----------------------------------------------------------------------------------------------------------------------
\section{Spectral and Photometric Analysis}\label{analysis}

Spectral templates created from observations of L, T and M-type objects were created to identify the spectral type of HD~984~B. L and T-type spectra were created from the NIRSPEC Brown Dwarf Spectroscopic Survey \citep{mclean}.  The brown dwarf spectra were binned to GPI resolution.  Templates were produced for T6, T5, T2, T1, T0, L8, L7, L6, L5, L4, L2, L1, L0, M9, and M8 spectra.  Additional templates for M-type objects M7, M6.5, M6, M5.5, M5, M4, M3, M2, and M1 in the J and H bands were created using spectra from the IRTF library \citep{cushing,rayner}.  Brown dwarfs and low mass stars used for each template are listed in Table~\ref{spec}.  
%Given the absence of GPI K-band data, K-band templates were adapted to SINFONI resolution for all spectral types.    

%----------------------------------------------------------
\begin{deluxetable}{llcc}
\tabletypesize{\scriptsize}
\tablewidth{0pt}
\setlength{\tabcolsep}{3pt}
\tablecaption{Spectral Types}
\tablehead{{Type} &{Model Object}}% &{Fitting Result (H band)}&{Fitting Result (J Band)}}
\startdata
\hline
T6&2MASS 2356-15 \\
T5&2MASS 0559-14 \\
T2&SDSS 1254-01 \\
T1& SDSS 0837-00\\
T0& SDSS 0423-04\\
L8& Gl 337c\\
L7& DENIS 0205-11AB\\
L6& 2MASS 0103+19\\
L5& 2MASS 1507-16\\
L4&GD 165B \\
L2& Kelu-1\\
L1& 2MASS 1658+70\\
L0&2MASS 0345+25 \\
M9&LHS 2065 \\
M8& VB 10\\
M7&Gl 644C \\
M6.5& GJ 1111\\
M6& Gl 406\\
M5.5& HD 94705\\
M5& Gl 51\\
M4 & Gl 499 \\
M3 & Gl 388 \\
M2 & HD 95735 \\
M1 & HD 42581 \\

\enddata
\label{spec}
%\tablecomments{}
\end{deluxetable}
%----------------------------------------------------------

%Templates were produced for T6 (2MASS 2356-15), T5 (2MASS 0559-14),T2 (SDSS 1254-01), T1 (SDSS 0837-00),  T0 (SDSS 0423-04), L8 (Gl 337c), L7 (DENIS 0205-11AB), L6 (2MASS 0103+19), L5 (2MASS 1507-16), L4 (GD 165B), L2 (Kelu-1), L1 (2MASS 1658+70), L0 (2MASS 0345+25), M9 (LHS 2065) and M8 (VB 10). spectra.  Additional templates for M-type objects M7 (Gl 644C), M6.5 (GJ 1111), M6 (Gl 406), M5.5 (HD 94705) and M5 (Gl 51) in the H and J bands were created using spectra from the IRTF library \citep{cushing,rayner}.  Given the absence of GPI K-band data, K-band templates were adapted to SINFONI resolution for all spectral types.  

We now compare our extracted spectra with other field brown dwarfs and low mass stars. IFS observations often produce spectral noise correlation \citep{greco16}, and GPI data cubes are also known to suffer this effect, especially at small separations close to the focal plane mask.  Before any comparisons to model spectral types can be made, the correlation needs to be analyzed to avoid biasing the $\chi ^2$ analysis. Given the high SNR of detections, it may be possible to fit higher frequency structures in the spectrum independently to the low frequency envelope. The noise characteristics, especially the spectral noise correlation, may differ with spectral frequencies, with the low frequencies mostly limited by highly correlated speckle noise slowly moving over the planet as a function of wavelength, while higher frequencies could be mainly limited by read or background noises, thus being mostly decorrelated between wavelength channels.

\begin{figure}[h!]
\epsscale{.55}
\plotone{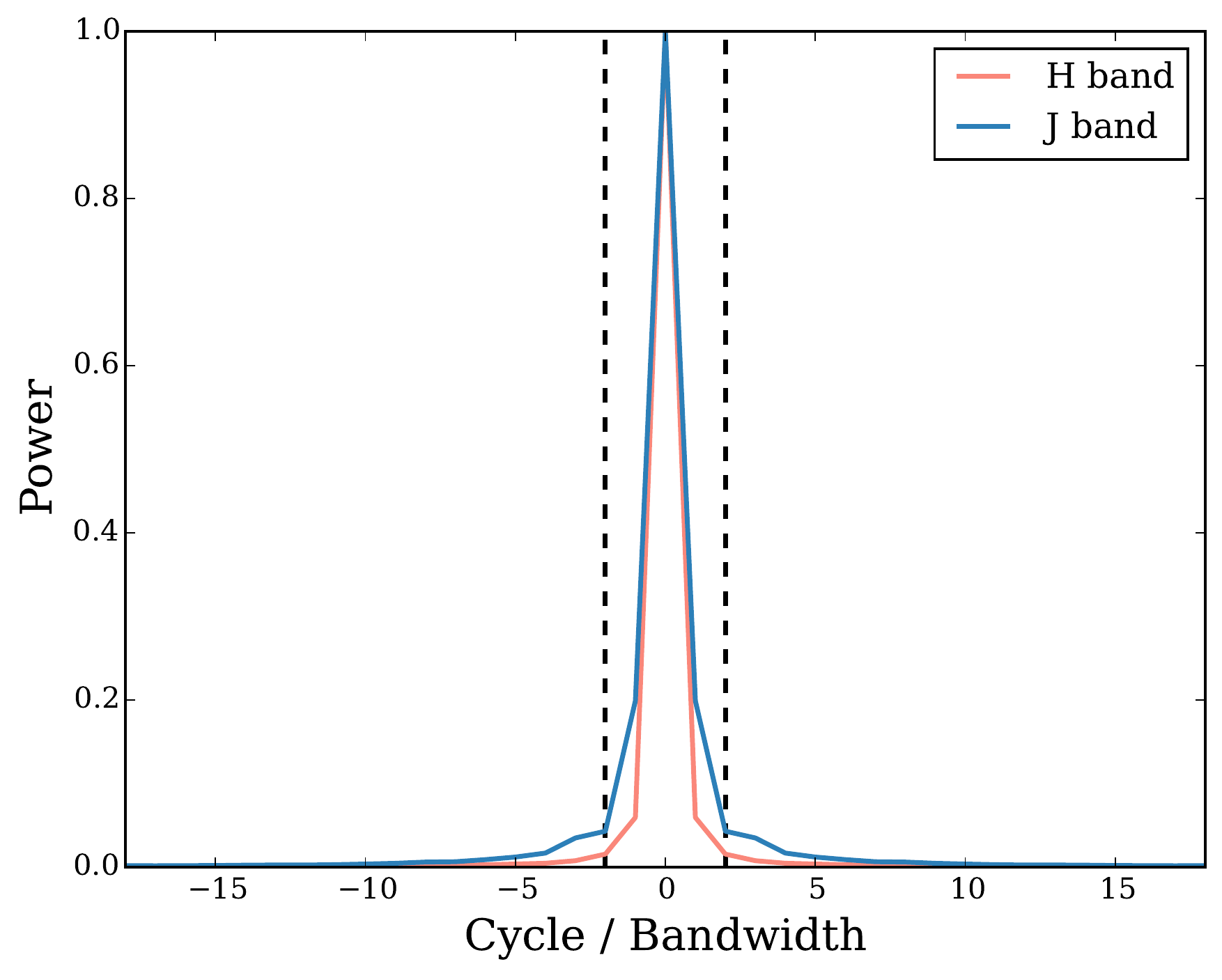}
\caption[Power Spectrum]{Power spectrum as a function of cycles per bandwidth for each bandpass. The H band has a bandwidth of $0.0084 \mu$m and J band has a bandwidth of $0.0065 \mu$m.  Vertical dashed lines indicate the boundary between low (between dashed lines) and high (outside dashed lines) frequency spectra.}
\label{pspec}
\end{figure}

\begin{figure}[h!]
  \centering
  \begin{tabular}[b]{@{}p{0.45\textwidth}@{}}
   \includegraphics[width=1.0\linewidth]{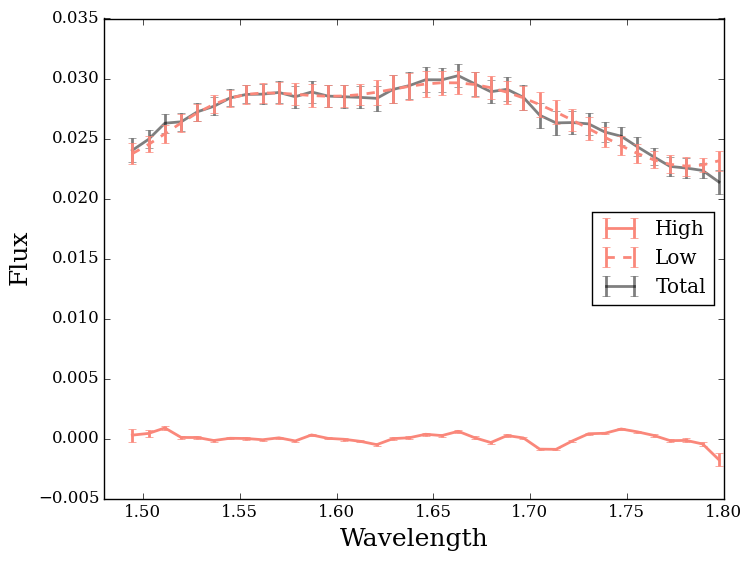} \\
    \centering\small (a)  
  \end{tabular}%
  \quad
  \begin{tabular}[b]{@{}p{0.45\textwidth}@{}}
    \includegraphics[width=1.0\linewidth]{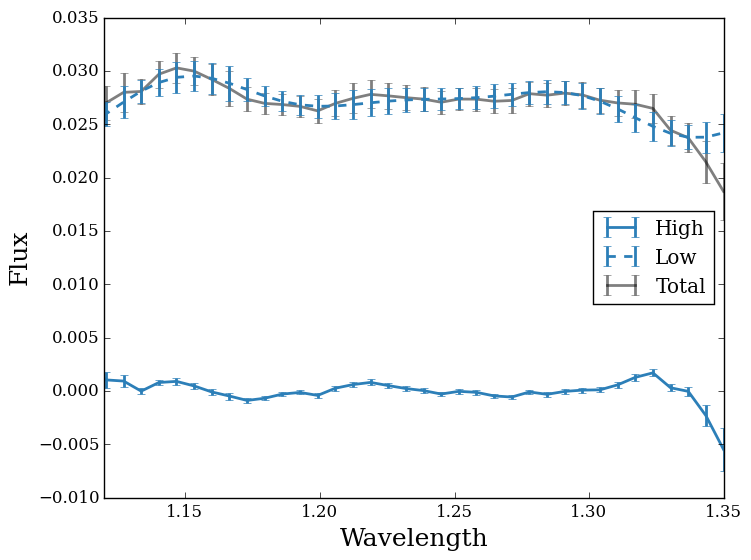} \\
    \centering\small (b) 
  \end{tabular} \\
  \caption[Spectral Spliting]{H band (a) J band (b) spectra split between low (dashed line) and high (solid line) frequencies.  Solid dark lines are the original spectra.}
  \label{splitspec}
\end{figure}

\begin{figure}[h!]
\epsscale{.55}
\plotone{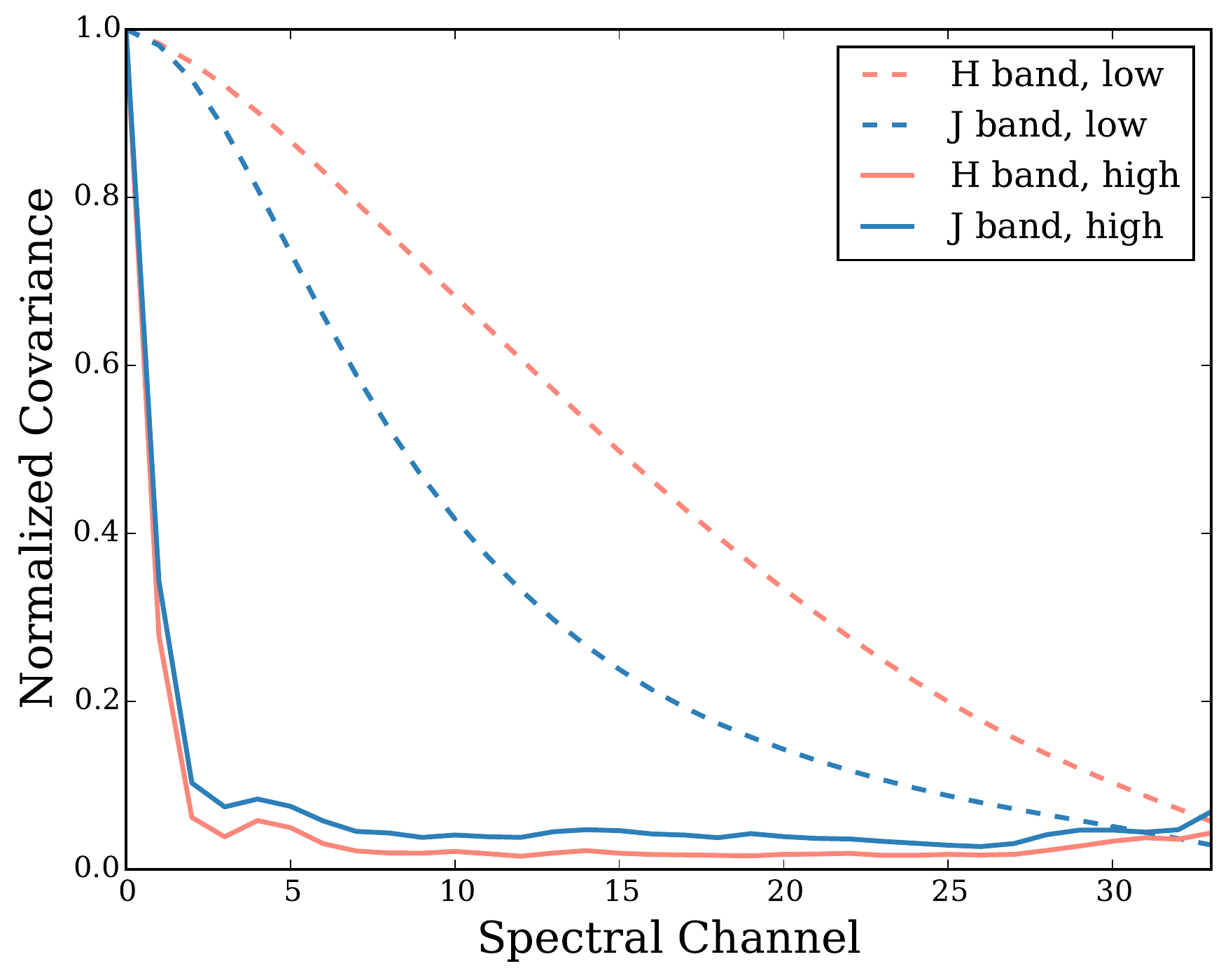}
\caption[Spectral Correlation]{Spectral correlation as a function of wavelength channels for each filter.  The low frequency spectra (dashed lines) show a much higher covariance than the high frequency spectra (solid lines).}
\label{correlation}
\end{figure}

The objects' spectra (HD 984 B and the field brown dwarfs) were split into low and high frequencies by taking the Fourier transform of the spectra (see Figure~\ref{pspec}).  Any frequencies between -2 and 2 cycles/bandwidth are considered low frequencies, and anything outside of that range is designated as high frequencies.  The errors were propagated by taking a noise ratio of the high to low spectra and splitting the errors at the same ratio between the high and low spectra. The split spectra are shown in Figure~\ref{splitspec}. Testing the spectral noise correlation for the high frequency spectra separately found correlations covering just three wavelength channels for both bands as expected from read and background noises (see Figure~\ref{correlation}); this spectral correlation is consistent with GPI's pipeline over spectral sampling.  The low frequency spectra was still highly correlated to fifteen and eight wavelength channels for J and H bands respectively (correlation to reach 50\% correlation). 
%----------------------------------------------------------
% Figure 
%----------------------------------------------------------
%----------------------------------------------------------
% Figure: Final Images
%----------------------------------------------------------

%----------------------------------------------------------
% Figure 
%----------------------------------------------------------

\begin{figure}[h!]
\epsscale{.55}
\plotone{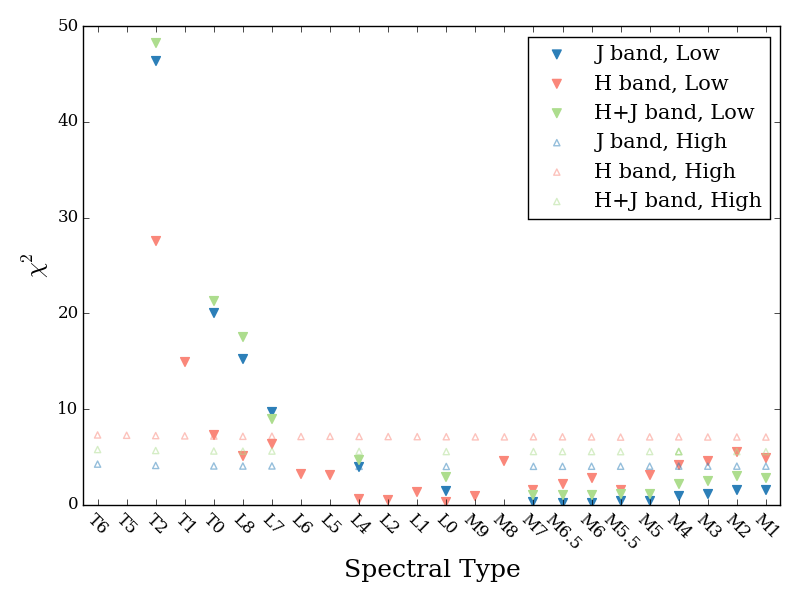}
\caption[$\chi$$^2$ Spectral Type Matching]{Reduced $\chi$$^2$ values for each model tested. The low frequency J spectra matched M type models well, with a best fit of M6$^{+3}_{-1}$. Low frequency H band spectra found a best match of type L0$^{+2}_{-1}$.  High frequency spectra are inconclusive. Combined H and J band low frequency spectra fitting found a best fit type M7$\pm$2.}
\label{chi}
\end{figure}

Since the wavelengths are correlated, each spectra can be binned accordingly to avoid biasing comparisons with model spectral types.  High frequency spectra was binned by averaging three adjacent channels into twelve groups, with the last group averaging four channels to incorporate the leftover channel.  H band low frequency spectra was binned into two groups of twelve channels each and one group of thirteen channels, and J band low frequency spectra was binned into five groups, three of seven channels each and two of eight channels each.  Each model spectra was binned in the same manner for a reduced $\chi$$^2$ comparison.  The results are shown in Figure~\ref{chi}.  High frequency spectra, though showing some spectral features, could not be well fit by the $\chi ^2$ analysis.  As the spectral type models did not have uncertainties, this could not be factored into the analysis and could potentially explain how the small high frequency error gave large $\chi ^2$ values.  $\chi ^2$ values were consistent across high frequency spectral matches and so a best fit match could not be determined from the high frequency spectra alone.  The low frequency J spectra matched M type models well, with a best fit of M6$^{+3}_{-1}$. Uncertainties in spectral type matching were chosen by taking the spectral type one $\sigma$ from the minimum $\chi ^2$.  Low frequency H band spectra found a best match of type L0$^{+2}_{-1}$. When the J and H band low resolution spectra were fit together, the best fit was a M7$\pm$2, in agreement with the \citet{meshkat} result of a type M6.0$\pm$0.5.  Using the spectral type-to-temperature conversion from \citet{stephens}, T$_{eff}=2673^{+175}_{-267}$, in fairly good agreement with \citet{meshkat}.  The uncertainty in effective temperature come from the uncertainty in spectral type.

%----------------------------------------------------------
% Figure 
%----------------------------------------------------------

%----------------------------------------------------------
% Figure: Final Images
\begin{figure}[h!]
  \centering
  \begin{tabular}[b]{@{}p{0.45\textwidth}@{}}
   \includegraphics[width=1.0\linewidth]{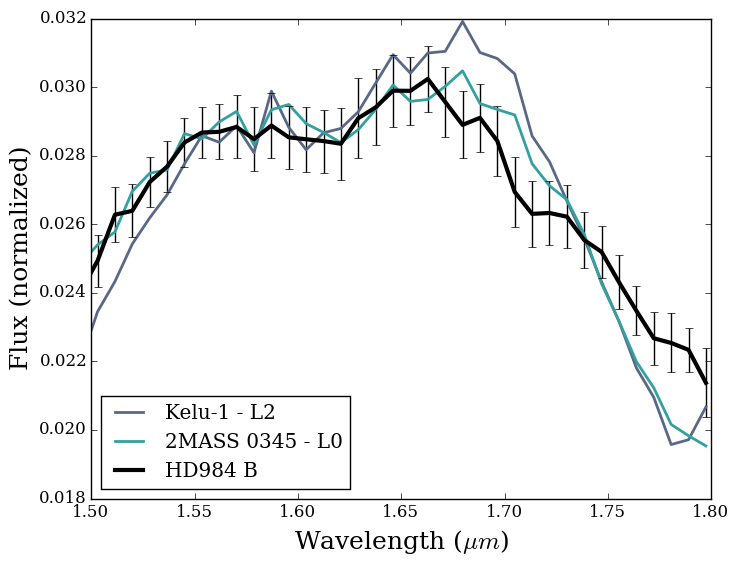} \\
    \centering\small (a)  
  \end{tabular}%
  \quad
  \begin{tabular}[b]{@{}p{0.45\textwidth}@{}}
    \includegraphics[width=1.0\linewidth]{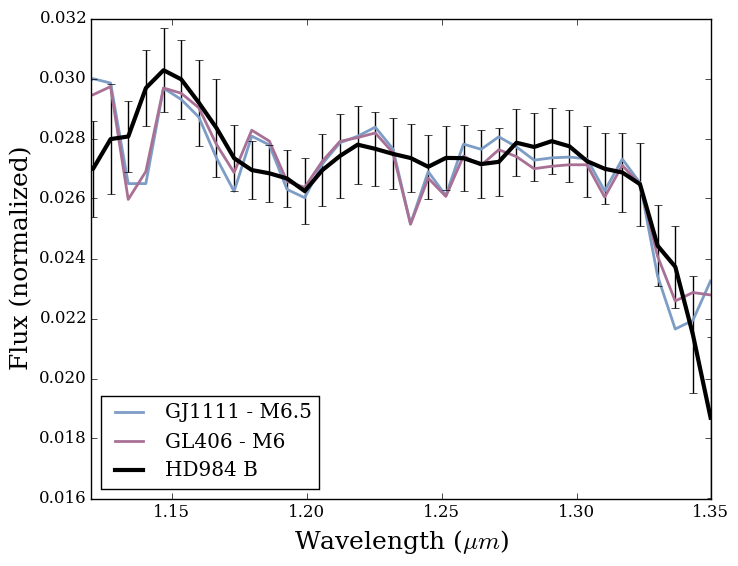} \\
    \centering\small (b) 
  \end{tabular} \\
  \caption[Spectral Models for H and J Spectra]{H band (a) J band (b) spectra plotted with the two most similar spectral type matches as well as the most similar type found from fitting both bands simultaneously (M7). }
  \label{hjspec}
\end{figure}
%----------------------------------------------------------
%----------------------------------------------------------
%% Figure 
%\begin{figure}[h!]
%\epsscale{1}
%\plotone{Figures/spectra.png}
%\caption[Spectra for HD 984 B]{Spectra for HD 984 B shown across all bands.  Solid coloured lines indicate where the model is a best fit for that band and dashed lines indicated the best fit model for a different band.}
%\label{spec}
%\end{figure}
%----------------------------------------------------------
%----------------------------------------------------------

\begin{figure}[h!]
\epsscale{.60}
\plotone{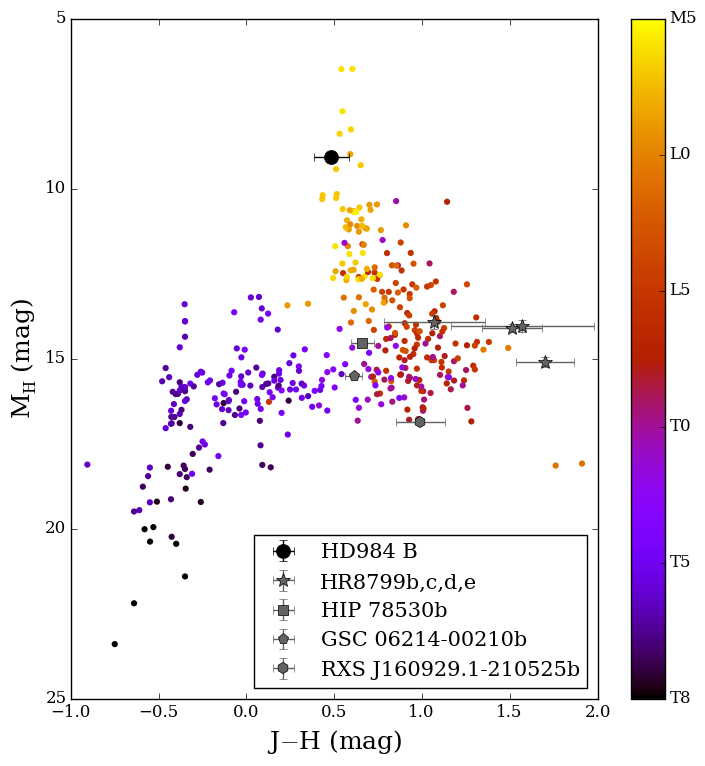}
\caption[CMD for HD 984 B]{J-H colour magnitude diagram showing HD984 B relative to other known brown dwarfs and low mass stars \citep{dupuy}.  Brown dwarf and low mass star spectral types are colour coded on a spectrum from dark purple (T types) to yellow (M types).  HD984 B is shown as a black circle and is located on the late M/early L dwarf cooling sequence.  Photometry for other planets from \citet{zurlo} (HR8799b,c,d,e), and \citet{lachapelle} (HIP 78530b, GSC, RXS).}
\label{cmd}
\end{figure}

Magnitudes for the companion object are calibrated by integrating the companion-to-star 
spectra, and correcting for the GPI filter transmission profile and Vega zero points \citep{derosa16}.  The J and H band apparent magnitudes were calculated to be 13.28$\pm$0.06 and 12.60$\pm$0.05, respectively.  Assuming a distance to the star of 47.1$\pm$1.4 pc \citep{van}, the absolute magnitudes of the object in J and H bands are 9.92$\pm$0.09 and 9.23$\pm$0.08, respectively.  These magnitudes were also compared to other brown dwarfs and low mass stars using a colour-magnitude diagram (see Figure~\ref{cmd}).  When compared with literature brown dwarfs from \citet{dupuy}, these magnitudes further corroborate the spectral type matching result of a late M-type object.
%----------------------------------------------------------
% Figure 
%----------------------------------------------------------

\begin{figure}[h!]
\epsscale{.65}
\plotone{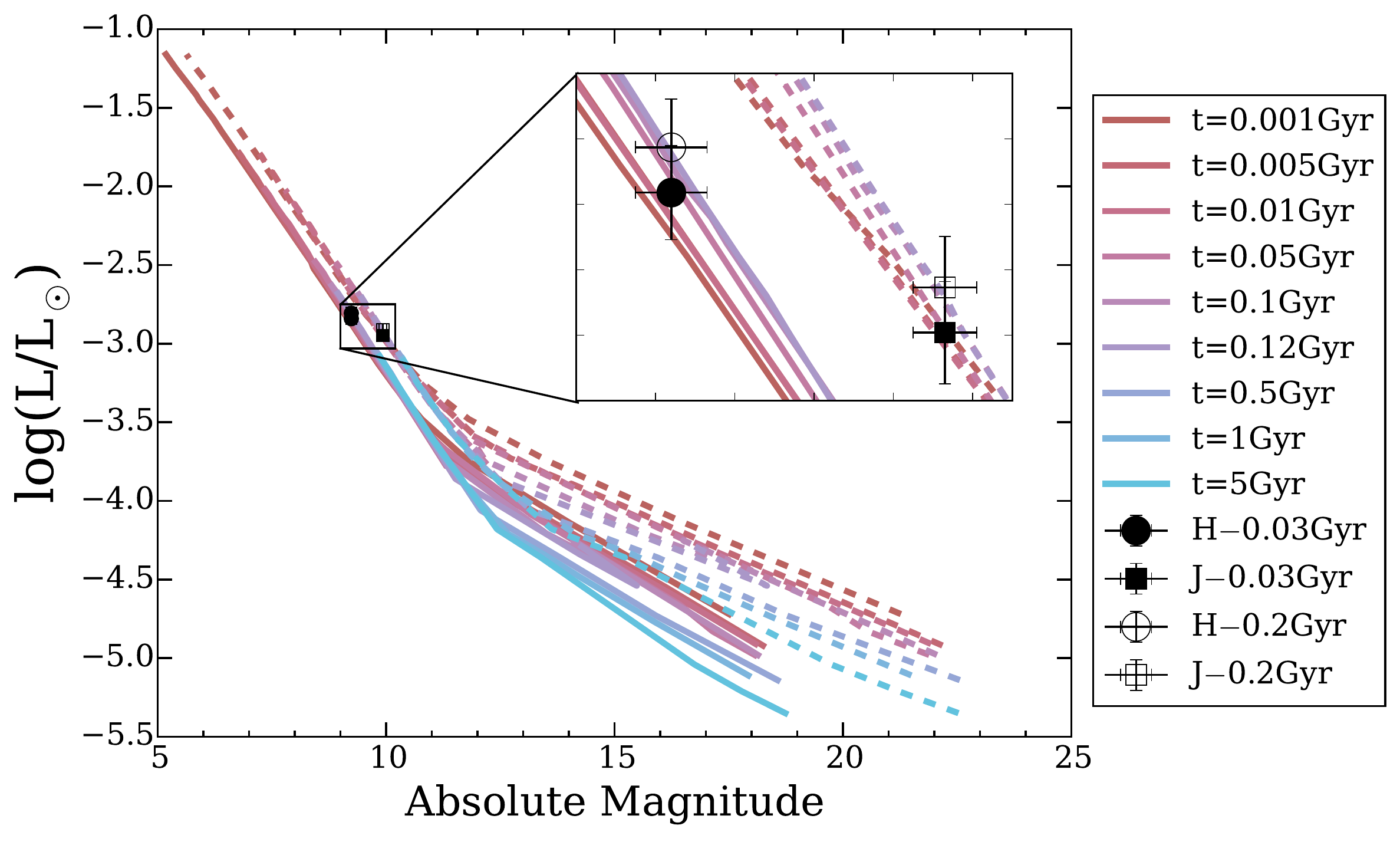}
\caption[Luminosity Models]{DUSTY luminosity models for HD 984 B. Isochrone models are colour coded by age with blue being older models and red younger models.  J band models are shown by dashed lines and H by solid. HD 984 B is show by black points for an age of 30 Myr and grey points for 200 Myr.  Circles represent the H band and squares the J band. The derived luminosities at 30 Myr are log(L/L$_{\sun}$) = $-$2.84$\pm$0.04 and $-$2.95$\pm$0.04  for H and J bands, respectively.  At 200 Myr the derived luminosities for H and J bands are log(L/L$_{\sun}$) = $-$2.81$\pm$0.04 dex and $-$2.91$\pm$0.04 dex. }
\label{lummodel}
\end{figure}

\begin{figure}[h!]
  \centering
  \begin{tabular}[b]{@{}p{0.42\textwidth}@{}}
   \includegraphics[width=1.0\linewidth]{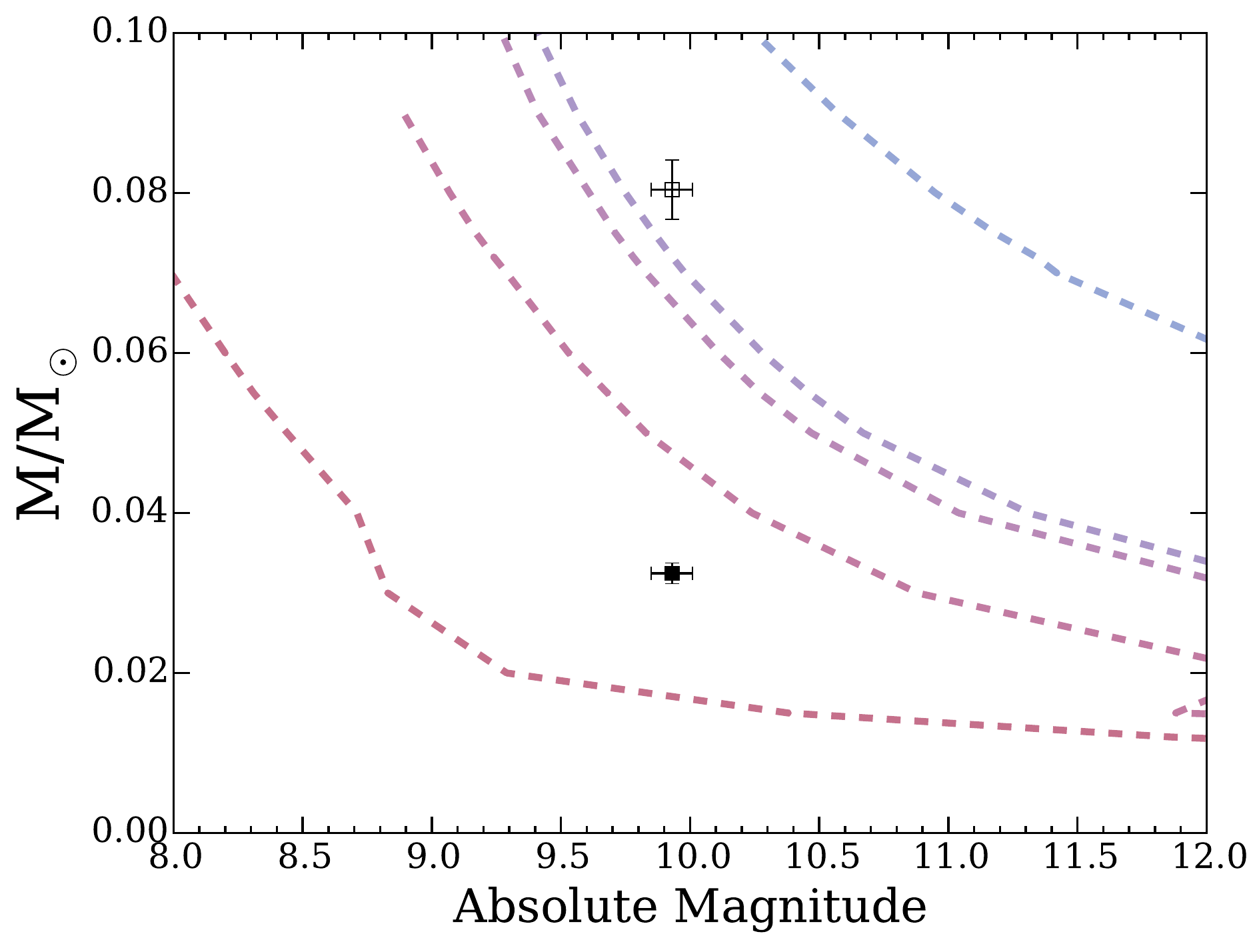} \\
    \centering\small (a)  
  \end{tabular}%
  \quad
  \begin{tabular}[b]{@{}p{0.42\textwidth}@{}}
    \includegraphics[width=1.0\linewidth]{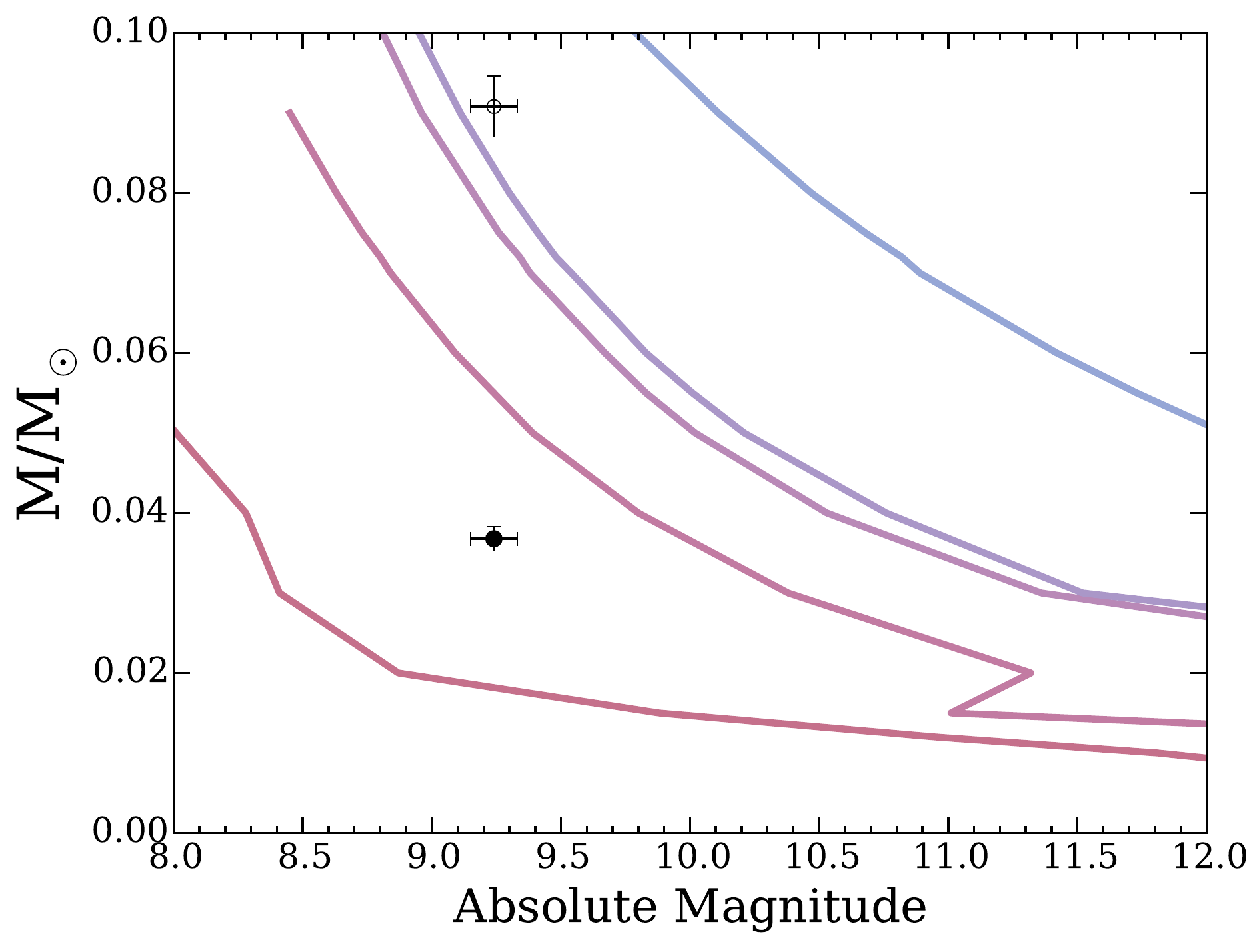} \\
    \centering\small (b) 
  \end{tabular} \\
  \caption[Mass Models]{DUSTY mass models for J band (a) and H band (b) along with the interpolated mass of HD 984 B at 30 Myr and 200 Myr. Colours and symbols the same as Figure~\ref{lummodel}. Masses are found to be 39$\pm$2 M$_{\mathrm{Jup}}$ at 30 Myr to 94$\pm$4 M$_{\mathrm{Jup}}$ at 200 Myr for the H band. The J band yields masses of 34$\pm$1 M$_{\mathrm{Jup}}$ and 84$\pm$4 M$_{\mathrm{Jup}}$ for the same ages. }
  \label{massmodel}
\end{figure}

Using DUSTY isochrone models \citep{chabrier}, we interpolated the luminosity in both bands.  Using the detailed analysis from \citet{meshkat}, we adopt the same age range, 30 -- 200 Myr, for HD 984 B.  Luminosity and mass uncertainties are propagated from uncertainties in the absolute magnitudes.  Although the age of the system is inconsequential when computing luminosity (see Figure~\ref{lummodel}), it is highly differential when estimating mass (see Figure~\ref{massmodel}).  The luminosity, accounting for the age range and both bands is calculated to be log(L$_{\mathrm{Bol}}$/L$_{\sun}$)$=-2.88\pm0.07$ dex, in agreement with \citet{meshkat}. 
%Assuming an age of 30 Myr, HD 984 B would have H and J band luminosities of log(L) = $-$2.84$\pm$0.04 L/L$_{\sun}$ and $-$2.95$\pm$0.04 L/L$_{\sun}$.  At 200 Myr the luminosities for H and J bands are log(L) = $-$2.81$\pm$0.04 L/L$_{\sun}$ and $-$2.91$\pm$0.04L/L$_{\sun}$.  
Using the same technique to calculate the masses we find the H band absolute magnitude corresponds to a range of masses from 39$\pm$2 M$_{\mathrm{Jup}}$ at 30 Myr to 94$\pm$4 M$_{\mathrm{Jup}}$ at 200 Myr. The J band yields masses of 34$\pm$1 M$_{\mathrm{Jup}}$ and 84$\pm$4 M$_{\mathrm{Jup}}$ for the same ages.  Temperature analysis using the same DUSTY models found object temperatures of 2458$\pm$32K to 2800$\pm$37K for J band over the same age range.  The H band magnitude give temperatures of 2545$\pm$28K to 2896$\pm$31K.
%0.091$\pm$0.004 M/M$_{\sun}$ 
%0.080$\pm$0.003 M/M$_{\sun}$

%----------------------------------------------------------
% Figure 
%----------------------------------------------------------
%----------------------------------------------------------
% Figure: Final Images
%----------------------------------------------------------

%----------------------------------------------------------
\begin{deluxetable}{lccc}
\tabletypesize{\scriptsize}
\tablewidth{0pt}
\setlength{\tabcolsep}{3pt}
\tablecaption{System Properties}
\tablehead{{Property} &{HD 984} &{HD 984 B$^1$}& {HD 984 B$^2$}}
\startdata
\hline
Distance &  47.1$\pm$1.4 pc \\
$\mu_{\alpha\ast}$ & 102.79$\pm$0.78 mas/yr\\
$\mu_\delta$ & -66.36$\pm$0.36 mas/yr \\
Age & 30 to 200Myr\\
%$\Delta$ H & -- &6.28$\pm$0.05\\
%$\Delta$ J & -- &6.53$\pm$0.06\\
m$_\mathrm{H}$ & 6.170$\pm$0.038   & 12.58$\pm$0.05 &12.60$\pm$0.05\\
m$_\mathrm{J}$& 6.402$\pm$0.023 & &13.28$\pm$0.06\\
Spectral Type  &  F7V &  M6.0$\pm$0.5 &M7$\pm$2 \\
Temperature (K)  & 6315$\pm$89  & 2777$^{+127}_{-130}$&$2673^{+175}_{-267}$ \\
log(L$_{\mathrm{Bol}}$/L$_{\sun})$ &0.346$\pm$0.027 & $-2.815\pm0.024$&$-2.88\pm0.07$ \\
Mass (M$_{\mathrm{Jup}}$) & $\sim$1.2M$_\sun$ & 33$\pm$6 to 94$\pm$10 &34$\pm$1 to 95$\pm$4  \\
Semi Major Axis (AU)$^3$ &&& 18 [12,27]\\
Period (years)$^3$ &&&  70 [37,132] \\
Inclination$^3$&&& 118$\degree$ [112,127]\\
Eccentricity$^3$ &&&0.241 [0.083,0.495]\\

\enddata
\label{props}
\tablecomments{$^1$Results from \citet{meshkat}; $^2$New results presented in this paper; $^3$Ranges listed encapsulate the 68\% confidence interval.}
\end{deluxetable}
%----------------------------------------------------------

%----------------------------------------------------------------------------------------------------------------------
%----------------------------------------------------------------------------------------------------------------------
\section{Conclusion}\label{conc}

Through new observations of HD 984 B with the Gemini Planet Imager we are able to confirm and add to the findings reported in \citet{meshkat}.  We find a best match spectral type of M7$\pm$2, and J and H band magnitudes of 13.28$\pm$0.06 and 12.60$\pm$0.05, respectively, in agreement with the discovery results and with known field brown dwarfs.  Furthermore, we found a separation and position angle of 216.3$\pm$1.0 mas and 83.3$\pm$0.3$\degree$ (H band) and 217.9$\pm$0.7 mas and 83.6$\pm$0.2$\degree$ (J band), which allowed us to perform the first orbital fitting of the companion when combined with astrometry from 2012 and 2014 epochs. This new epoch gave orbital fitting results of an $\sim$18 AU, 70 year orbit with an eccentricity of 0.24 and inclination of 118$\degree$.  Analysis of the magnitudes found a luminosity of log(L$_{\mathrm{Bol}}$/L$_{\sun}$) = $-2.88\pm0.07$ dex, using DUSTY models.  H band mass estimates, again from DUSTY models, gave an age dependant mass of 39$\pm$2 M$_{\mathrm{Jup}}$ at 30 Myr to 94$\pm$4 M$_{\mathrm{Jup}}$ at 200 Myr.  J band magnitudes give masses of 34$\pm$1 M$_{\mathrm{Jup}}$ and 84$\pm$4 M$_{\mathrm{Jup}}$.  DUSTY models of H band magnitudes gave a temperature of 2545$\pm$28K to 2896$\pm$31K (for an age range of 30 -- 200 Myr), while a spectral type-to-temperature conversion gave T$_{eff}=2673^{+175}_{-267}$.  J band magnitudes yield a DUSTY model temperature of  2458$\pm$32K to 2800$\pm$37K over the same age range.

Although spectral noise correlation rendered much of the wavelength channels dependent, especially at H-band, the spectra were split into high and low spatial frequencies to confirm that spectral correlation is spectral frequency dependent.  While the lesser correlated high frequency spectra could not be used to identify a spectral match in this case, this method may prove useful to match spectral features with K band data, where narrow spectral features, such as CO, can be identified and fitted.  Splitting the spectra in these cases will allow for proper noise statistics and improved $\chi ^2$ analysis.  In addition to this frequency splitting, the covariance method detailed in \citet{greco16} can also be performed per frequency bin to fit the spectra.
\newpage

\chapter{Identifying Damped Lyman $\alpha$ Hosts with Angular Differential Imaging}\label{dla}

%----------------------------------------------------------------------------------------------------------------------
%----------------------------------------------------------------------------------------------------------------------

The advantages of ADI and AO have long been known for their assets in the field of high contrast direct imaging of exoplanets.  However, their application has been previously untested in imaging the host galaxies of damped Lyman $\alpha$ systems (DLA).   This chapter presents a pilot study of the first application of ADI to directly imaging the host galaxy of a DLA.

%----------------------------------------------------------------------------------------------------------------------
%----------------------------------------------------------------------------------------------------------------------

\section{Introduction}
Damped Lyman alpha systems are inferred through their signatures in quasar spectra, but their host galaxies are often hard to see as they are out-shined by the quasars.  Just as it is challenging to see exoplanets when they are overwhelmed by light from their host star, it is difficult to see DLA host galaxies who reside at close impact parameters from a quasar.  The stage laid by the exoplanet community is thus primed for the detection of dim DLA host galaxies near bright quasars.  Before we see how this application can work, let us first consider the basics of DLA systems. 

\subsection{A Short History of DLAs}

\begin{figure}[h!]
\epsscale{.85}
\plotone{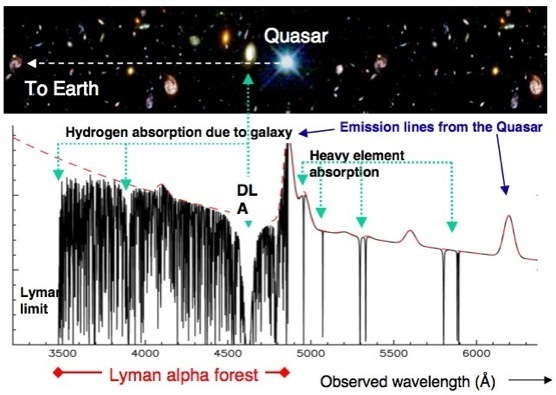}
\caption[Quasar Absorption Systems]{Quasar spectra can be riddled with absorption lines caused by systems along the line of sight.  These absorption lines are extremely helpful in determining the chemical and physical properties of the early universe where emission lines are too faint to probe. Figure from \href{http://academics.smcvt.edu/jomeara/Research_files/droppedImage.jpg}{\url{http://academics.smcvt.edu/jomeara/Research_files/droppedImage.jpg}}.}
\label{qals}
\end{figure}

Quasars, which are typically found at high redshift \citep{hewitt,paris}, are some of the brightest objects in our universe; some exceed the Milky Way's luminosity 100 times \citep{greenstein}.  A subclass of active galaxies, quasars are powered by the infall of gas and dust onto a central supermassive black hole \citep{carroll}.  When a quasar's line of sight is intercepted by intervening gas, the resulting absorption lines can be measured to gather information about the chemical and physical properties in the early universe (See Figure~\ref{qals}).  Since chemical abundances typically are only directly measurable for the brightest stars in our galaxy and its neighbours, quasar absorption systems are an obvious way to investigate chemical abundances outside our local group \citep[e.g.][]{pettini00,prochaska,dessauges-zavadsky}.  Furthermore, a deeper inspection of these systems can yield physical properties of the system like temperature  \citep{kanekar14b,kanekar13,kanekar,york} and gas kinematics  \citep[e.g.][]{prochaskawolfe,ledoux06}, in addition to the chemical abundances.   

%---------------------------------------------------------
%---------------------------------------------------------

Quasar absorption systems are classified by their absorption lines.  Some are detectable through neutral hydrogen absorption \citep[\ion{H}{1}, e.g.][]{wolfe,lanzetta,ellison01,kanekar02,prochaska04,prochaska05,wolfe2,noterdaeme12}, others through ionized magnesium absorption \citep[\ion{Mg}{2}, e.g.][]{sargent,lanzetta,bergeron,steidel,churchill,nielsen13}.  \ion{H}{1} systems are easily identifiable in quasar spectra due to the absorption lines that are generated as neutral hydrogen transitions between the ground state (n=1) and first excited state (n=2) \citep{carroll}. The plethora of absorption lines caused by clouds of neutral hydrogen at different redshifts is called the Lyman-alpha forest \citep{lynds}.  \ion{H}{1} systems are typically classified by their neutral hydrogen column density (log N(\ion{H}{1})). Those with the lowest column densities (log N(\ion{H}{1})$\textless$ $17$ atoms$/$cm$^2$) are known as Ly$\alpha$ forest clouds and are associated with absorption in the intergalactic medium \citep{sargent80,sargent82,fang96,rauch96,rauch98,webb09}.  Lyman limit systems (LLS) have $17$ $\textless$ log N(\ion{H}{1}) $\textless$ 20 atoms$/$cm$^2$, and the highest column density absorbers with log N(\ion{H}{1}) $\ge$ 20.3 atoms$/$cm$^2$ are the so-called damped Lyman $\alpha$ systems \citep[DLAs,][]{wolfe2}. The `damping' in these systems is due to the high column density which results in broadening of the wings of the absorption profile.  

%---------------------------------------------------------
\begin{figure}[h!]
\epsscale{.85}
\plotone{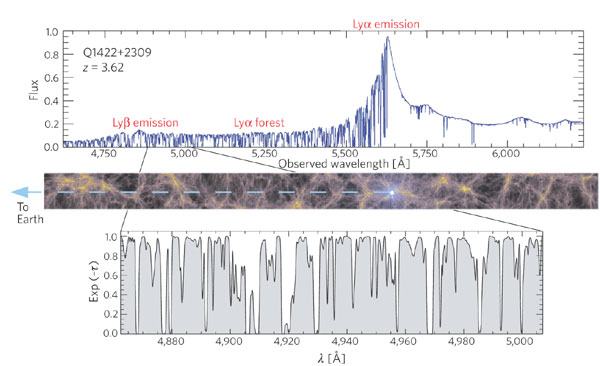}
\caption[Lyman Alpha Forest]{Spectrum of a quasar at z $=3.62$ with an inset of the Lyman $\alpha$ forest produced by intervening neutral gas along the line of sight. Inset and middle image were created using hydrodynamical simulations which have been found to reproduce observed spectra with high accuracy.  Figure from \citet{springel}. }
\label{forest}
\end{figure}
%---------------------------------------------------------
It is generally thought that DLAs are caused by galaxies, and LLS and sub-DLAs, with column densities less than DLAs, are due to galaxy halos and inflows and outflows.  However, high resolution imaging of M31 shows clumping of \ion{H}{1} demonstrating that a sub-DLA could also reside in the centre of a large galaxy \citep{braun}.  Similar studies have been conducted for other nearby galaxies \citep{hoffman,zwaan05}.

 The abundance of the DLA atomic gas reservoir is little changed over many Gyrs at intermediate and high redshifts \citep[e.g.][]{chrighton,sanchez-ramirez,neeleman} indicating that DLAs have gas available for star formation  over a significant fraction of cosmic time.  Combined with the chemical enrichment associated with DLAs, which manifest a wide range of metallicities from a less than 1/100 to in excess Z$_{\odot}$ metallicity \citep[e.g.][and references therein]{rafelski,berg}, we expect these high column density absorbers to represent a broad cross-section of galaxies and hence provide a window into galaxy evolution at these epochs.  Indeed, there is a large body of research spanning the last two decades and beyond that have studied many aspects of DLAs, ranging from their elemental ratios \citep[e.g.][]{pettini00,prochaska,dessauges-zavadsky}, molecular fraction \citep[e.g.][]{ledoux03,noterdaeme}, kinematics \citep[e.g.][]{prochaskawolfe,ledoux06}, ionization properties \citep[e.g.][]{vladilo01,milutinovic} and dust depletion in the interstellar medium \citep[ISM, e.g.][]{pettini94,pettini97,vladilo11}.  Despite these advances, some of the most fundamental properties of the absorbing galaxies, such as their luminosities, stellar masses and morphologies remain unknown for the vast majority of the DLA population.   For many years, most DLAs with identified host galaxies were at relatively low-to-intermediate redshift \citep[e.g.][]{rao, bowen,chen,lebrun},  but there are now a growing number detected at $z>2$ \citep[e.g.][]{djorgovski,weatherley,fynbo,peroux12,kashikawa,hartoog,mawatari}.  Nonetheless, the identification of wholesale numbers of host galaxies for DLAs remains one of the outstanding challenges in the field.
 
%-------------------------------------------------------------------------------------------------------------------

\subsection{Properties of DLAs}
As an increasing number of DLAs are being discovered, an understanding of their physical nature is slowly forming.  Additionally, now that numerous DLAs have been found at a range of redshifts \citep[e.g.][]{djorgovski,bowen,chen,fynbo,rao,mawatari}, it is also possible to determine how the population of DLAs has evolved over time.

\begin{figure}[h!]
\epsscale{.65}
\plotone{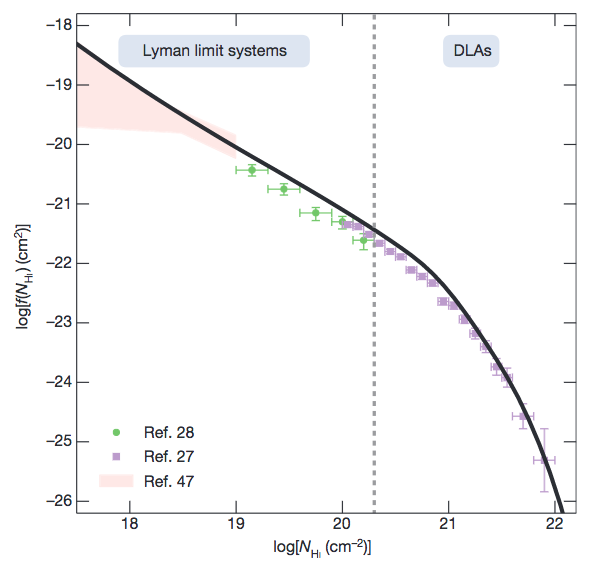}
\caption[Column Density Distribution Function]{The distribution of absorption systems is heavily weighted towards Lyman limit systems even though it is in fact the damped Lyman alpha systems which contain most of the neutral gas.  Observations of absorption systems at  z=3 from \citet[][green]{zafar}, \citet[][purple]{noterdaeme09}, and \citet[][pink]{prochaska10} compared with simulated galaxies from a hydrodynamical simulation \citep{vogel}.}
\label{cddf}
\end{figure}

\begin{figure}[h!]
\epsscale{.65}
\plotone{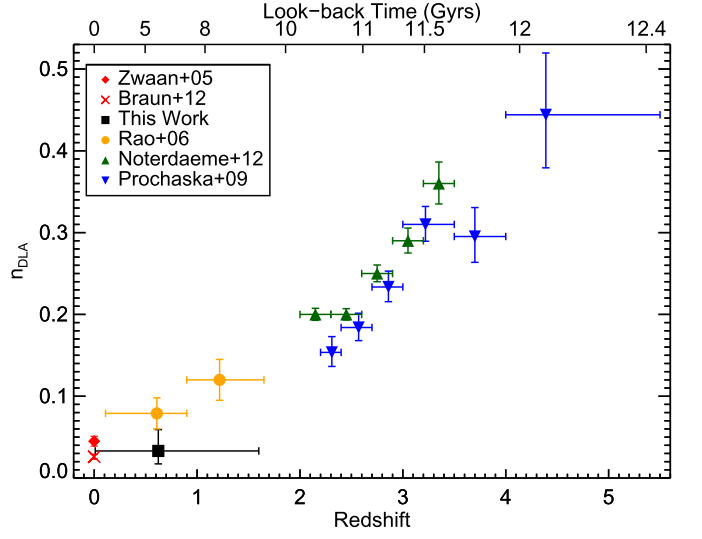}
\caption[Evolution of DLAs]{\citet{neeleman} has found a decrease in the number density of DLAs over time.}
\label{h1evolution}
\end{figure}

\begin{figure}[h!]
\epsscale{.65}
\plotone{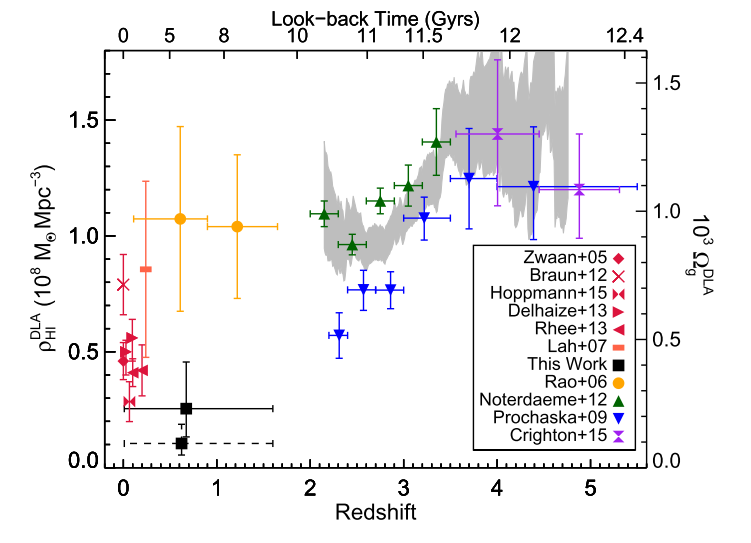}
\caption[Evolution of $\Omega_{\mathrm{HI}}$]{\citet{chrighton,sanchez-ramirez,neeleman} have all seen a decrease in $\Omega_{\mathrm{HI}}$ over time. Note $\Omega_{\mathrm{HI}}$ was converted to $\rho^{\mathrm{DLA}}_{\mathrm{HI}}$ assuming DLA gas accounts for  85\% of the cosmic \ion{H}{1}. Figure from \citet{neeleman}.}
\label{omegah1}
\end{figure}

%HI Content
Given the slope of the column density distribution function of high redshift absorption line systems (see Figure~\ref{cddf}), it is the DLAs, rather than the more numerous Ly$\alpha$ forest clouds, or Lyman limit systems, that contain the bulk of the neutral gas available for star formation \citep[e.g.][]{tytler,sargent,peroux01,noterdaeme12}. In addition, a decrease in the number of DLAs since the formation of the universe has been noticed \citep[See Figures~\ref{h1evolution}, \ref{omegah1};][]{neeleman}.  Similarly, measurements of $\Omega_{\mathrm{HI}}$, the mass density of atomic hydrogen gas scaled to the critical density, have a noticeable downward trend at low redshift \citep[see Figure~\ref{omegah1};][]{chrighton,sanchez-ramirez,neeleman}.
%---------------------------------------------------------
%---------------------------------------------------------
%---------------------------------------------------------
%---------------------------------------------------------
%---------------------------------------------------------
%---------------------------------------------------------'
%METALLICITY

The great advantage of DLAs seen in quasar spectra is the high resolution at which the spectra can been observed thanks to the quasars' extreme brightness. This allows for characterization of metallicity abundances in the early universe \citep[e.g.][]{pettini94,pettini97,prochaska02,rafelski,berg}.  Comparing the amount of metals relative to the amount of hydrogen in galaxy or gas cloud, allows for a measure of the metal enrichment which is indicative of age and galactocentric radius \citep[e.g.][]{aller,searle,berg13,magrini}.  Metallicity is computed related to the solar values through the equation:
\begin{equation}
\mathrm{[X/H]} = \mathrm{log(X/H)}  - \mathrm{log(X/H)}_\sun
\label{eq:metallicity}
\end{equation}

where (X/H)$_\sun$ is the solar abundance ratio of the number of X atoms to H atoms ($n_{\mathrm{x}}/n_{\mathrm{H}}$).  Metallicities of DLAs have long been studied \citep[e.g.][]{morton,meyer87,meyer,pettini90}. With the advent of 8-10m telescopes and high resolution spectrometers, metal lines could be more accurately measured \citep[e.g.][]{prochaskawolfe,prochaska,dessauges-zavadsky,akerman}.  Studies of metallicity evolution show an increase in metallicity with decreasing redshift \citep[e.g.][]{savaglio00,rafelski}.

%MZR

\begin{figure}[h!]
\epsscale{.65}
\plotone{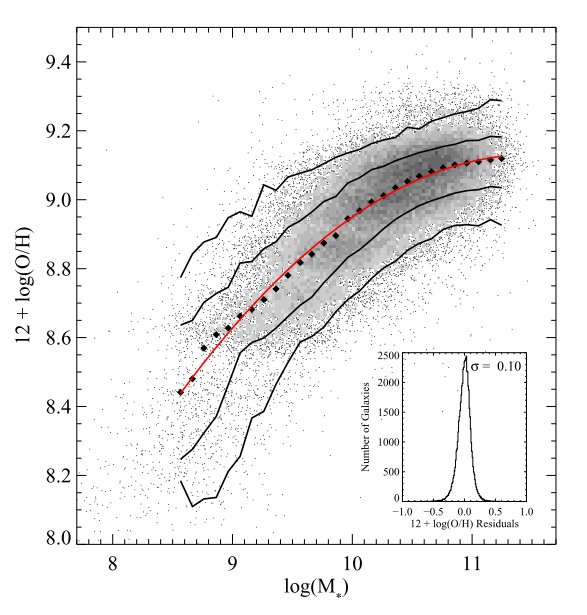}
\caption[MZR from $\sim$53,400 star-forming galaxies]{MZR from $\sim$53,400 star-forming galaxies \citep{tremonti}.  The large black dots represent the median in bins of stellar mass.  Black lines are contours enclosing 68\% and 95\% of the data, and the red line is a polynomial fit to the data. Residuals to the fit are shown in the inset plot.   }
\label{tremontiMZR}
\end{figure}

\begin{figure}[h!]
\epsscale{1.0}
\plotone{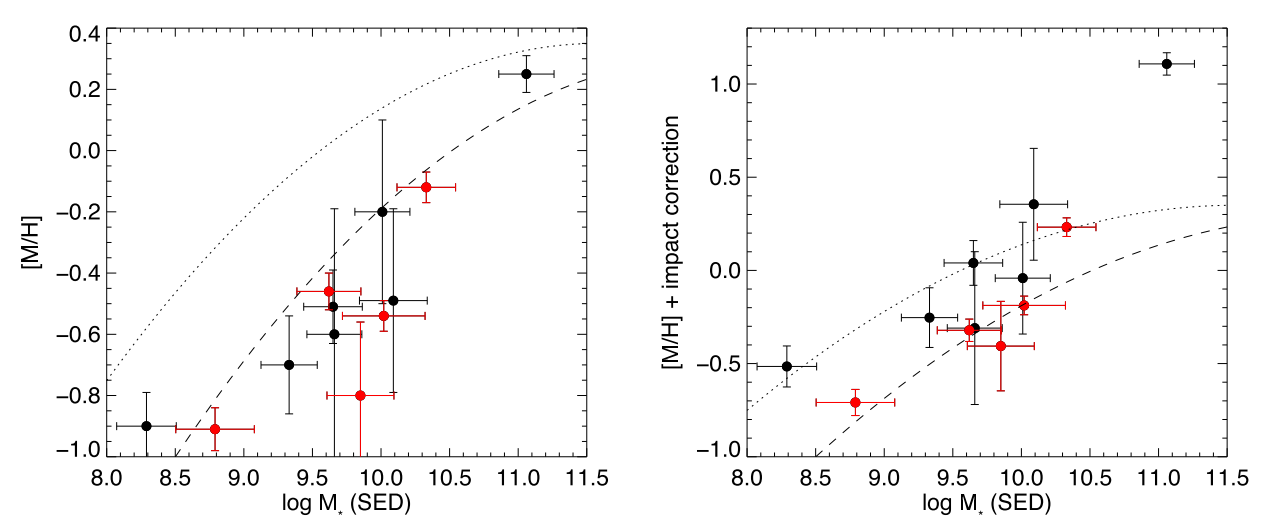}
\caption[MZR from \citet{christensen}]{A DLA MZR from \citet{christensen}.  Black dots show DLAs at z \textless 1, and red dots are DLAs at z \textgreater 2.  The dotted and dashed lines show MZR from \citet{maiolino} for galaxies at z$=0.7$ and z$=2.2$ respectively.  The righthand plot shows the DLAs once their metallicities have been corrected for by a factor of $0.022b$, where $b$ is the impact parameter.   }
\label{christensenMZR}
\end{figure}

%Direct measurements of the mass of a DLA host galaxy are challenging.   When multi-band imaging is available, it is possible to fit spectral energy distributions (SED) to find the masses \citep[e.g.][]{peroux,krogager13}, but if only metallicities are available, masses can be inferred by comparing the abundances of heavy metals to hydrogen \citep{christensen}.  
Metallicity is also indicative of the galaxy mass.  Heavy elements, which are produced in massive stars, are dispersed into the interstellar medium (ISM) via solar winds and other mass loss events.  Thus, the chemical abundances of heavy elements in the ISM is correlated to stellar mass in star-forming galaxies.  Since oxygen is one of the strongest lines in emission line spectra of \ion{H}{2} regions of galaxies, it is often used as a proxy for metallicity.  The correlation between mass and metallicity was first demonstrated by \citeauthor{lequeux} in 1979 by using a sample of nearby galaxies. More recently, \citet{tremonti} used a catalogue of $\sim$53,400 star-forming galaxies to find a tight mass-metallicity relation (MZR) with a spread of only $\sim$0.1 dex (Figure~\ref{tremontiMZR}).  With the availability of datasets from large surveys in the last decade, the MZR has been studied extensively.  It has been extended down to $\sim$ $10^6$ M$_\sun$ \citep{lee,zahid12,berg12}, and out to z $\sim$ 3 \citep{maiolino,erb,zahid,zahid14,savaglio}.  The MZR has also been studied for DLAs \citep{maiolino, christensen,moller}.  Since oxygen can be challenging to measure in quasar absorption spectra, other metals can be used.  Zinc is a common choice as it is less prone to depletion than other metals, like iron. Sulphur can also be used as a proxy as it traces oxygen \citep{berg13}.  Using a sample of previously discovered DLAs, \citet{christensen} determine a MZR that is dependant on impact parameter, as shown in Figure~\ref{christensenMZR}.   Accounting for the impact parameter they find the following MZR:
\begin{equation}
\mathrm {log(M_\star^{DLA}/M}_\sun)=1.76(\mathrm{[M/H]}+0.022b+0.35z+5.04) ,
\label{eq:mzr}
\end{equation}

where [M/H] is the metallicity, $b$ is the impact parameter and $z$ is the redshift. The notation [M/H] used for metallicity is in reference to the solar scale, as described in Equation~\ref{eq:metallicity}.  Notably, this relation can also be used to infer a host galaxy's mass when only the metallicity and redshift are known.  

%---------------------------------------------------------
%---------------------------------------------------------
%---------------------------------------------------------
%---------------------------------------------------------
%Spin Temp
Neutral hydrogen is a ubiquitous component of low-density regions of the ISM.  As the spins of the proton and electron flip from parallel to antiparallel, a photon is emitted at the hyperfine 21-cm line \citep{carroll}.  By combining observations of the 21-cm optical depth and N(\ion{H}{1}) determined from Ly$\alpha$, it is possible to determine the spin temperature (T$_s$), which indicates the fraction of warm ($\sim$$500-5000$K) and cool ($\sim$$40-200$K) neutral medium of the intervening DLA:

\begin{equation}
\mathrm{N(HI)} = 1.823 \times 10^{18} \bigg( \frac{T_s}{f}\bigg)\int \tau dV ,
\label{eq:spintemp}
\end{equation}

where $f$ is the covering factor (fraction of the radio flux density covered by the DLA which is measured at frequencies near the 21-cm line using very long baseline interferometric observations) and $\tau$ is the optical depth \citep{york,kanekar03b}.

The majority of DLAs for which spin temperature measurements exist exhibit values of T$_s$ $>$ 1000 K, with relatively few low spin temperature measurements, particularly at high redshifts   \citep[e.g.][for a rare example of a low T$_s$, high $z$ absorber]{kanekar14b,kanekar13,kanekar,york}.   An anti-correlation between the gas phase metallicity and T$_s$ has been proposed by \citet{kanekar}, whereby the abundance of metals allow effective cooling in the ISM \citep {kanekar01}.  In other words, galaxies with lots of metals have more radiation pathways for gas cooling which would result in a higher cold gas fraction and thus a lower spin temperature.  The most recent compilations of abundance and spin temperature measurements seem to support this anti-correlation \citep{ellison12,kanekar14}, although sample sizes remain small.  It has been suggested that galaxies with low spin temperatures tend to be massive, luminous spiral types while those with high spin temperatures are associated with dwarf galaxies \citep{york}.  Additionally, \citet{kanekar14} find a significant redshift evolution in spin temperature with an increase in temperature at higher redshifts.  

%Kinematics

%Molc Content

%Host Galaxies

\subsection{Host Galaxy Identification}
Despite the abundant identification of DLAs in quasar spectra \citep{noterdaeme12}, only a handful of host galaxies to these systems have been positively identified with direct imaging \citep[e.g.][]{omeara06,bouche,fum,peroux,hamilton}.  The fundamental challenge for the identification of a DLA host galaxy is the overwhelming brightness of the quasar relative to the galaxy, which is expected to be found at low impact parameter from the quasar \citep{rao03}.  A few methods have been developed to circumvent the blinding quasar light, and to allow the observer to search for galaxy hosts close to the quasar line of sight.  One such method is the 'Double-DLA' technique \citep{fum,omeara06}, in which a quasar exhibits multiple high column density absorbers.  The higher redshift system is used as a natural blocking filter to eliminate (rest-frame) far ultraviolet emission from the quasar so that the continuum emission from the lower redshift DLA can be directly measured. 

An alternative approach has been to search for absorbers in the spectra of the optical afterglow of gamma-ray bursts (GRBs) \citep[e.g.][]{vreeswijk,chen05,prochaska07}.  GRB afterglows can shine brighter than quasars, but they fade rapidly \citep{kann}.  Spectra taken along the sightline of the GRB can reveal intervining DLA, sub-DLA and other absorbing systems \citep{schulze}.  After the afterglow has faded, follow up imaging and spectroscopy can be used to identify the host galaxy \citep[e.g.][]{masetti, pollack,vree,schulze,ellison06,chen09,chen10}.  It is thought that GRB DLAs probe the interstellar medium of a galaxy \citep{schady}.  Again, this technique is limited to a relatively small number of absorbers.  

Finally, spectroscopy with integral field units (IFUs) has been used to successfully identify DLA hosts at z $\sim$ $1-2$ \citep[e.g.][]{peroux,bouche}.  This technique uses narrow band images generated from IFU data cubes to search for H$\alpha$ or other emission lines along quasar sightlines, and is applicable when the DLA redshift optimally places emission lines between the bright sky lines in the IR.  Similarly, using a prism, like that of the Advanced Camera for Surveys (ACS) on Hubble, can spread the quasar light in a quantifiable manner allowing for its subtraction \citep{hamilton}.

\begin{figure}[h!]
\epsscale{.95}
\plotone{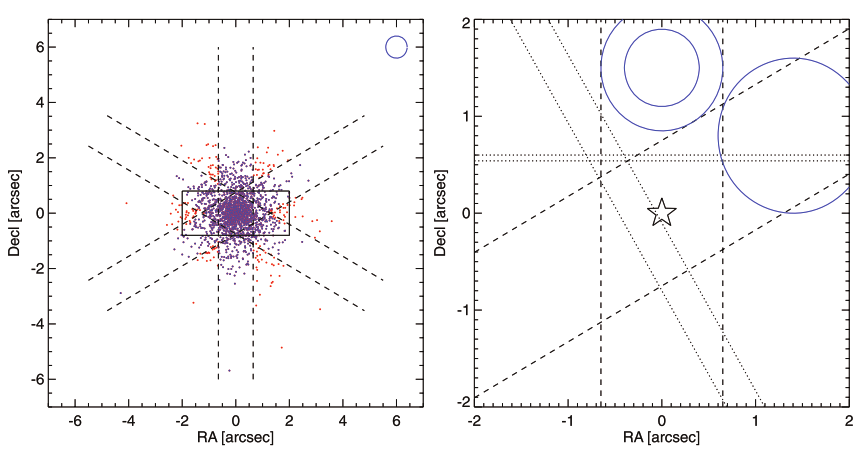}
\caption[DLA Triangulation]{At left, \citet{fynbo} simulate how many DLAs would be visible with three X-Shooter slit orinetations.  Of 5000 simulated DLAs (dots), slightly over 90\% would be visible with this positioning (blue dots).  On the right, the two orientations (dashed lines) with their respective seeing (blue circles) used for observing Q 2222-0946 are shown.  The dotted lines show 1$\sigma$ regions for the measured impact parameters in both slits.}
\label{b}
\end{figure}

%impact parameter
Even when the host cannot be directly imaged, it is still possible to determine the impact parameter with a careful application of long-slit spectroscopy triangulation.  \citet{fynbo} find that by using multiple observations with slits at different angles, they can find overlap between observations which can identify the host galaxy's position angle and separation from the quasar line of sight (see Figure~\ref{b}).  Previous DLA host galaxy studies have found impact parameters are typically a few tens of kpc with higher log N(\ion{H}{1}) systems being systematically closer to the quasar \citep[e.g.][]{moller98,chen,cooke,monier}.  \citet{rao} find a median impact parameter for DLA hosts of 17.4 kpc and \citet{reeves} note that \ion{H}{1} DLA discoveries with impact parameters greater than 20 kpc are 'extremely rare'.  Larger separations have been concluded to more likely be associated with multiple absorbers \citep{ellison07}. 

%---------------------------------------------------------
%---------------------------------------------------------

%AO advantages

AO has an obvious application in the search for DLA host galaxies, thanks to the improved diffraction-limited angular resolution and deep contrast that can be achieved.  Despite its obvious benefits, relatively few AO-aided searches for DLA hosts, outside IFU usage, have been attempted in the past \citep{chun06,chun10}.   Part of the historical challenge of AO observations has been the limited sky coverage of natural guide stars.  Moreover,  AO imaging techniques require fairly complex analysis methods, to deal with a highly time and space sensitive PSF for example.  Luckily, much time and work has gone into characterizing and developing data reduction techniques, and the introduction of laser guide star technology opens the night sky to the application of AO.

In particular, the AO technique ADI, which was developed for directly imaging exoplanets \citep[see Section~\ref{adi},][]{marois06}, holds promise for the field of DLAs.   The subtraction of the reference PSF removes off-axis light which increases the sensitivity, allowing for the detection of fainter objects at lower impact parameters, which is where DLAs are expected to be found.  Previous AO imaging of DLAs have largely used multi-step methods of azimuthal PSF subtraction \citep{chun10,chun06}.  Applying ADI to AO imaging of DLAs can greatly simplify the reduction process and potentially increase the limits of detection.

\subsection{Overview}
This chapter presents the pilot study of the first application of ADI with AO to directly imaging the host galaxy of the DLA  (z$_{abs}$=0.602, log N(HI) = 21.2$\pm$0.1 $/$cm$^2$) seen towards the quasar J1431+3952. This chapter is organized as follows: Section~\ref{overview} describes the DLA that is the target of this pilot study. Observations are presented in Section~\ref{qsoobsv} and reduction methods in Section~\ref{qsoredu}.  In Section \ref{qsoresults} we describe the determination of candidate positions and magnitudes (\ref{qsosimul}), our calculations of stellar mass (\ref{qsomass}), and the detection limits of this study (\ref{qsocurves}). Section~\ref{qsodisc} discusses candidate DLA host galaxies in the context of known scaling relations and directions for future work. Section~\ref{qsoconc} summarizes our conclusions.  Throughout, a $\Lambda$ cold dark matter cosmology is assumed with $H_0$ = 68 km s$^{-1}$ Mpc$^{-1}$,  $\Omega_\Lambda$ = 0.70 and $\Omega_M$ = 0.30.

%----------------------------------------------------------------------------------------------------------------------
%----------------------------------------------------------------------------------------------------------------------

\section{Target selection and overview}\label{overview}

AO requires a bright point source (usually a star) close to the target in order to compute corrections to the incoming wavefront.  When there is no star bright enough close to the target, it is necessary to use a LGS.    Although the LGS is useful in fields without bright guide stars, a source is still needed for tip/tilt corrections.  If the target is bright enough (R-band apparent magnitude m$_R$\textless17.5), it can be used for tip/tilt corrections.  For this pilot study, we therefore selected a relatively bright QSO with a known DLA: J1431+3952  (z$_{em}$=1.215, m$_K$=14.03, m$_R$=16.07).  The only previous attempts to identify the galaxy counterpart of this DLA relied on shallow, relatively poor seeing quality\footnote{The typical seeing for SDSS, given by the PSF width, is 1.43" \citep{sdss}.} Sloan Digital Sky Survey (SDSS) imaging.  Both Ellison et al. (2012) and Zwaan et al. (2015) identified a galaxy at an impact parameter of $\sim$ 5 arcsec; although the relative proximity makes this galaxy an appealing host galaxy candidate, its photometric redshift was determined by Ellison et al. (2012) to be $z=0.08\pm0.02$, inconsistent with the absorption redshift  ($z_{abs}=0.602$).  

In addition to the advantageous brightness of the background QSO, we selected J1431+3952 for our pilot observations because of extensive characterization of its DLA by UV, optical and radio spectroscopy.  Initially selected as a candidate DLA based on its strong MgII lines, \citet{ellison12} used UV spectroscopy with the Cosmic Origins Spectrograph (COS) on the Hubble Space Telescope (HST) to determine an HI column density of log N(HI) = 21.2$\pm 0.1$.  Thanks to its radio brightness, J1431+3952 has also been the target of 21cm absorption measurements (Ellison et al. 2012; Zwaan et al. 2015). By combining the 21cm optical depth and N(HI) determined from Ly$\alpha$, it is possible to determine the spin temperature (T$_s$), which indicates the fraction of warm and cool neutral medium of the intervening DLA.  The majority of DLAs for which spin temperature measurements exist exhibit values of T$_s$ $>$ 1000 K, with relatively few low spin temperature measurements, particularly at high redshifts   \citep[e.g.][for a rare example of a low T$_s$, high $z$ absorber]{kanekar14b,kanekar13,kanekar,york}. Ellison et al. (2012) determined that the DLA towards J1431+3952 exhibits one of the lowest spin temperatures yet measured (T$_s$ = 90$\pm$23 K), and Zwaan et al. (2015) report an even lower value  (T$_s$ = 65$\pm$17 K)\footnote{Likely due to
a larger measured integrated optical depth of the abosrption feature, although \citet{zwaan} do also estimate a larger covering factor 3 times larger than \citet{ellison12}.}.  An anti-correlation between the gas phase metallicity and T$_s$ has been proposed by \citet{kanekar}, whereby the abundance of metals allow effective cooling in the ISM \citep{kanekar01}.  The most recent compilations of abundance and spin temperature measurements seem to support this anti-correlation \citep{kanekar14}, although sample sizes remain small.  The general association between high elemental abundances and low T$_s$ is observed in the DLA towards J1431+3952:  Ellison et al. (2012) determine a relatively high (compared with the general DLA population, e.g. Berg et al. 2015) metallicity of [Zn/H] $ = -0.80$$\pm$$0.13$.  

A further prediction of \citet{kanekar} is that low spin temperature DLAs should be hosted by relatively luminous galaxies.  A stellar mass--metallicity relation is observed over a wide range in redshifts, such that more metal rich galaxies are hosted by more massive galaxies  \citep[e.g.][]{tremonti,erb,zahid14}, which should in turn be more luminous.  Indeed, the small number of low spin temperature DLAs which have galaxy counterparts tend to be relatively luminous spirals \citep{kanekar02,kanekar03}.  However, very few DLAs have the full complement of data that permit an investigation between metallicity, spin temperature and identified galaxy counterpart to more completely characterize the inter-relation of these properties.  The selection of a low T$_s$ DLA, with high metallicity (that is hence predicted to be hosted by a relatively bright galaxy), and also towards a bright background QSO, therefore makes an excellent target for our pilot test of the ADI technique applied to the search for DLA hosts.  A complete summary of the DLA's properties can be found in Table~\ref{dlaprops}.

%----------------------------------------------------------------------------------------------------------------------
%Table: DLA Properties
\begin{table}[h!]
\centering
\centering\caption{Properties of the DLA associated with QSO J1431+3952}
\begin{tabular}{ccc}
    \toprule
    \midrule
    \bfseries Property & 
    \bfseries Value & 
    \bfseries Reference \\
    \midrule
z$_{abs}$ &0.60190 &1 \\
logN(HI)&21.2$\pm$0.1&1\\
{[Fe/H]}&$-$1.50$\pm$0.11&1\\
{[Zn/H]}&$-$0.80$\pm$0.13& 1\\
{[Cr/H]}&$-$1.31$\pm$0.13&1\\
{[Mn/H]}&$-$1.61$\pm$0.12&1\\
{[Ti/H]}&$-$1.41$\pm$0.12&1\\
T$_s$(K)&65$\pm$17, 90$\pm$23&2,1\\
Covering Fraction &0.32,  0.95&1,2\\
    \midrule
   \toprule
\multicolumn{3}{c}{References: 1 - \citet{ellison12}; 2 - \citet{zwaan}. } \\
\multicolumn{3}{c}{Abundances determined using solar values from \citet{asplund}.} \\
\end{tabular}   
\label{dlaprops}
\end{table}
%----------------------------------------------------------------------------------------------------------------------

%Subsection
\section{Observations}\label{qsoobsv}

The quasar J1431+3952 was observed using the f/32 camera on the Near InfraRed Imager and Spectrometer \citep[NIRI,][]{hodapp} at Gemini North over three nights in 2013 during program GN-2013A-Q-17. NIRI was used with the Gemini AO system, ALTAIR \citep[ALTtitude conjugate Adaptive optics for the InfraRed,][]{herriot}.  For the AO system, a LGS was used in conjunction with the quasar, the latter of which was sufficiently bright for tip/tilt corrections.  The NIRI pixel scale with the f/32 field lens is 0.0214 arcsec/pixel.  The typical Strehl ratio is $\sim$10\%.  

The images were all taken in the K-short band ($1.99-2.30\mu$m) with 60 second exposures of one coadd at the time when the object was closest to transit, so as to maximize the field-of-view rotation.  Images were obtained for three nights, however,  one night showed a strong elongation in the light from the quasar, most likely due to the errors locking the AO system, and so could not be used in this study.  Of the images used, 19 were from April 27, 2013 and 123 were from April 26, 2013. These 142 images were combined using a signal to noise ratio weighting scheme as described in Section ~\ref{qsoredu}.  A full list of observing conditions can be found in Table~\ref{observations}.
%----------------------------------------------------------------------------------------------------------------------
%Table 2: Observations
\begin{deluxetable}{ccc}
\tabletypesize{\scriptsize}
\tablewidth{0pt}
\setlength{\tabcolsep}{3pt}
\tablecaption{Observational Details}
\tablehead{{Date}&{26 April 2013} &{27 April 2013}}
\startdata
Number of Images & 123 & 19\\
FOV Rotation & 86.0\degree & 15.3$\degree$ \\
Average airmass & 1.13 & 1.07 \\
Atmospheric Seeing &  0.41$\pm$0.06 & 0.30$\pm$0.04\\
\enddata
\label{observations}
 \tablecomments{Atmospheric Seeing is listed as the average with the error representing the standard deviation in seeing during the course of the night.}
\end{deluxetable}
%----------------------------------------------------------------------------------------------------------------------

%Subsection
\section{Reduction Methods}\label{qsoredu}

The images from the two nights in April were each processed separately using a routine developed specifically for this project.  A master dark image was made from a median combination of all dark images taken that night, and a master flat was constructed from the median of all the flat images, after they had been dark subtracted and flux normalized to one.  Each quasar science image was dark subtracted and divided by the master flat field image. All science images from April 26 had vertical striping in the lower left quadrant due to an error in detector readout.  Corrections were made by taking the first ten rows in each quadrant, calculating the median value of each column and subtracting that column median value from each row over the entire quadrant.  Similarly, the images from both nights had distinct horizontal readout bias from the detector chips.  This was corrected by taking the first hundred columns in each row, calculating the median value, and subtracting that median value from each pixel in the whole row.   Additional corrections were made for instrumental distortion using a distortion map by R. Galicher (private communication).  After corrections, the image was unsharp masked with a large median box of 50 $\times$ 50 pixels (1.07 $\times$ 1.07 arcseconds) to remove the background flux and yet large enough to avoid removing flux for a resolved and diffuse galaxy.  Each image was registered to the common centre with sub-pixel accuracy by fitting a gaussian to the quasar's PSF with the IDL routine \texttt{gauss2dfit.pro}.  Registration was executed with a custom IDL routine using the IDL function 'rot' with a cubic convolution interpolation method and an interpolation parameter of -0.5.

While very advanced PSF subtraction techniques exist, such as LOCI \citep{lafreniereb}, SOSIE \citep{marois10}, KLIP  \citep{soummer}, TLOCI \citep{marois14} and LLSG \citep{gomez}, we decided to use a simple subtraction algorithm to minimize self-subtraction since the DLA host is likely to be resolved.  For our method, a reference PSF for the quasar was created by taking the median of all the science images for each night.  This PSF was subtracted from each individual image, significantly removing the quasar's signal, and allowing the detection of low impact parameter galaxies.  ADI has the advantage of offering high correlation between PSFs in consecutive exposures, enabling a precise reconstruction of the overall PSF.  Any off-axis point-source light, such as light from a DLA host galaxy, will be mostly median averaged out in the reference PSF and will not be significantly subtracted from the final image. Extended objects, however, will be subject to some self-subtraction.  As the quasar image is not moving on the detector during the entire observing sequence, PSF subtraction simultaneously doubles for sky subtraction, resulting in an increased signal to noise over the whole image, not just at the centre where the PSF is subtracted.  This allows for substantial time savings during observations since sky images are not required.  After PSF subtraction, the images were rotated to north up using the IDL function 'rot' which used the cubic convolution interpolation method with an interpolation parameter of -0.5. The rotated images were then median combined.  Since the two nights have different seeing and number of images, it is important to combine them in a way that weights the noise from each night to maximize the detection limit.  To achieve this, a final combined image was created from a weighted mean of the images from the two nights.  This was done by scaling the flux of the images to be the same and then combining the images with a weight determined by their signal to noise ratios:

\begin{equation}
image = \frac{(a*snr_a^2)+(b*snr_b^2)}{snr_a^2+snr_b^2},
\label{eq:weight}
\end{equation}

where $a$ refers to images from one night, and $b$ the other.  The signal to noise ratios for each image ($snr_a$ and $snr_b$) are calculated by dividing the maximum value of the quasar by the standard deviation in an empty annulus around the quasar. The SNR ratio for the 27th to the 26th was 0.599.

\section{Results}\label{qsoresults}

\begin{figure}[h!]
  \centering
    \begin{tabular}[b]{@{}p{0.32\textwidth}@{}}
   \includegraphics[width=1.0\linewidth]{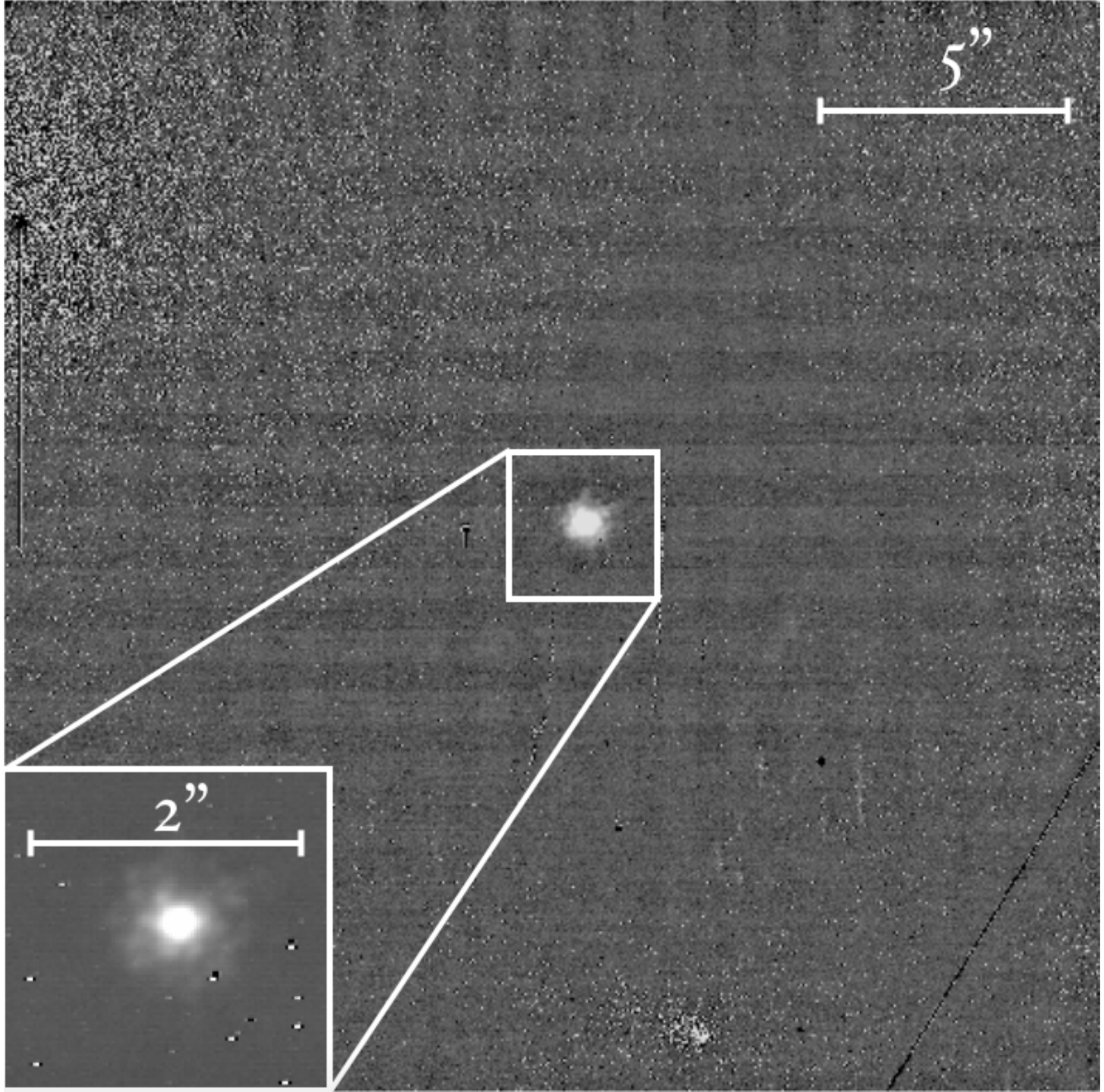} \\
    \centering\small (a) Quasar PSF 
  \end{tabular}%
 % \quad
  \begin{tabular}[b]{@{}p{0.36\textwidth}@{}}
   \includegraphics[width=1.0\linewidth]{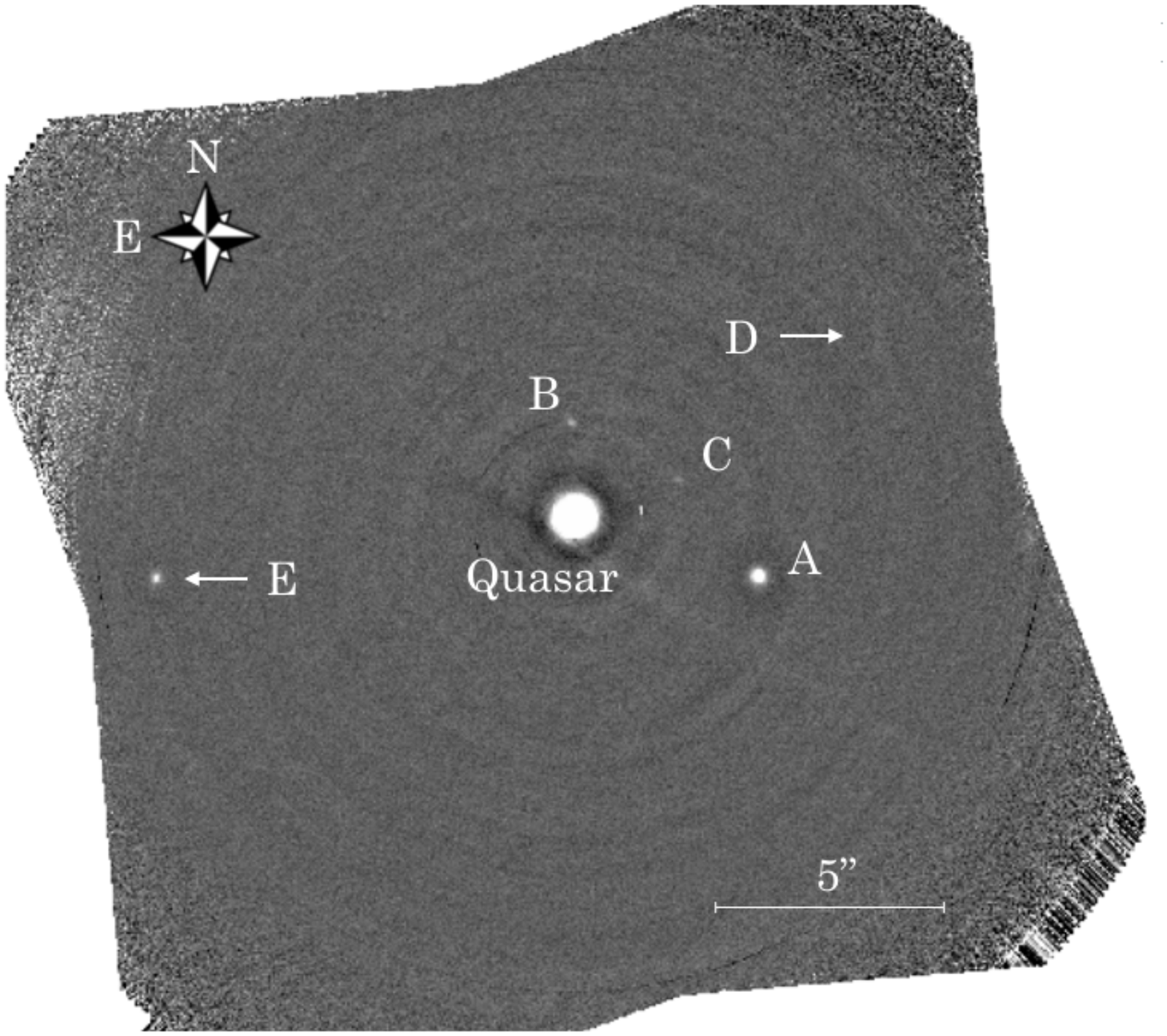} \\
    \centering\small (b)   Image without PSF subtraction
  \end{tabular}%
  %\quad
  \begin{tabular}[b]{@{}p{0.36\textwidth}@{}}
    \includegraphics[width=1.0\linewidth]{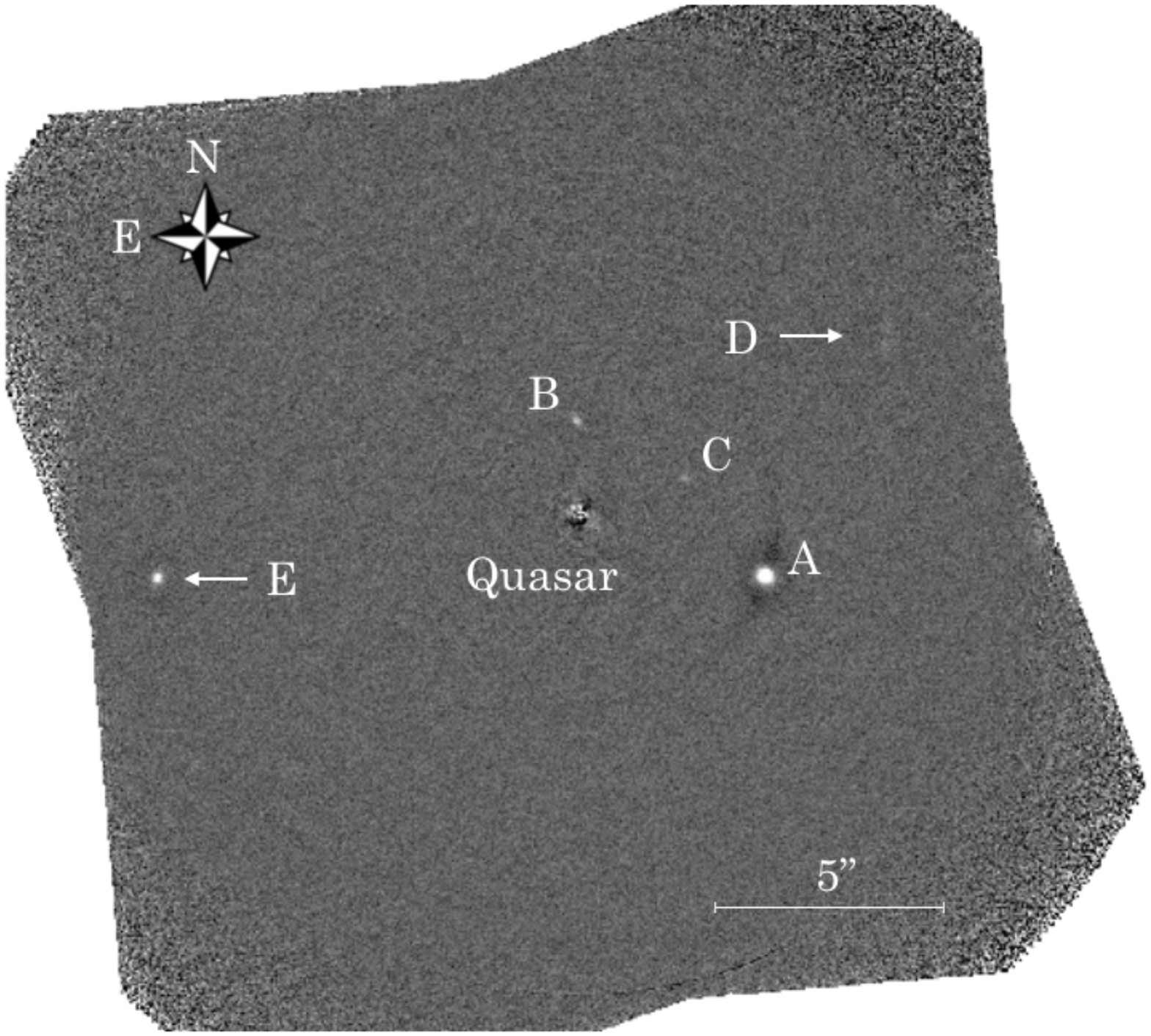} \\
    \centering\small (c) Final Image with PSF subtraction
  \end{tabular}
  \caption[Final Images]{Quasar PSF (a), and final reduced image without (b) and with (c) PSF subtraction. Objects considered for this study are marked A, B and C.  Two other new objects (D and E) at higher impact parameters are indicated with arrows, but are not considered further in this paper.  The central object labeled 'Quasar' in frame (c) is the residual image of the quasar after PSF subtraction. The dark arc around the brightest object in frame (c) is caused by the residual signature from the median of the object as it moved in PA.  The faint diagonal line slightly below the quasar in frame (b) is from a detector artifact that was not well subtracted by the dark frame.  No large scale structure is seen around objects A, B or C.}
  \label{fig1}
\end{figure}

The final reduced and combined NIRI image is shown in Figure~\ref{fig1}, in which 5 objects (A, B, C, D and E) are detected with angular separations from the QSO ranging from 2 to 10 arcsec.  Based on previous DLA host galaxy studies, impact parameters are typically a few tens of kpc with higher log N(HI) systems being systematically closer to the QSO \citep[e.g.][]{moller98,chen,cooke,monier}.  \citet{rao} find a median impact parameter for DLA hosts of 17.4 kpc and \citet{reeves} note that DLA discoveries with impact parameters greater than 20 kpc are 'extremely rare'.  Larger separations have been concluded to more likely be associated with multiple absorbers \citep{ellison07}.  On the basis of these studies, and the high log N(HI) of QSO J1431+3952,  an upper limit of $b$ = 30 kpc, or $\sim$ 5 arcsec at a redshift of $z=0.6$, will be used to distinguish candidate host galaxies from field interlopers in this study.  Three of the five objects in our reduced image (A, B and C) fulfill this impact parameter criterion.  Object D is located to the northwest of the quasar at 8.1$\pm$0.1" (55.8$\pm$0.7 kpc) and object E to the east (9.70$\pm$0.07", 66.8$\pm$0.5 kpc). Given their large impact parameters, D and E are excluded from further study in this work.   

\begin{figure}[h!]
\epsscale{.65}
\plotone{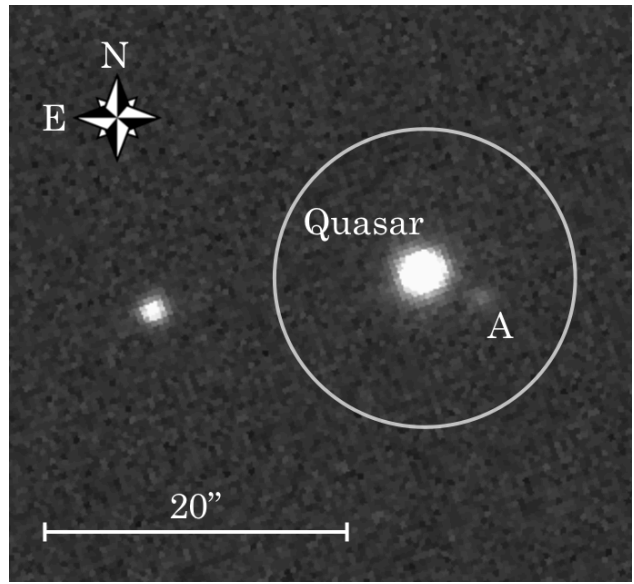}
\caption[SDSS Image]{Section from an SDSS Image, field 391, run 3813, camcol 3, i-band.  The NIRI field-of-view visible for all angles of rotation is indicated by the white circle, which is centred on the quasar.  The bright object east of the quasar is not visible in our field of view.  Object A in this figure is the same object as the one labelled "A" in Figure~\ref{fig1}.   North is up and East is left. }
\label{sdss}
\end{figure}

The brightest of the three objects considered in this study, A (at an impact parameter of $\sim$ 30 kpc), was faintly visible before the raw images had been processed and is also visible in archival SDSS imaging, which we show for comparison in Figure~\ref{sdss}.  This is the same galaxy identified by Ellison et al. (2012) and Zwaan et al. (2015).  However, as mentioned above, the photometric redshift of galaxy A ($z=0.08$) is inconsistent with that of the DLA ($z_{abs}=0.602$).  Galaxies B and C (at impact parameters $\sim$ 15 and 17 kpc respectively) are identified in our NIRI image for the first time.  We return to the discussion of the most likely absorber in Section~\ref{qsodisc}.  The darker arc around object A is due to a residual signature from the median of the object as it is moved in position angle and the dark circular halo is due to the unsharp mask.

%---------------------------------------------------------------------------------------------------------------------

%----------------------------------------------------------------------------------------------------------------------
%----------------------------------------------------------------------------------------------------------------------
% Figure 2: SDSS Field

%----------------------------------------------------------------------------------------------------------------------

%Subsection
\subsection{Candidate Galaxy Simulation and host parameters}\label{qsosimul}

Having made basic identifications of three candidate absorbing galaxies within our impact parameter search radius, more precise measurements of their magnitudes, locations and FWHM were determined through forward modelling.  This technique involved creating replicate objects with varying parameters, inserting them in the individual images and then determining which of the simulated objects best matched the real objects after going through image subtraction and combination.  The advantage of this method is that it accounts for any missing flux that is removed with combining the images and PSF subtraction.  For example, if an extended object overlaps in adjacent images, some flux would be lost in the final image, and negative wings would be apparent.  However, if a model was created for each image, not just the combined, it would need to exactly match the actual object to precisely reproduce the final image.

First, the initial parameters of the objects were determined by fitting a 2D gaussian to the three candidate galaxies in the final reduced image.  Simulated images were then created by placing 2D gaussian objects created from the initial parameters at the same location as the candidate galaxies in each exposure in an empty frame with the same dimensions as the images.  In order to account for nonlinear rotation during the 60 second exposure, instead of placing one gaussian at the object's location, five gaussians were placed to match the rotation of the parallactic angle, as five was sufficient to span the spread in parallactic angle.  This made the model slightly smeared along the direction of rotation.  While this wasn't necessary for the modelling of the fainter objects, B and C, at low impact parameters, it was critical for the brightest object, A, which was at the highest impact parameter.  These simulated gaussians, whose total flux was normalized to one, were convolved with the PSF of the quasar.  Each simulated frame was then processed in the same manner as the real exposures were -- PSF subtraction over the entire image, rotate north up, median combine all exposures from each night and combine nights with a SNR weight.  Each object was multiplied by a scaling factor to best match its intensity to its pair's real intensity.  The simulated image was subtracted from the original image.  If the objects had been modelled perfectly, there would be no residual image left and the noise would match the background level.  To test this, the standard deviation in a 1.5$\lambda$/d aperture was found at the location of the object and compared to the standard deviation in the noise at a similar sized annulus around the aperture.  Each object's location, FWHM, and scaling factor was adjusted until the noise was minimized in the aperture.  For each object, optimization of the parameters reached the background noise level.  The parameters from the models that minimized the noise were then taken to calculate the magnitudes and locations for each object.   Position angles (PA) were calculated counterclockwise from the North axis. 

\begin{figure}[h!]
\epsscale{.85}
\plotone{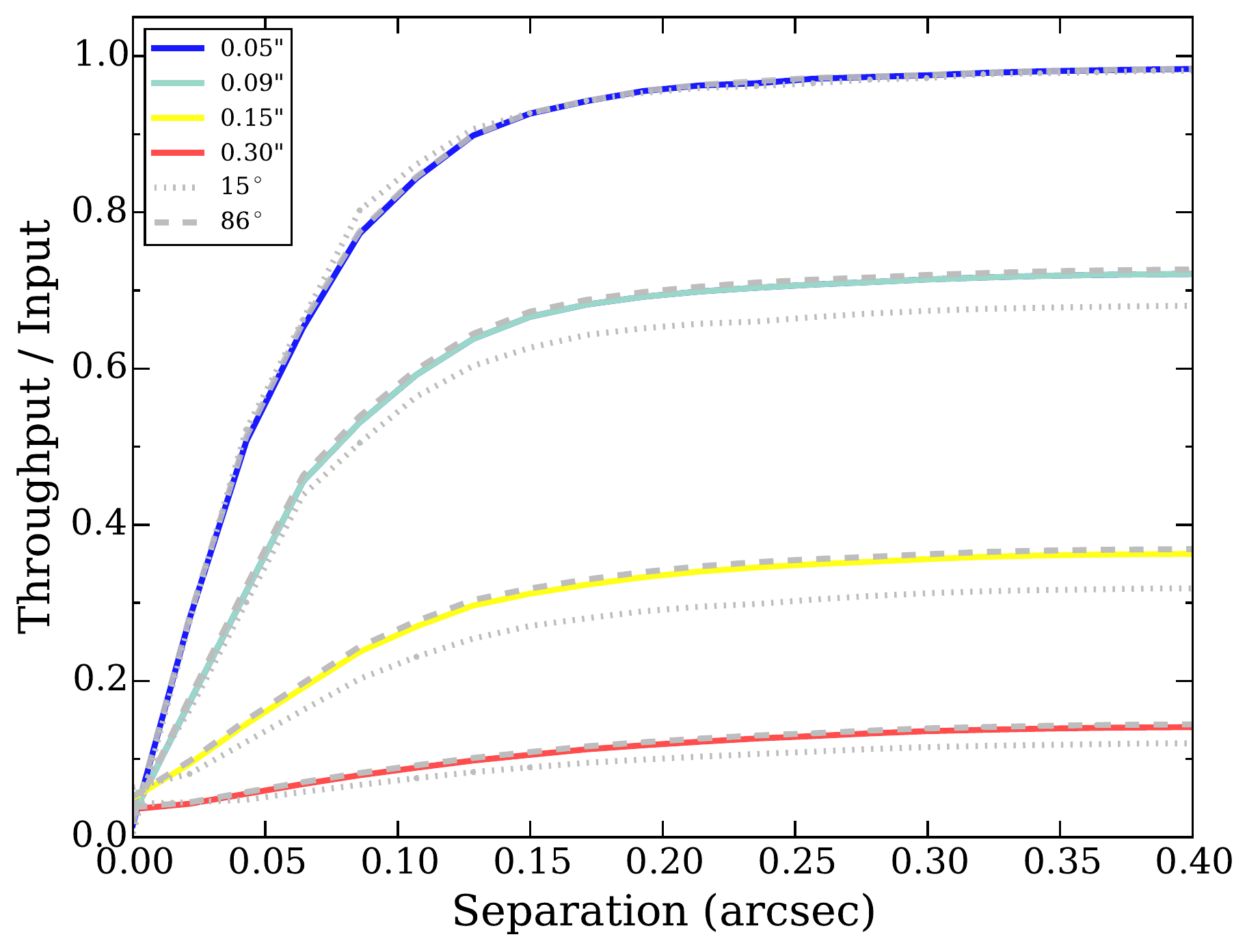}
\caption[Throughput Modelling]{Throughput as a function as radius and model object size for both nights.   Grey dashed and dotted lines indicate the separate throughput for each night (different FOV rotation) and the coloured line shows the combined throughput.   The throughput varies sightly for the different field-of-view rotations. The throughput only drops below 90\% for the inner 0.13 arcsec (0.88 kpc) when the model FWHM is 0.05 arcsec.  For larger models, the throughput drops significantly with a maximum of 37\% for a 0.15" FWHM object.}
\label{throughput}
\end{figure}

Since throughput can decrease at low impact parameters, simply measuring the flux at small annuli can be misleading.  To adjust for the reduction in throughput due to partial self-subtraction, a calibration was performed by inserting artificial unresolved (FWHM of 0.05") model galaxies (gaussians normalized to one) at different separations from the centre.  These models were inserted in the same manner as described above, except that instead of placing the models in the same locations as the objects, they were placed at the same PA with different separations.  Measuring the models' peak flux in the final image, as opposed to the peak flux of the original inserted model gives the percentage of throughput.  The throughput was measured for each night separately as the nights had different field-of-view rotations, which effects the throughput.  The throughput as a function of radius and model object size can be seen in Figure~\ref{throughput}.  A polynomial was fit to the throughput percentage as a function of separation and was used to adjust adjust each pixel in the original image based on the pixel's separation from centre of the image.  These adjustments were applied to the stacked image from each night before they were combined together.  Once each night had been adjusted, they were combined with a weighted mean and the dispersion was calculated at each annulus.  The throughput corrections only affected the inner 1 kpc of the image when correcting with a 0.05" FWHM model galaxy.  For the final images, a throughput correction using a 0.05" model was applied, although others were tested (Figure~\ref{throughput}).  The unsharp mask removes the low spatial frequencies of the larger, diffuse models, significantly reducing the flux throughput.  However, the simulated models for the throughput are smooth gaussians, they are not perfect galaxy analogues; galaxies typically have a higher concentration of light at the core. Thus, these throughput models represent the worse-case-scenario of an extremely diffuse galaxy with no central bulge.  Any real galaxy likely has a more compact profile, and even more structure, like spiral arms, which would increase the throughput and detectability. A more detailed study of throughput using real galaxy images is beyond the scope of this study.
%----------------------------------------------------------------------------------------------------------------------
% Figure: Throughput

%----------------------------------------------------------------------------------------------------------------------

To determine the uncertainties associated with the astrometric measurements of location, FWHM and the intensity scaling factor from the simulated images, 10 model templates were created in the same manner as described above, except instead of placing the models at the same position angles as the objects, they were placed at different position angles and the separations of the original galaxies.  Again, exposures and nights were combined and scaled in the same manner and subtracted from the final image.  The locations and FWHM for each of the 10 images were recovered and measured, and their standard deviation was taken as the error.  For the magnitude uncertainty, an additional intensity scale model (at the same scaled intensity of the model that matched the object) was subtracted from the models at different PA.  The magnitude uncertainty from modelling was added in quadrature with the uncertainty of the quasar's magnitude for the total magnitude uncertainty.  

The three candidate host galaxies (A, B and C) were characterized using the measurements from the simulated images.  To find the galaxies' magnitudes, the simulated image was flux calibrated using an image combined without the PSF subtraction and the quasar magnitude from a photometric catalogue \citep{schneider}. The apparent K-band magnitudes of the galaxies A, B and C were 16.74$\pm$0.06, 19.8$\pm$0.1 and 20.4$\pm$0.2 respectively.  The impact parameters of the galaxies are  30.370$\pm$0.004, 14.7$\pm$0.4, 17.3$\pm$0.3 kpc, assuming the redshift of the DLA.  A full list of the model-derived parameters can be found in Table~\ref{galxprops}. 

The FWHM of the gaussian modelled to each object was used to determine the angular size of the object.  By taking the FWHM found through forward modelling, we have allowed the FWHM to be a free parameter and thus are able to empirically derive the object size.  The FWHM of the quasar was 0.050" $\times$ 0.056", and is taken as the final angular resolution achieved with AO, consistent with a diffraction limited PSF produced by the AO system (theoretical value of 0.067").  Objects B and C were larger (0.058$\pm$0.006" $\times$ 0.079$\pm$0.008" and 0.075$\pm$0.007" $\times$ 0.057$\pm$0.006") than the angular resolution but only in one direction, which suggests these objects may be elongated. Object A  was significantly smaller (0.01$\pm$0.03" $\times$ 0.01$\pm$0.03") than the resolution and thus is clearly unresolved.  The FWHM sizes of objects B and C  correspond to $\sim$$0.3-0.5$kpc.  Although these sizes seem quite small for a galaxy at the redshift of the absorber, it is likely this measurement corresponds to only the bulge of the galaxy which is bright enough for our detection; the actual galaxy could be much larger.

\subsection{Mass Calculations}\label{qsomass}

The K-band magnitudes of the galaxies were converted to approximate stellar masses using the luminosity to stellar mass relation detailed in \citet{long}.  The conversion derived in this paper is to deal specifically with early type galaxies and assumes a formation redshift of 4.  To convert the magnitudes into masses for a certain wavelength, $\lambda$, the following equation is used:
\begin{equation}
log(M_{gal}) = log(M/L_\lambda)+0.4kcor_\lambda +2 log(d_{pc}) - 2.0+0.4M^\sun_\lambda - 0.4m_\lambda ,
\label{eq:mass}
\end{equation}
where $M/L_\lambda$ is the mass to light ratio solar units, $kcor_\lambda$ is the k-correction, $d_{pc}$ is the distance in parsecs, $M^\sun_\lambda$ is the absolute magnitude of the Sun, and $m_\lambda$ is the apparent magnitude of the galaxy.  $kcor_\lambda$ is calculated with the absorber redshift and parameters detailed in \citet{long}.  For the K-band,  $M^\sun_\lambda$ is 3.41 \citep{allen}.  The mass to light ratio assumes the redshift  ($z$) derived from the DLA spectra and various initial mass functions (IMF):

\begin{equation}
(M/L_\lambda)=a_0 +a_1z +a_2z^2 +a_3z^3 ,
\label{eq:ml}
\end{equation}

where the coefficients $a_i$ are parameterized for six different IMF models.  For a Chabrier IMF and  \citet{bruzual} models, the coefficients in order are: 0.03, 0.10, -0.008, 0.0004. The Chabrier model based on code from \citet{bruzual} is used throughout this paper for comparison with models from \citet{christensen} and \citet{maiolino}.  Log stellar masses of 11.1, 9.9 and 9.7 M$_\sun$ are calculated for objects A, B and C respectively.  Systematic uncertainties in mass-luminosity relationships due to stellar population synthesis (SPS) and IMF models can be hard to quantify.  Previous studies have documented uncertainties to be  $\sim$0.2 dex \citep{bell,mobasher, mendel} to upwards of $\sim$0.3 dex \citep{conroy}.  The difference in masses for the objects presented in this paper, calculated with different IMF models and codes, yield a spread in $0.18-0.19$ dex, which is comparable to previous works. 
 \citet{long} derive their mass retrievals for their mass estimators using two mock galaxy catalogues.  For the K-band mass estimates used in this paper, \citet{long} find 1$\sigma$ errors of $\sim$$30\%$ from their mock galaxy catalogues. These errors are similar though slightly larger to the results of other studies, perhaps due to \citeauthor{long}'s (\citeyear{long}) use of only one band whereas other studies looked at multiband SED fitting.  We adopt a conservative $30\%$ systematic error which is added in quadrature to the photometric error in calculating the final mass with uncertainty.  A list of the masses and their uncertainties for each object can be found in Table~\ref{galxprops}. 
 
\begin{figure}[h!]
\epsscale{.75}
\plotone{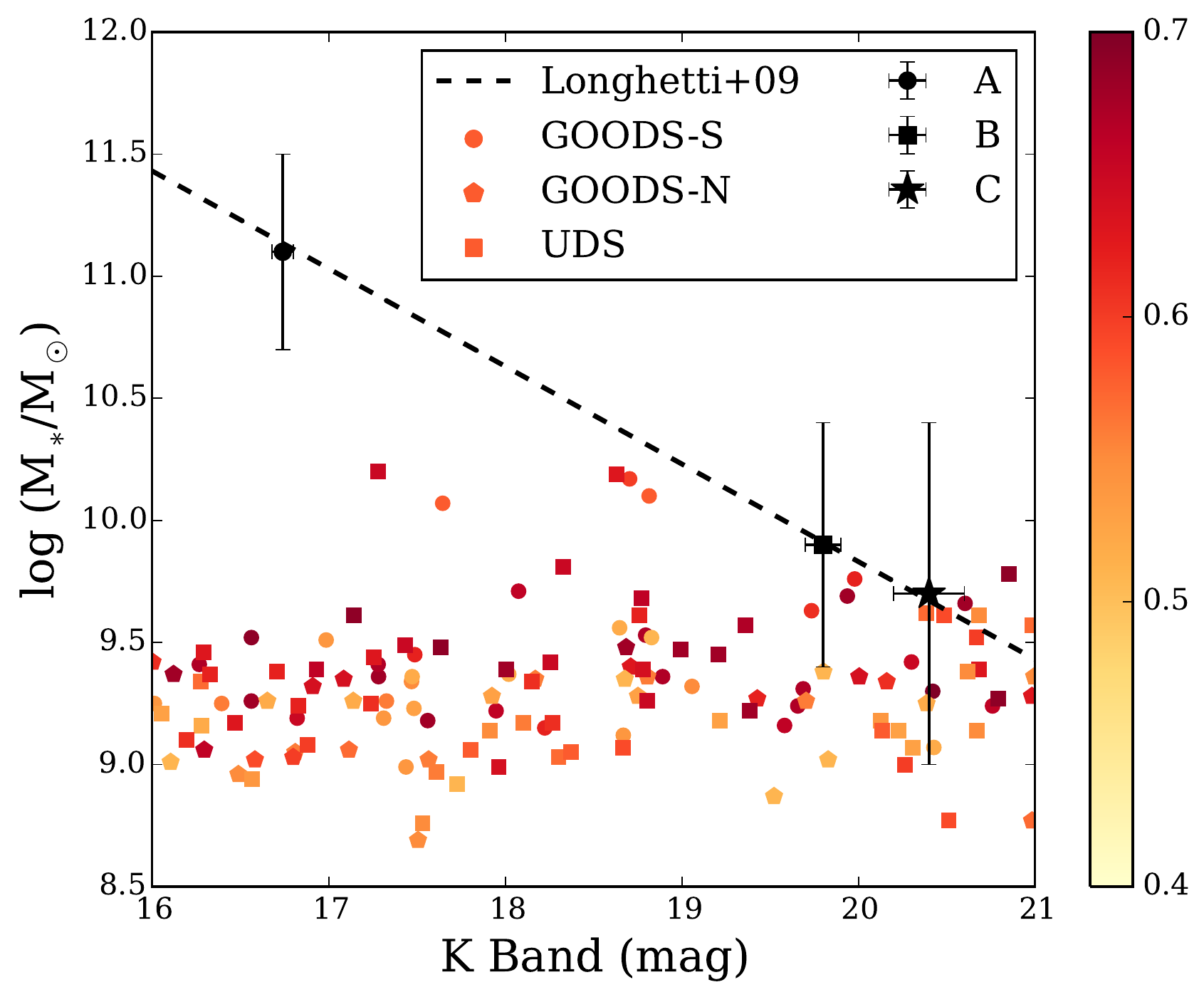}
\caption[Galaxy Mass Comparison]{Objects from this study shown with galaxies from the 3D-HST catalogue.  Galaxy colours indicate the redshift of the galaxy.  Shapes are indicative of the different fields from the catalogue. }
\label{catcompare}
\end{figure}

To check if the assumptions under the \citet{long} were valid, the masses and magnitudes obtained for each object were checked with galaxies in the GOODS-S, GOODS-N and UDS fields in the 3D-HST catalogue \citep{skelton,brammer}.  Galaxy masses in this catalogue are calculated with the FAST code \citep{kriek} using the \citet{bruzual} stellar population synthesis model library with a Chabrier IMF at solar metallicity.  Spectroscopic redshifts were used when available, and photometric redshifts when spectroscopic data was unavailable for dim objects.  Compared with the \citet{long} estimates, the 3D-HST catalogue shows lower masses at brighter magnitudes (see Figure~\ref{catcompare}).  However, objects B and C are consistent with the 3D-HST galaxies, suggesting that the mass estimate is within reason.

%----------------------------------------------------------------------------------------------------------------------
% Figure 
%---------------------------------------------------------------------------------------------------------------------

Our derived sizes and masses can be compared to known galaxy scaling relations.  In the local universe, galaxies with masses similar to objects B and C (log(M$_*$) = 9.9 and 9.7  M$_\sun$ respectively) are found to have effective radii of 1 -- 1.5 kpc, in contrast to the measured FWHM values of 0.3 -- 0.5 kpc.  However, at higher redshifts, galaxies tend to be more compact, with sizes up to a factor of two smaller at $0.5<$ z $<1$ than in the local universe \citep[][and references therein]{fan}.  The sizes measured for our galaxies are therefore plausibly in the range expected at these redshifts.  Indeed, in \citet{van14} there are several cases of $\sim$ 10$^{10}$ M$_\sun$ galaxies with sizes less than 0.5 kpc at intermediate redshifts.  We conclude that the sizes derived from our observations are not in conflict with galaxy size-mass relations at $z \sim 0.6$, supporting their candidacy as the absorbing galaxy.

\subsection{Sensitivity Curves}\label{qsocurves}

With formulations connecting K-band magnitudes and stellar mass, it is possible to assess the sensitivity of our observations to detections as a function of M$_{\star}$ and impact parameter. The detection limits were found by calculating the dispersion in a 1.5$\lambda$/D width annulus using the IDL \texttt{robust\_sigma.pro} routine using the final image that had been throughput corrected (see Section~\ref{qsosimul}) so as to accurately reflect the sensitivity at low impact parameters. The dispersion was flux normalized to the quasar's flux and converted into apparent magnitude.  A five sigma detection level, or five times the dispersion of the background flux, is assumed to ensure detection. 

\begin{figure}[h!]
\epsscale{.75}
\plotone{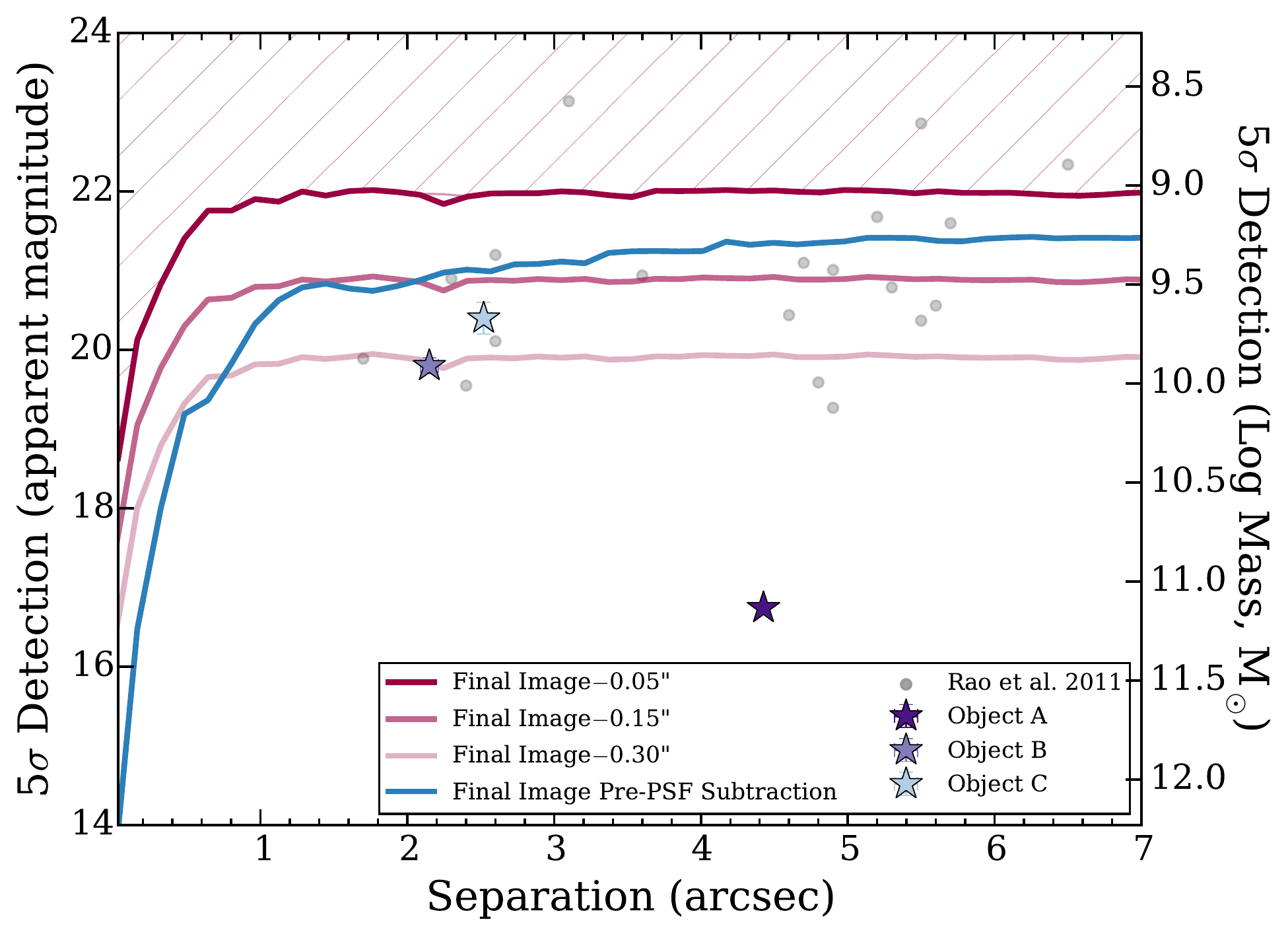}
\caption[Sensitivity Curves]{Parameter space sensitivity for DLA host detection using the methods presented in this paper.    The red curve indicates the detection limits of this study, such that objects with combinations of K-band magnitude and impact parameter in the hashed region could not have been detected.  The three red lines indicate the detection limits after correcting throughput for different model galaxy FWHM sizes.  The blue curve shows the detection limit for the final image that has not been PSF subtracted and with a throughput correction of a 0.05" model galaxy.  The quasar's PSF limits the detection level, particularly at low impact parameters.  Masses on the right hand y-axis were converted from the apparent magnitudes on the left hand y-axis using the method of \citet{long} described in Section~\ref{qsomass}.   The three detected galaxies within our impact parameter threshold are shown at their measured magnitudes (and stellar mass equivalent) and separations as star symbols.  All magnitudes are in the K-band; photometric errors are typically smaller that the symbol.    For comparison, we also show previously identified DLA host galaxies compiled in \citet{rao}.    }
\label{contrast}
\end{figure}

The sensitivity curve is shown in Figure~\ref{contrast}, were the hashed region shows the combination of K-band magnitude and impact parameter of galaxies that could not have been detected in our final image.  For images that have not been PSF subtracted, the limit for detection is noticeably shallower, especially at low impact parameters, i.e. the PSF subtraction dramatically helps increasing detectability at low impact parameters where DLA hosts are most likely to be.  The PSF subtraction additionally helps at larger radii where it acts as a background sky subtraction, which is why the PSF subtracted sensitivity curve never overlaps the non-PSF subtracted curve.  The three galaxies (A, B and C) within our impact parameter threshold are shown as star symbols, where the errors are typically smaller than the symbol size.    We note that the range of M$_{\star}$ and impact parameter space in which candidates could have been detected extends to even lower values than those exhibited by the galaxy candidates detected in this study.  For example, we could have detected a $10^{9.2}$ M$_{\odot}$ galaxy (approximately 1/10 the mass of the Milky Way) as close as 0.5 arcsec (3.4 kpc).  At impact parameters beyond $\sim$ 6 kpc, our mass sensitivity plateaus at $\sim$ log $10^{9.0}$ M$_{\odot}$.  However, a caveat to this method of deriving sensitivity curves is that it is more accurate for galaxies that are close to being unresolved point sources, as extended galaxies with more diffuse emission are harder to detect after PSF subtraction.  This is shown with the three red curves which use model galaxies of 0.05" to 0.30" FWHM to correct for throughput (see Section~\ref{qsosimul}).  For throughput correction with a larger (0.30"), diffuse model galaxy, the detection limit decreases by an of magnitude.  Although this may seem significant, this is a worse-case scenario as most galaxies will not be entirely diffuse and would still have a visible dense core or spiral arms.  For comparison, Figure~\ref{contrast} also shows the positions of DLA host galaxies detected in \citet{rao} for comparison.  Host galaxies were only selected from \citet{rao} if they had K-band magnitudes, b $\textless$ 10 arcsec, and have been confidently identified using spectroscopic redshift, photometric redshift or colours. 

%----------------------------------------------------------------------------------------------------------------------
% Figure 
%---------------------------------------------------------------------------------------------------------------------

\section{Discussion}\label{qsodisc}

Until now, the only object previously identified within 50 kpc of the J1431+3952 quasar sightline was object A  \citep{ellison12,zwaan}, see Figure \ref{sdss}.  Since it has been previously proposed  \citep{kanekar}  that DLAs with low spin temperatures and high metallicities such as this one are associated with relatively luminous disk galaxies, object A, which we have estimated in this work to have a stellar mass log (M$_{\star}$/M$_{\sun}$) $\sim$ 11 might initially seem like a compelling candidate.  However, its low photometric redshift \citep[$z=0.08$,][]{ellison12} is inconsistent with the DLA redshift ($z_{abs}=0.602$)\footnote{Results in \citet{abazajian} indicate that SDSS galaxies assigned a photometric redshift of 0.1 never have spectroscopic redshifts above 0.3.}.  The photometric redshift was based on limited SDSS photometry, so it is nonetheless interesting to further explore the possibility of A as the DLA host. With the identification of additional candidates B and C, we can further explore the likelihood of each as the absorbing galaxy, by considering scaling relations of various physical properties. 

\begin{figure}[h!]
\epsscale{.75}
\plotone{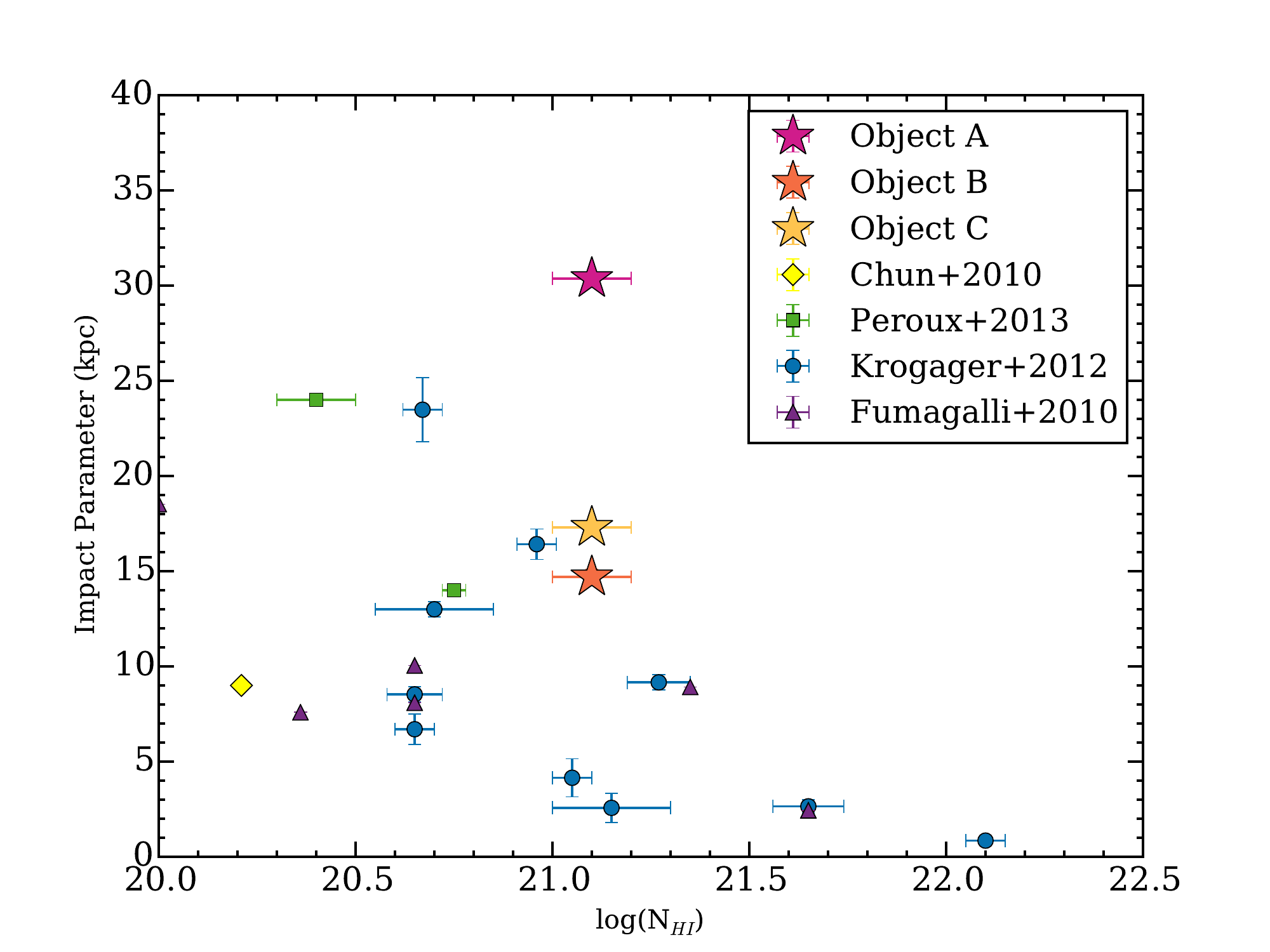}
\caption[Impact parameter -- N(HI) Column Density Relation]{Impact parameter plotted with log HI column density for the three objects in this study as well as four other studies \citep{chun10,peroux13,krogager,fum}}
\label{nhi-b}
\end{figure}

We begin by exploring the trend between impact parameter and neutral hydrogen column density of DLA hosts.  This trend has previously been noted by several studies which find an inverse correlation between the neutral hydrogen column density and impact parameter \citep[e.g.][]{krogager,moller98,peroux,monier, rao,christensen07}.  For example, \citet{rao} report the anticorrelation at a 3$\sigma$ level significance.  Figure~\ref{nhi-b} shows the results from our study combined with previously identified and confirmed  DLA hosts  \citep{chun10,peroux,krogager,fum}.  Objects B and C are consistent with the general anti-correlation seen in previous studies, whereas object A is at a significantly higher impact parameter than would be expected for its N(HI), even given the high spread in the trend.   In addition to its inconsistent photometric redshift, Figure~\ref{nhi-b} therefore adds further evidence against object A being the DLA host.

%----------------------------------------------------------------------------------------------------------------------
% Figure 
%----------------------------------------------------------------------------------------------------------------------

\begin{figure}[h!]
\epsscale{.75}
\plotone{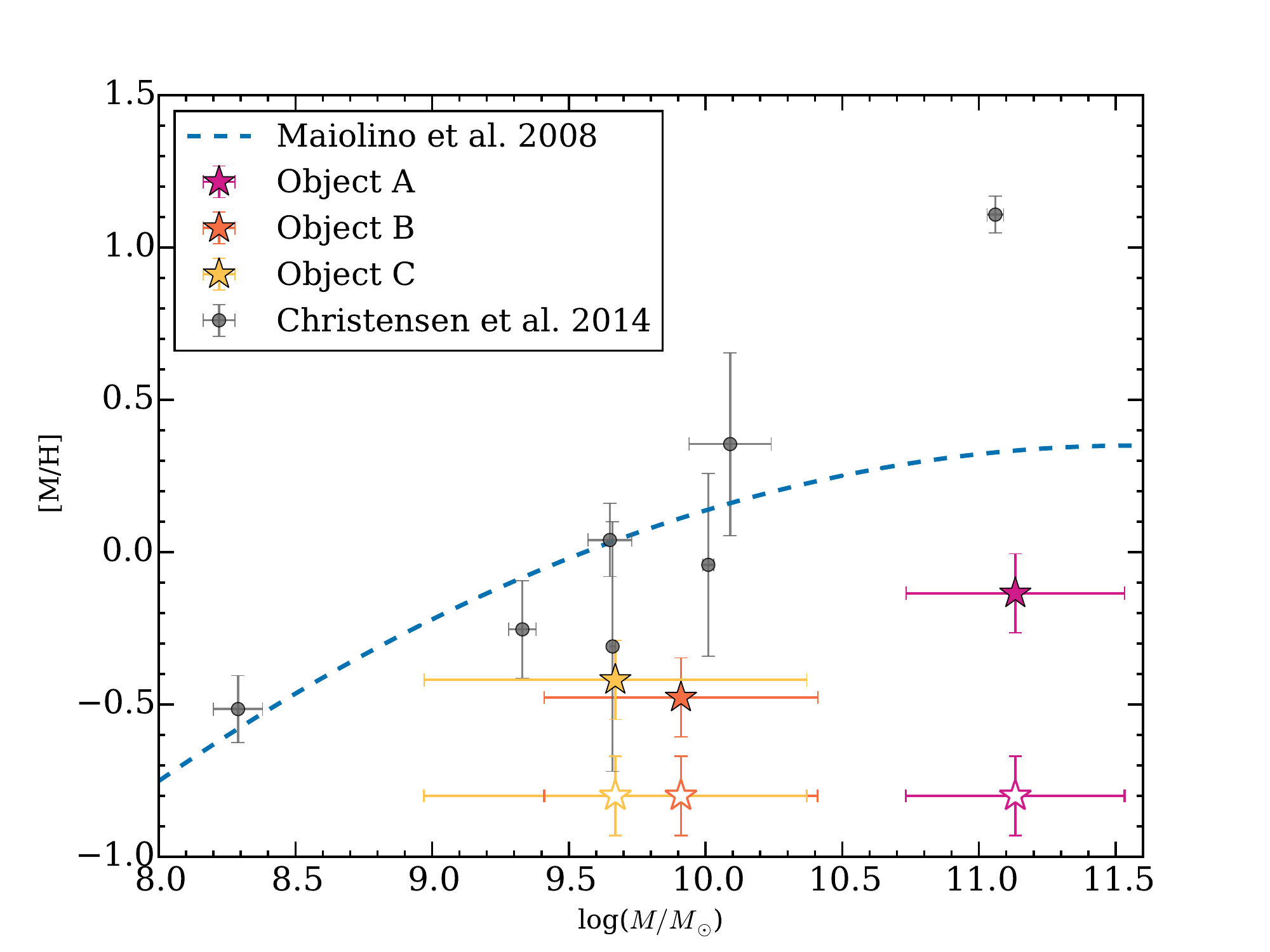}
\caption[Mass-Metallicity Relationship]{MZR fits from \citet{maiolino} plotted with the objects from this study (open stars).  Once impact parameter adjustments from \citet{christensen} have been applied (filled stars), the objects are closer to the established MZR.  Zinc measurements from \citet{ellison12} are used as a proxy for metallicity for the three objects.  DLAs from \citet{christensen} with z $<$ 1 (median z=0.52) are shown as grey circles for comparison.}
\label{mzr-mh}
\end{figure}

Luminosity or stellar mass-metallicity (MZR) relations have been well studied across a range of redshifts starting with \citeauthor{lequeux} in 1979. Based on the high $z$ mass-metallicity relation described in \citet{maiolino}, \citet{christensen} established the relationship for DLAs at a range of redshifts and derived a correction for the impact parameter.  Specifically,  \citet{christensen} add 0.022$b$ (where $b$ is the impact parameter in kpc) to the absorption line metallicity to account for the impact parameter.  Figure~\ref{mzr-mh} shows the three candidate DLA galaxies from our study compared with the mass-metallicity relation from \citet{maiolino}.  We adopt the common approach of using zinc as our elemental tracer of metallicity, using the metallicity reported in \citet{ellison12}.    Without any correction for impact parameter, the absorption line metallicity of the DLA combined with the stellar masses of the 3 candidate host galaxies identified in this study all fall below the mass-metallicity relation presented in \citet{maiolino}, as shown by open stars in  Figure~\ref{mzr-mh}.    After the application of the impact parameter correction to the metallicity, all of the candidate host galaxies move closer to the expected MZR, although none are a particularly good match. \citet{christensen} find a scatter in their relation of 0.39 dex in log $M^{DLA}_{*}$ which encapsulates only object C.

%----------------------------------------------------------------------------------------------------------------------
% Figure 
%----------------------------------------------------------------------------------------------------------------------

\begin{figure}[h!]
\epsscale{.75}
\plotone{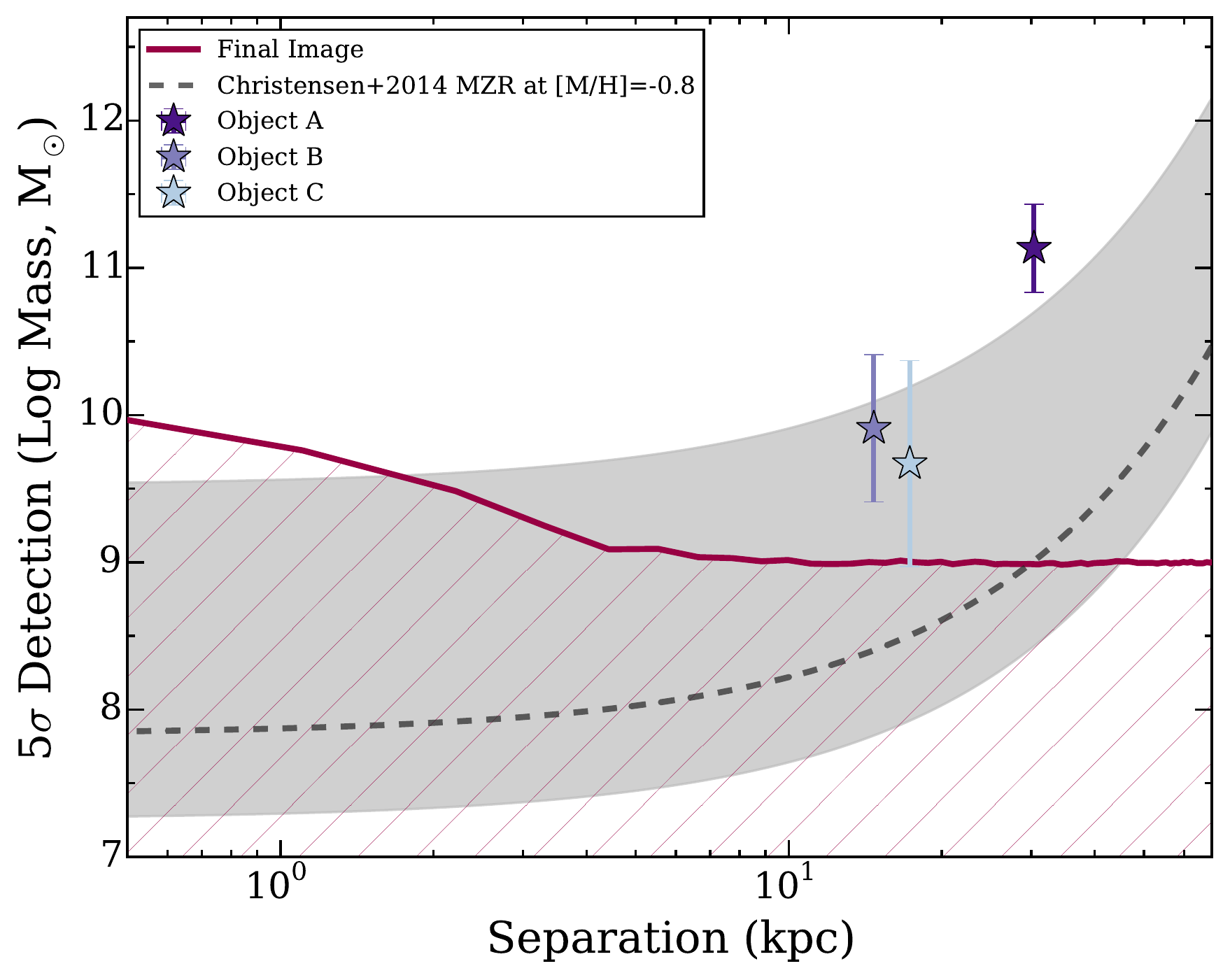}
\caption[Sensitivity Curve MZR ]{If we look at the sensitivity curves as a function of physical separation, we can see that assuming the DLA host follows the \citet{christensen} MZR with the redshift and metallicity fixed to the DLA (grey dashed line), the host would only have been detected in this study if its stellar mass log (M$_{\star}$/M$_\sun$) was greater than 9 and separation greater than 30 kpc.  The grey shaded region shows a range of MZR for other known DLA hosts in the redshift range $z=0.3-0.9$ \citep[DLA catalogue from][]{berg}.  DLA hosts following the MZR could be detected with the techniques described in this paper at impact parameters low as 1.7 kpc. For the redshift range selected, changes in the conversion of arcsec to kpc is negligible for the sensitivity curves.}
\label{contrast2}
\end{figure}

An alternative way to use the MZR relation is to assume that the host galaxy follows the relation derived by Christensen et al. (2014), and use the absorption line metallicity to determine the predicted stellar mass as a function of impact parameter.  This calculation is shown in Figure ~\ref{contrast2} as a dashed line, over-plotted with our sensitivity curve.    Figure ~\ref{contrast2} shows that \textit{if} the host galaxy follows the impact-parameter dependent MZR derived by Christensen et al. (2014), it would have been detectable only if log M$_{\star}$/M$_{\odot} >$ 9, with corresponding impact parameter $>$ 30 kpc.  Although we do have two further detections in the NIRI field at larger impact parameters (objects D and E), we have shown in Figure \ref{nhi-b} that DLAs with column densities as high as the sightline studied here (log N(HI)=21.1) are rarely found at high impact parameters.  

Although the dashed line in Figure \ref{contrast2} may seem to imply that our study is not particularly sensitive to galaxies on the MZR for this DLA, significant scatter may be expected.  We therefore use the sample of DLAs  in the redshift range $z=0.3-0.9$ with zinc detections (for metallicity comparison) from \citet{berg} to gauge the possible scatter around Christensen et al. (2014)'s best fit relation.  The result is shown in the gray shaded region in Figure \ref{contrast2}, in which it can be seen that host galaxies with sufficiently high metallicities following the MZR would be detectable to separations as low as 1.7 kpc for true point source galaxies.  Moreover, galaxies B and C lie inside this shaded region, implying that they are plausible candidates for the DLA host.  

The sensitivity limits shown here were derived for a 16th magnitude (R-band) quasar observed with an 8-meter telescope.  Observing fainter quasars would increase detection limits in the region limited by quasar PSF residual (i.e. the inner one arcsecond).  However, the current limits of AO require a 17.5 magnitude quasar or nearby star for tip-tilt corrections.  The rest of the image outside of one arcsecond, which is background noise limited, could be improved with longer integrations, as contrast increases with the square root of integration time.  As the next generation of thirty-meter telescopes become available, larger apertures will increase resolution and allow for quasars 16 times fainter to be observed with AO, allowing for detections of host DLA at 3 times lower impact parameters and 16 times fainter (3 mag deeper contrast) in the background-limited regime.  

%----------------------------------------------------------------------------------------------------------------------
% Figure 
%----------------------------------------------------------------------------------------------------------------------

%----------------------------------------------------------------------------------------------------------------------
\begin{table}[h!]
\centering
\caption{Candidate Galaxy Characteristics}
\begin{tabular}{rccc}
    \toprule
   \midrule
    \bfseries Object & 
    \bfseries A & 
    \bfseries B & 
    \bfseries C \\ 
   \midrule
    Image  & \includegraphics[width=2cm,height=3cm,keepaspectratio]{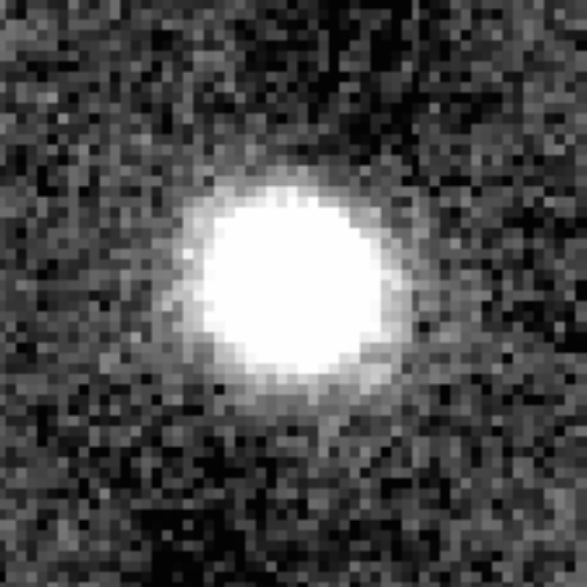}& \includegraphics[width=2cm,height=3cm,keepaspectratio]{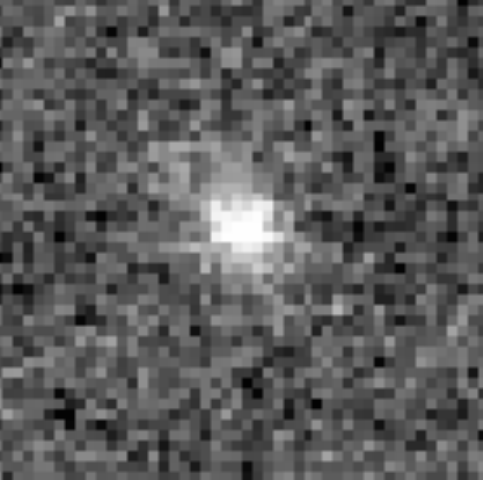} & \includegraphics[width=2cm,height=3cm,keepaspectratio]{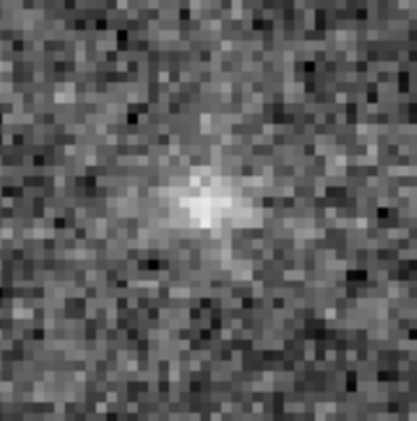}\\
  Magnitude (K band) & 16.74$\pm$0.06 & 19.8$\pm$0.1& 20.4$\pm$0.2 \\
Impact Parameter (arcsec) & 4.4260$\pm$0.0006 & 2.14$\pm$0.06 & 2.52$\pm$0.05\\
Impact Parameter (kpc) & 	30.370$\pm$0.004	 & 14.7$\pm$0.4 &17.3$\pm$0.3	\\
Position Angle (deg) &251.701$\pm$0.009 & 2.61$\pm$0.08 & 289.32$\pm$0.07 \\
Log Mass (M$_\sun$)& 11.1$\pm$0.4 & 9.9$\pm$0.5 & 9.7$\pm$0.7\\ 
  \midrule
   \toprule
   \tablecaption{Candidate Galaxy Characteristics}
\tablecomments{Impact parameter in kpc calculated assuming the redshift of the absorber.}
\end{tabular}
\label{galxprops}
\end{table}
%----------------------------------------------------------------------------------------------------------------------

From comparisons to scaling relations between impact parameter and N(HI), and stellar mass with metallicity, we have shown that galaxies B and C (log M$_{\star}$/M$_{\odot} =$ 9.9, 9.7 respectively) are plausible candidates for the host galaxy of the DLA towards J1431+3952.  Alternatively, the host could be an as yet unidentified galaxy with relatively low stellar mass, with the MZR constraining the limit to be log M$_{\star}$/M$_{\odot} <$ 9.  In none of these scenarios is the host galaxy particularly massive, falling several times below the mass of the Milky Way.  Kanekar \& Chengalur (2003) noted that low T$_s$ DLAs tend to be associated with luminous ($l \sim$ L$_{\star}$) galaxies.  The DLA towards J1431+3952 appears to be an exception to this trend.

We finish this discussion with a review of several caveats to the mass determinations presented here, such as the adopted mass-light ratios which rely on the assumption of an early type galaxy.  Early types have effectively no dust, so using the K-band magnitude can work well as an overall magnitude estimator (K. Thanjavur 2015, private communication).  Younger galaxies have more dust and are more prone to extinction effects which can lead to an underestimate of magnitude and thus an underestimate of the mass.  The derivations were also done for a specific formation redshift, though \citet{long} mention they find similar results for different formation redshifts out to $z_f$=6. There is also significant uncertainty in the mass calculation, most of which comes from uncertainty in the stellar modelling and the availability of only K-band photometry.  The noise is also a strong contributor for calculating the flux from less luminous candidate galaxies.  Ultimately, photometric observations in additional bands are necessary to ascertain the true mass of the galaxies through spectral energy distribution calculations and spectrometry is needed for accurate redshift assessment of the candidate host galaxies.

%----------------------------------------------------------------------------------------------------------------------
\section{Conclusion}\label{qsoconc}

The first application of ADI to direct imaging of DLA galaxies has resulted in three candidates (within 5 arcsec, or 30 kpc at the redshift of the absorber) for the $z_{abs}=0.602$ DLA seen in the quasar J1431+3952. Determination of the sensitivity curve for our observations indicates that we could have detected a galaxy whose stellar mass was as low as $10^{9.4}M_{\odot}$ at a separation of 3.4 kpc, or  $\sim 10^{9.0}M_{\odot}$ beyond $\sim$ 6 kpc. Based on the K-band photometry of our NIRI observations, we determine stellar masses of log (M$_{\star}$/M$_{\odot}$) $= 9.9, 9.7 $ and $ 11.1$ for the three candidates, which are located at impact parameters of 15, 17 and 30 kpc respectively.   The two galaxies at the lowest impact parameters are new detections in our NIRI data.  Based on a photometric redshift of $z=0.08$ (Ellison et al. 2012), the unresolved nature of the object, and inconsistency with the N(HI) -- impact parameter relation (e.g. Krogager et al. 2012; Christensen et al. 2014), we conclude that the DLA is not associated with the highest mass, largest  separation of the three candidates.  The remaining two galaxies are consistent with these scaling relations and therefore represent plausible candidates for the DLA host.  Follow-up spectroscopy is required to confirm the redshifts of the remaining two candidates and observations in just one additional band would allow for additional mass-luminosity constraints \citep{bell01}.    Our results indicate that despite its low spin temperature, the host galaxy of this DLA is unlikely to be of high stellar mass (or luminosity).  

Since current technology limits our seeing to dwarfs in the Milky Way's local group, we have a limited understanding of dwarf galaxies across time.  Even if a bright galaxy is imaged at the right redshift and assumed to be a DLA host, the DLA could instead be originating in a satellite dwarf that is far below the threshold for detection.  Although much progress has been made in identifying DLA hosts, there is still a long way to go in understanding the full picture of DLAs.  

\newpage

\chapter{Summary}\label{sum}

High contrast regions in our universe hold much to be discovered and understood.  Through the use of techniques like angular differential imaging, we can begin to delve into these areas of study.  The work presented here has shown the applications of angular differential imaging ranging from nearby stars to the distant universe.  This chapter presents a summary of these diverse applications.

The original application of ADI had the intent of discovering new planets.  Chapter~\ref{tloci} presented new ways to optimize exoplanet detection with the TLOCI algorithm.  In a large survey, like GPIES, it is important to be able to detect the full spectrum of planet types with efficiency.  Parameter optimization tests were run with TLOCI to identify the optimal settings.  Results indicated that a standard matrix inversion, a moderate aggressivity (0.7) and a small unsharp mask (5 pixels) were the best settings to optimize planet detection.  At least two planet spectra (T2 and T8) should be used in iteration with the TLOCI algorithm in order to boost the SNR needed to detect faint new planets.  Although these settings have been found to generally give the best SNR results, factors which change between datasets, such as field-of-view rotation and seeing conditions, can effect which settings will optimize the SNR in that particular dataset.  The optimized parameters have already been incorporated into TLOCI reductions of GPIES data.  Future versions of TLOCI may include annuli-specific parameter optimization to help adjust for intra-image variations. 

Stepping away from the optimizations of TLOCI brings us to work done with GPIES.  In Chapter~\ref{bd}, new observations of the substellar object HD 984 B were presented.  HD 984 was observed with GPI in August 2015 and the substellar companion was observed in H and J bands.  The separation and position angles were found to be 216.3$\pm$1.0 mas and 83.3$\pm$0.3$\degree$ (H band), and 217.9$\pm$0.7 mas and 83.6$\pm$0.2$\degree$ (J band).  The astrometry from these observations combined with archival astrometry from 2012 and 2014 epochs helped constrain the companion's motion to an $\sim$18 AU, 70 year orbit, at a 68\% confidence interval between 12 and 27 AU, with an eccentricity of 0.24 and inclination of 118$\degree$.  Spectral analysis found the best match type to be a M7$\pm$2, in agreement with the best match (M6.0$\pm$0.5) reported in the discovery paper \citep{meshkat}.  The J and H band magnitudes of 13.28$\pm$0.06 and 12.60$\pm$0.05 were used to calculate the luminosity (log(L$_{\mathrm{Bol}}$/L$_{\sun}$) = $-2.88\pm0.07$ dex) and age dependent mass (34$\pm$1 M$_{\mathrm{Jup}}$ at 30 Myr to 94$\pm$4 M$_{\mathrm{Jup}}$ at 200 Myr) using DUSTY models.  DUSTY models of H band magnitudes gave a temperature of 2545$\pm$28K to 2896$\pm$31K (for an age range of 30 -- 200 Myr), while a spectral type-to-temperature conversion gave T$_{eff}=2673^{+175}_{-267}$.  J band magnitudes yield a DUSTY model temperature of  2458$\pm$32K to 2800$\pm$37K over the same age range.  Additionally, we presented a method of splitting the spectra into low and high spatial frequencies to reduce spectral covariance and allow for proper noise statistics and improved $\chi ^2$ analysis.  This method may prove useful with future K band data, where narrow spectral features, such as CO, can be identified and fitted.

ADI has clearly proven to be an effective tool in finding small, dim objects in our galaxy, but as Chapter~\ref{dla} shows, ADI is also a powerful way to discover new galaxies outside our local group.  This chapter presented the first application of ADI to direct imaging of DLA galaxies.  Often the overwhelming brightness of a quasar makes it impossible to see the host DLA galaxy that is evident in the quasar spectrum.  The quasar J1431+3952 was imaged in the K band using ADI and a laser guide star system at Gemini North and processed with a simple PSF subtraction.  The application of ADI was shown to be successful with three candidates discovered within 5 arcsec (30 kpc at the redshift of the absorber).  Stellar masses of log (M$_{\star}$/M$_{\odot}$) $= 9.9, 9.7 $ and $ 11.1$ were found for the three candidates, which are located at impact parameters of 15, 17 and 30 kpc respectively.  From inconsistencies with the photometric redshift, mass--metallicity and N(HI)--impact parameter relations, it is unlikely that the most massive object is associated with the DLA.  However, follow-up spectroscopy is required to identify the true host absorber.  Sensitivity curves generated from the imaging data indicate that future detections in similar systems could detect a galaxy whose stellar mass was as low as $10^{9.4}M_{\odot}$ at a separation of 3.4 kpc, or  $\sim 10^{9.0}M_{\odot}$ beyond $\sim$ 6 kpc.

As the next generation of extremely large telescopes come online, future observations of exoplanets, brown dwarfs and DLAs will be able to reach to deeper contrasts and smaller impact parameters.  The Thirty Meter Telescope is projected to see first-light in 2024 an the James Webb Space Telescope (JWST) expected to launch in 2018.  JWST will have unmatched infrared sensitivity and will be capable of transit spectroscopy with the goal of looking for atmospheres similar to Earth's.  It will additionally be equipped with a coronagraph to enable direct imaging of exoplanets and near and mid-infrared detectors to probe a wide range of spectral signatures in the exoplanetary atmospheres.  

From near to far and small to big, ADI has proven to be a robust tool for new discoveries in our universe.  GPIES continues to make observations using ADI and process data with TLOCI and other PSF subtraction algorithms.  And now that ADI has shown its worth in imaging DLA hosts, it can be used in future DLA imaging.  As the next generation of thirty-meter telescopes come online, larger apertures will allow for the detection of even fainter objects.  Teasing apart the light in these high contrast regions allows us to slowly understand more about these subjects in the never-ending quest to better understand the universe.    
\newpage

% The Bibliography 
	\TOCadd{Bibliography}

	\bibliographystyle{apj_8}
	%\addcontentsline{toc}{chapter}{\numberline{}\sf\bfseries{Bibliography}}

\end{document}

%% file: macros/style.tex
\usepackage[top=1.3in,left=1.5in,bottom=1in,right=1in,headsep=0.5in]{geometry}

\usepackage{setspace}
\usepackage{fancyhdr}

\newcommand\TOCadd[1]{\newpage\phantomsection\addcontentsline{toc}{chapter}{#1}}

\usepackage{subfig}	% MT: Supports subfigures
\usepackage[titles,subfigure]{tocloft} % MT: I added the square brackets
\newcommand{\HRule}{\rule{\linewidth}{0.5mm}}

%% file: macros/use_packages.tex
\usepackage{longtable}

\usepackage{amsmath}
\usepackage{amsthm}
\usepackage{amssymb}

\usepackage{xspace}
\usepackage{textcase}

\usepackage{wasysym}
\usepackage{graphics}
\usepackage{graphicx}   % A package to allow insertion of